\pdfoutput=1
\documentclass[a4paper,11pt]{article}
\usepackage{heppub}
\usepackage{xcolor}

\graphicspath{{plots/}}

\newcommand{\WidthTwoSubfigs}{0.5\textwidth}

\newcommand{\GeV}{\,\mathrm{GeV}}
\newcommand{\TeV}{\,\mathrm{TeV}}

\newcommand{\nn}{\nonumber}

\newcommand{\pts}{\mathrm{pts}}

\newcommand{\dd}{\mathrm{d}}

\newcommand{\ms}{\mskip 1.5mu}
\newcommand{\bs}{\mskip -1.5mu}

\newcommand{\chili}{\textsc{ChiliPDF}}

\allowdisplaybreaks[1]

\newcommand{\rev}[1]{#1}

\title{ChiliPDF: Chebyshev Interpolation for Parton Distributions}

\author[a]{Markus Diehl,}
\author[b]{Riccardo Nagar,}
\author[a]{and Frank J.~Tackmann}

\affiliation[a]{Deutsches Elektronen-Synchrotron DESY, Notkestr.~85, 22607 Hamburg, Germany}
\affiliation[b]{Universit\`a degli Studi di Milano-Bicocca \& INFN Sezione di Milano-Bicocca,\\
Piazza della Scienza~3, Milano 20126, Italy
}

\emailAdd{markus.diehl@desy.de, riccardo.nagar@unimib.it, frank.tackmann@desy.de}

\abstract{%
Parton distribution functions (PDFs) are an essential ingredient for theoretical
predictions at colliders. Since their exact form is unknown, their handling and
delivery for practical applications relies on approximate numerical methods. We
discuss the implementation of PDFs based on a global interpolation in terms of
Chebyshev polynomials. We demonstrate that this allows for significantly higher
numerical accuracy at lower computational cost compared with local interpolation
methods such as splines. Whilst the numerical inaccuracy of currently used local
methods can become a nontrivial limitation in high-precision applications, in
our approach it is negligible for practical purposes. \rev{This holds in particular
for differentiation and for Mellin convolution with kernels that have end point
singularities. We illustrate our approach for these and other important numerical
operations, including  DGLAP evolution, and find that they are performed accurately
and fast.}  Our results are implemented in the C++ library \chili.
}

\date{v1: December 17, 2021 \\ v2: April 1, 2022}

\preprint{\vbox{%
\hbox{DESY 19-110}}
}

\arxivnumber{2112.09703}

\begin{document}

\maketitle

\section{Introduction}
\label{sec:intro}

Theoretical predictions at hadron colliders require parton distribution
functions (PDFs), which describe the partonic content of the colliding hadrons.
PDFs are nonperturbative objects and their exact form is unknown, such that
their handling and delivery in practical applications requires
approximate numerical methods. Currently available tools use local interpolation
over a finite grid of function values, such as splines. For example, the LHAPDF
library~\cite{Buckley:2014ana}, which has de facto become the standard interface
with which PDFs are provided to the community, implements a cubic spline
interpolation.

In this paper, we demonstrate a different approach for the numerical
representation of PDFs, which is based on a global, high-order interpolation
using Chebyshev polynomials, and which allows for significantly higher numerical
accuracy at lower computational cost than local interpolation methods. We have
implemented this approach in the C++ library \chili,%
\footnote{\underline{Ch}ebyshev \underline{I}nterpolation
\underline{Li}brary for \underline{PDF}s}
which is used for most of the numerical demonstrations in the following.
We think, however, that the methods presented in our paper are of interest
beyond their implementation in a specific software package.

There is a long history of using families of polynomials for handling PDFs or
related quantities.  Without any claim to completeness, let us mention a few
examples.  To express PDFs in terms of their Mellin moments, Bernstein
polynomials were proposed in \refcite{Yndurain:1977wz} and Jacobi polynomials in
\refcite{Parisi:1978jv}.  An explicit solution of the evolution equations was
obtained in \refscite{Furmanski:1981ja, Kobayashi:1994hy} by expanding PDFs and
splitting functions in Laguerre polynomials.  More references and
discussion can be found in \refscite{Chyla:1986eb, Blumlein:1989pd,
Krivokhizhin:1990ct}.  To compute analytically the complex Mellin moments of
parton luminosities, \refscite{Bonvini:2010tp, Bonvini:2014joa} used an
expansion in Chebyshev polynomials.  The latter also appear in the
parameterization of initial conditions in the PDF fits of
\refscite{Pumplin:2009bb, Glazov:2010bw, Martin:2012da, Harland-Lang:2014zoa}.
We find that all these methods
differ quite significantly from each other and from the method to be presented
in this work.

Let us emphasize that we are not concerned here with the question of how best to
parameterize input PDFs that are fitted to data, nor with the associated
systematic error or bias due to the choice of parameterization. For our
purposes, we consider the fitted input PDFs to be known ``exactly''. Our method
could be used to address this parameterization issue as well, but we leave this
for future exploration.

What we are concerned with is the numerical implementation and handling of PDFs
in practical applications, for instance when using PDFs evolved to some scale to
obtain quantitative predictions from analytic cross section formulae or with
Monte-Carlo event generators.  The interpolation of PDFs comes with an inherent
inaccuracy that is of purely numerical origin, similar to numerical integration
errors.  Such inaccuracies should at the very least be small compared to
uncertainties due to physics approximations such as the perturbative expansion.
But ideally, one would like such inaccuracies to be negligible or at least small
enough to be of no practical concern. We will show that this goal can indeed be
achieved with Chebyshev interpolation.

Several observations lead us to believe that the performance of available local
interpolation methods is becoming insufficient. Generically, the relative
accuracy of the interpolation provided by LHAPDF is expected (and
found to be) in the ballpark of $10^{-3}$ to $10^{-4}$. For high-precision
predictions this is already close --- perhaps uncomfortably close --- to the desired
percent-level theoretical precision. Moreover, PDFs typically appear in the
innermost layer of the numerical evaluation, whose output is then further
processed. For example, they enter at the maximally differential level, which is
subject to subsequent numerical integrations. Each step comes with some
loss of numerical precision, which means the innermost elements should have a
higher numerical accuracy than what is desired for the final result. If an
integration routine becomes sensitive to numerical PDF inaccuracies,
its convergence and hence the computational cost can suffer greatly, because the
integrator can get distracted (or even stuck) trying to integrate numerical
noise. We have explicitly observed this effect for quadrature-based integrators,
which for low-dimensional integrations are far superior to Monte-Carlo
integrators, but whose high accuracy and fast convergence strongly depends on
the smoothness of the integrand.
Another issue is that with increasing perturbative order, the convolution
kernels in cross section formulae become more and more steep or strongly
localized. This requires a higher numerical accuracy of the PDF, because its
convolution with such kernels is sensitive not just to its value but also to its
derivatives or its detailed shape. In \refcite{Dulat:2017prg}, it was indeed
observed that the limited numerical accuracy of PDFs provided by LHAPDF can
cause instabilities in the final result.

Many applications require an accurate and fast execution of nontrivial
operations on PDFs, such as taking Mellin moments, computing Mellin convolutions
with various kernels, taking derivatives, and so forth.
A prominent example is DGLAP evolution, which has been extensively studied in
the literature, see e.g.~\refscite{Miyama:1995bd, Ratcliffe:2000kp,
Pascaud:2001bi, Dasgupta:2001eq, DelDebbio:2007ee}, and for which there is a
variety of codes that solve the evolution equations either via Mellin moments
\cite{Weinzierl:2002mv, Vogt:2004ns, Candido:2022tld} or directly in $x$
space~\cite{Cafarella:2008du, Salam:2008qg, Botje:2010ay, Bertone:2013vaa,
Bertone:2017gds}. The latter require in particular the repeated Mellin
convolution of PDFs with splitting functions, in addition to interpolation in
$x$.

Another important application is the computation of beam functions and similar
quantities, which typically appear in resummation formulae.
In multiscale problems, beam functions can depend on additional
dynamic variables~\cite{Procura:2014cba, Lustermans:2019plv, Lustermans:2019cau,
Gaunt:2014xxa, Gaunt:2020xlc, Hornig:2017pud, Michel:2018hui, Bonvini:2015pxa,
Pietrulewicz:2017gxc}, in which case their evaluation necessitates fast
on-the-fly evaluation of Mellin
convolutions and in some cases their subsequent DGLAP evolution.
A further area is the calculation of subleading power corrections,
where the first and second derivatives of PDFs with respect to $x$ are explicitly
required~\cite{Moult:2016fqy, Moult:2017jsg, Ebert:2018lzn, Boughezal:2016zws,
Boughezal:2018mvf, Bhattacharya:2018vph, Ebert:2018gsn}.
With cubic splines, the first and second derivative
respectively correspond to quadratic and linear interpolation and hence
suffer from a poor accuracy. Numerical instabilities in derivatives computed from
spline interpolants were for instance reported in \refcite{Ball:2016spl}.

Last but not least, for double parton distributions \cite{Diehl:2017wew} or other
multi-dimensional distribution functions, local interpolation methods become
increasingly cumbersome due to the large number of variables and even
larger number of distributions. By contrast, our global interpolation
approach allows for a controllable numerical accuracy with reasonable memory and
runtime requirements. This will be discussed in future work.

Clearly, there is always a trade-off between the numerical accuracy of a method
and its computational cost, and the accuracy of different methods should be
compared at similar computational cost (or vice versa). In our case, the primary
cost indicator is the number of points on the interpolation grid, which controls
both the number of CPU operations and the memory footprint.%
\footnote{Of course, the inherent complexity of a method matters as well.
For instance, more complex and thereby more accurate splines tend to have a higher
computational cost per grid point. However, these are secondary effects we will
not focus on.}
As a result of its low polynomial order, local spline interpolation has an accuracy
that scales rather poorly with the grid size,
so that even a moderate increase in precision requires a substantial increase
in computational effort. This can be (partially) offset by
improving the performance of the implementation itself, see for example
\refscite{Bertone:2017gds, Carrazza:2020qwu}.
Such improvements can be made for any given method, but they do
not change the accuracy scaling of the method itself.

Global interpolation approximates a function over its domain by a
single, high-order polynomial. On an equispaced grid, this leads
to large oscillations near the edges of the interpolation interval,
such that the interpolant never converges and may in fact diverge
exponentially. This is well known as Runge's phenomenon~\cite{Runge:1901}. It
is often misinterpreted as a problem of polynomial interpolation
in general, which may be one reason why local interpolation methods such as
splines tend to be preferred. What is perhaps not sufficiently
appreciated~\cite{Trefethen11sixmyths} is the fact that Runge's phenomenon is
caused by the use of an \emph{equidistant} grid and can be completely avoided by
using a non-equidistant grid that clusters the grid points toward the edges of the
interval.  An example for this is Chebyshev interpolation, which
leads to a well-convergent approximation, i.e., one that can be made
arbitrarily precise by increasing the number of interpolation points. Moreover,
approximation with Chebyshev polynomials on a finite interval is
closely related to and thus as reliable as the Fourier series
approximation for periodic functions.
We find that Chebyshev interpolation for PDFs has an excellent accuracy scaling.
As we will show, it easily outperforms local spline interpolation by
many orders of magnitude in accuracy for the same or even lower number of grid
points.

In the \hyperref[sec:chebyshev]{next section}, we provide some mathematical
background on Chebyshev interpolation as we require it. In
\sec{pdf_interpolation}, we present the global Chebyshev interpolation of PDFs
used in \chili\ and compare its accuracy with local spline interpolation. We
also discuss methods for estimating the numerical accuracy, as well as the basic
operations of taking derivatives and integrals of PDFs. In
\secs{mellin_convolution}{dglap_evolution}, we describe the implementation and
accurate evaluation of Mellin convolutions and of DGLAP evolution with \chili. We
conclude in \sec{conclusions}.  In an \hyperref[app:rungekutta]{appendix}, we
discuss the performance of different Runge-Kutta algorithms for solving the
evolution equations.

\section{Chebyshev interpolation}
\label{sec:chebyshev}

In this section, we give a brief account of Chebyshev interpolation, i.e.\ the
interpolation of a function that is discretized on a grid of Chebyshev points.
This includes the topics of differentiation, integration, and of estimating the
interpolation accuracy.  A wealth of further information and mathematical
background can be found in \refcite{Trefethen}.\footnote{The first chapters of
this book are available on \url{https://people.maths.ox.ac.uk/trefethen/ATAP}.}

Throughout this section, we consider functions of a variable $t$ restricted to
the interval $[-1, 1]$.  The relation between $t$ and the momentum fraction $x$
of a PDF is specified in \sec{pdf_interpolation}.

\paragraph{Chebyshev polynomials.}
The Chebyshev polynomials of the first and second kind, $T_k(t)$ and $U_k(t)$,
are defined by
\begin{align} \label{eq:TU-def}
T_k(\cos \theta) &=  \cos k \ms \theta
\,, &
U_{k}(\cos \theta) &= \frac{\sin(k+1) \ms \theta}{\sin\theta}
\end{align}
for integer $k \ge 0$.  They are related by differentiation as $\dd T_k (t) /
\dd t = k \ms U_{k-1} (t)$, and they are bounded by $|T_k(t)| \le 1$ and
$|U_{k}(t)| \le k+1$.  The relations
\begin{align}
V_{k}(-t) &= (-1)^k\, V_{k}(t)
\,, &
V_{0}(t) &= 1
\,, &
V_{k+2}(t) &= 2 t\ms V_{k+1}(t) - V_{k}(t)
\end{align}
hold for both $V = T$ and $V = U$. They show that $T_k(t)$ and $U_k(t)$ are
indeed polynomials, which may not be immediately obvious from \eq{TU-def}.
Both families of polynomials form an orthogonal set, i.e.\ for $k, m \ge 0$
they satisfy
\begin{align} \label{eq:contin_ortho}
\int_{-1}^1 \frac{\dd t}{\sqrt{1 - t^2}} \; T_k (t) \, T_m (t)
&= \frac{\alpha_k \ms \pi}{2} \, \delta_{k m}
\,, &
\int_{-1}^1 \! \dd t \, \sqrt{1 - t^2} \; U_k (t) \, U_m (t)
&= \frac{\pi}{2} \, \delta_{k m}
\,,\end{align}
where $\alpha_0 = 2$ and $\alpha_k = 1$ otherwise.

For given $N$, the \emph{Chebyshev points} are given by
\begin{align} \label{eq:cheb_points}
t_j &= \cos \theta_j
\,, \qquad
\theta_j = \frac{j \pi}{N}
\qquad
\text{with}
\quad
j = 0, \ldots, N
\,.\end{align}
They form a descending series from $t_0 = 1$ to $t_N = -1$ and satisfy the
symmetry property $t_{N-j} = - t_j$.  The polynomial $T_N(t)$ assumes its maxima
$+1$ and minima $-1$ at the Chebyshev points.  We call the set of Chebyshev
points a \emph{Chebyshev grid}.  Using \eq{TU-def} and expressing sines and
cosines as complex exponentials, one readily derives the discrete orthogonality
relations
\begin{align} \label{eq:discr_ortho}
\sum_{j=0}^{N}\, \beta_j\ms T_{k}(t_j)\, T_{m}(t_j)
&= \frac{N}{2 \beta_k} \, \delta_{k m}
&& \text{for}\quad k, m = 0, \ldots, N,
\nn \\
\sum_{j=1}^{N-1} \bigl( 1 - t_j^2 \ms\bigr)\, U_{k-1}(t_j)\, U_{m-1}(t_j)
&= \frac{N}{2} \, \delta_{k m}
&& \text{for}\quad k, m = 1, \ldots, N-1,
\end{align}
where $\beta_0 = \beta_N = 1/2$ and $\beta_j = 1$ otherwise.

Notice that the density of Chebyshev points increases from the center toward the
end points of the interval $[-1, 1]$. This feature is crucial to avoid Runge's
phenomenon for equispaced interpolation grids, as discussed in the
\hyperref[sec:intro]{introduction}.

\paragraph{Chebyshev interpolation and Chebyshev series.}
We now consider the approximation of a function $f(t)$ by a finite sum of
Chebyshev polynomials $T_{k}(t)$.  Using $T_k(t_j) = T_j(t_k)$ and the first
relation in \eq{discr_ortho}, one finds that the sum
\begin{align} \label{eq:cheb_interp}
p_N(t) &= \sum_{k=0}^{N}\, \beta_k\ms c_k\ms T_{k}(t)
\end{align}
with the interpolation coefficients
\begin{align} \label{eq:cheb_interp_coeff}
c_k &= \frac{2}{N} \sum_{j=0}^{N}\, \beta_j \ms f(t_j)\, T_k(t_j)
\end{align}
satisfies
\begin{align}
p_N(t_j) &= f(t_j)
\qquad
\text{for}\quad
j = 0, \ldots, N
\,.\end{align}
In other words, $p_N(t)$ is the unique polynomial of order $N$ that equals
the function $f(t)$ at the $N+1$ Chebyshev points $t_0, \ldots, t_{N}$. We
therefore call $p_N(t)$ the \emph{Chebyshev interpolant}.

A sufficiently smooth function $f(t)$ can be expanded in the \emph{Chebyshev
series}
\begin{align} \label{eq:cheb_series}
f(t) &= \lim_{N \to \infty} f_N(t)
\,, &
f_N(t) &= \sum_{k=0}^N a_k\ms T_{k}(t)
\,,\end{align}
whose series coefficients
\begin{align} \label{eq:cheb_series_coeff}
a_k &= \frac{2}{\alpha_k \ms \pi}  \int_{-1}^1 \frac{\dd t}{\sqrt{1 - t^2}}\;
       f(t)\, T_{k}(t)
\end{align}
readily follow from the first orthogonality relation in \eq{contin_ortho}.
Substituting $t = \cos\theta$ and using \eq{TU-def}, we recognize in
\eqs{cheb_series}{cheb_series_coeff} the Fourier cosine series for the function
$F(\theta) = f(\cos \theta)$, which is periodic and even in $\theta$. The
Chebyshev series on the interval $[-1, 1]$ is thus nothing but the Fourier series
of a periodic function in disguise, with the same
excellent convergence properties for $N\to\infty$. The Chebyshev series $f_N(t)$
is not immediately useful in practice, because its coefficients can only be
obtained by explicitly carrying out the integral in \eq{cheb_series_coeff}. The
Chebyshev interpolant $p_N(t)$, however, can be computed very efficiently (see
below) and only requires evaluating $f(t_j)$ at the $N+1$ Chebyshev points. The
key property of interpolating in the Chebyshev points is that the resulting
interpolation coefficients $c_k$ approach the series coefficients $a_k$ in the
limit $N\to\infty$. Their precise relation can be found in \cite[chapter 4]
{Trefethen}. It is therefore guaranteed that $p_N(t)$ approaches $f_N(t)$
and thus $f(t)$ for $N\to\infty$.

\paragraph{Interpolation accuracy.}
How accurately $p_N(t)$ approximates the function $f(t)$ depends on the
smoothness of $f$ and its derivatives.  We give here a convergence statement
that is useful for the interpolation of PDFs.  For typical parameterizations,
input-scale PDFs are analytic functions of the momentum fraction $x$ for $0 < x
<1$, but nonanalytic at $x = 1$, where they behave like $(1-x)^\beta$
with noninteger $\beta$.  We anticipate that this corresponds to a behavior like
$(1+t)^\beta$ at $t= -1$ when we map an $x$ interval onto a Chebyshev grid in
$t$.

Suppose that on the interval $[-1,1]$ the function $f$ and its derivatives up to
$f^{(\nu - 1)}$ with $\nu \ge 1$ are Lipschitz continuous,\footnote{The
following corresponds to theorem 7.2 in \refcite{Trefethen}, with the condition of
absolute continuity being replaced by the stronger condition of Lipschitz
continuity, which we find sufficient for our purpose and easier to state.} and
that the derivative $f^{(\nu)}$ is of bounded variation $V$.
We recall that a function $F(t)$ is Lipschitz continuous on $[-1,1]$ if there
exists a constant $C$ such that $|F(s) - F(t)| \le C \ms |s - t|$ for all $s, t
\in [-1,1]$.  A differentiable function $F(t)$ is of bounded variation $V$ on
$[-1,1]$ if the integral
\begin{align}
V &= \int_{-1}^1 \dd t \, |F'(t)|
   = \sum_{k=0}^{K} |F(t_{k+1}) - F(t_{k})|
\end{align}
is finite, where in the second step we have split the interval $[-1,1]$ into $K$
subintervals with boundaries $t_0, \ldots, t_{K+1}$ such that on each
subinterval $F'(t)$ has a definite sign. Under these conditions, one has
\begin{align} \label{eq:cheb_convergence}
\big| f(t) - p_N(t) \big| & \le \frac{4 V}{\pi \nu (N - \nu)^\nu}
\end{align}
for all $N > \nu$ and all $t \in [-1,1]$.  We note that the Chebyshev series has
a similar convergence property, which is obtained by replacing $p_N(t)$ with
$f_N(t)$ and $4 V$ with $2 V$ in~\eq{cheb_convergence}.

For the example of a function $f(t) = (1+t)^{n + \delta}$ with integer $n \ge 1$
and $0 < \delta < 1$, the above statement holds with $\nu = n$.   The function
and its derivatives up to $f^{(n-1)}$ are Lipschitz continuous on $[-1,1]$.  The
$n$th derivative is proportional to $(1+t)^{\delta}$ and not Lipschitz
continuous but of bounded variation.  The $(n+1)$st derivative is proportional
to $(1+t)^{-1+\delta}$ and thus not of bounded variation, because it diverges at
$t= -1$.

It is important to note that \eq{cheb_convergence} provides a bound on the
maximal \emph{absolute} interpolation error anywhere in the interval. Often the
absolute interpolation error will be much smaller over most of the interval.
In the vicinity of a point where $f(t)$ goes to zero, the \emph{relative} error
can still remain large, and the convergence $p_N(t) / f(t) \to 1$ for $N \to
\infty$ is in general not uniform over the full interval.  We will indeed see
this for the interpolation of PDFs close to zero crossings or to the end point
 $x=1$.

\paragraph{Barycentric formula.}
A simple and efficient way to compute the Chebyshev interpolant is given by the
\emph{barycentric formula}
\begin{align} \label{eq:barycentric}
p_N(t) &= \sum_{j=0}^{N} f(t_j) \, b_j(t)
\,,\end{align}
where $t_j$ denotes again the Chebyshev points, and the barycentric basis
functions are given by
\begin{align} \label{eq:bary_basis}
b_j(t) &= \, \frac{\beta_j \, (-1)^{j}}{t - t_j}
  \Bigg/ \sum_{i=0}^{N} \frac{\beta_i \, (-1)^{i}}{t - t_i}
\,.\end{align}
$b_j(t)$ is a polynomial of order $N$, although this is not evident from
\eq{bary_basis}.  The number of operations for evaluating the barycentric
formula scales like $N$.  The formula is found to be numerically stable in the
interpolation interval.  We note that it is \emph{not} stable for extrapolating
the function $f(t)$ outside this interval \cite[chapter 5]{Trefethen}.

The representation given by \eqs{barycentric}{bary_basis} is a special case of
the barycentric formula for the polynomial $L(u)$ of order $N$ that interpolates
a function $f(u)$ given on a set of $N+1$ distinct points $u_0, \ldots, u_N\ms$:
\begin{align} \label{eq:bary_general}
L_N(u) &= \sum_{j=0}^{N} f(u_j) \, \ell_j(u)
\end{align}
with basis functions
\begin{align} \label{eq:bary_basis_general}
\ell_j(u) &= \frac{\lambda_j}{u - u_j}
  \Bigg/ \sum_{i=0}^{N} \frac{\lambda_i}{u - u_i}
\,, &
(\lambda_i)^{-1} &= \prod_{j\neq i} \, ( u_j - u_i )
\,.\end{align}
$L_N$ is called the \emph{Lagrange polynomial} for the pairs of values $\{ u_j,
f(u_j) \}$.  We will again use \eq{bary_general} below. The simple form of
\eq{bary_basis} comes from the fact that the Chebyshev points yield the very
simple weights $\lambda_i = \beta_i \ms (-1)^i\,2^{n-1}/n$.

\paragraph{Differentiation.}
Given the Chebyshev interpolant $p_N(t)$ for a function $f(t)$, one can
approximate the derivative $f'(t) = \dd f(t)/ \dd t$ by the derivative
$p'_N(t)$.  Note that in general $f'$ is not equal to $p'_N$ at the Chebyshev
points.  Obviously, one cannot compute the exact values of $f'(t_j)$ from the
function values $f(t_j)$ on the grid.

The derivative $p'_N(t)$ is a polynomial of degree $N-1$ and thus also a
polynomial of degree $N$ (with vanishing coefficient of $t^N$). It is therefore
identical to its own Chebyshev interpolant of order $N$, so we can compute
it on the full interval $[-1,1]$ by the barycentric formula
\begin{align} \label{eq:bary_deriv}
p'_N(t) &= \sum_{j=0}^{N} p'_N(t_j) \, b_j(t)
\,.\end{align}
To obtain the values of $p'_N(t_j)$, we take the derivative of \eq{cheb_interp}
using $T'{\!}_k = k \ms U_{k-1}$.  The resulting discrete sums are easily
evaluated using \eq{TU-def} and expressing the sine function in terms of
complex exponentials.  We then obtain the relation
\begin{align}
\label{eq:cheb_deriv}
p'_N(t_j) &= \sum_{k=0}^{N}\, D_{j k}\, f(t_k)
\end{align}
with $D_{0 0} = - D_{N N} = (2 N^2 + 1) /6$ and
\begin{align}
D_{j j} &= - \dfrac{\cos\theta_j}{2 \sin^2\theta_j}
\quad\text{for}\quad j \neq 0, N,
\quad
&
D_{j k} &= \dfrac{\beta_k}{\beta_j}\,
         \dfrac{(-1)^{j+k}}{t_j - t_k}
\quad\text{for}\quad j \neq k
\,.\end{align}
Note that the matrix multiplication \eqref{eq:cheb_deriv} maps a vector $f(t_k)
= \text{\it const}$ onto the zero vector, as it must be because the derivative
of a constant function is zero.

Higher derivatives of the Chebyshev interpolant $p_N(t)$ can be computed by
repeated multiplication of $f(t_k)$ with the differentiation matrix $D_{j k}$
and subsequent application of the barycentric formula.  Since each derivative
reduces the degree of the interpolating polynomial by $1$, the accuracy of
approximating $f^{(n)}_{}(t)$ by $p^{(n)}_N(t)$ gradually degrades with
increasing $n$.  This loss of accuracy can be accounted for by choosing a
sufficiently large order $N$ to start with.

\paragraph{Integration.}
Using \eq{cheb_interp} together with
\begin{align} \
\int_{-1}^1 \dd t\, T_{k}(t)
&= \begin{cases}
   2/ (1-k^2)  & \text{for even $k$,} \\
   0           & \text{for odd $k$,}
   \end{cases}
\end{align}
one readily obtains the integration rule
\begin{align} \label{eq:CC-rule}
\int_{-1}^1 \dd t \, f(t)
&\approx \sum_{\genfrac{}{}{0pt}{}{k=0}{\text{even}}}^{N}\,
   \frac{2 \ms \beta_k \ms c_k}{1-k^2}
  = \sum_{j=0}^{N}\, w_j\ms f(t_j)\;
\end{align}
with weights
\begin{align} \label{eq:CC-weights}
w_j &= \frac{4 \beta_j}{N}\,
   \sum_{\genfrac{}{}{0pt}{}{k=0}{\text{even}}}^{N}\,
   \beta_k\, \frac{\cos (k \ms \theta_j)}{1-k^2}
\,.\end{align}
This is known as \emph{Clenshaw-Curtis quadrature}.  A detailed discussion of
its accuracy (and comparison with Gauss quadrature) can be found in
ref.~\cite[chapter 19]{Trefethen} and \refcite{Trefethen:2008sia}.

\paragraph{Interpolation without end points.}
The interpolant \eqref{eq:cheb_interp} is a sum over Chebyshev polynomials
$T_k$.  Another interpolant can be obtained from the polynomials $U_{k}$, namely
\begin{align} \label{eq:cheb_interp_noend}
q_{N-2}(t) &= \sum_{k=1}^{N-1}\, d_k\ms U_{k-1}(t)
\end{align}
with coefficients
\begin{align} \label{eq:cheb_interp_coeff_noend}
d_k &= \frac{2}{N} \sum_{j=1}^{N-1}\,
   f(t_j)\, \bigl( 1 - t_j^2 \ms\bigr) \, U_{k-1}(t_j)
\,.\end{align}
Using the second relation in \eq{discr_ortho} together with $\sin \theta_j\,
U_{k-1}(t_j) = \sin \theta_k \, U_{j-1}(t_k)$, one finds that
\begin{align}
q_{N-2}(t_j) &= f(t_j)
\qquad
\text{for}\quad j = 1, \ldots, N-1
\,.\end{align}
This means that $q_{N-2}(t)$ interpolates $f(t)$ on the same Chebyshev points as
$p_{N}(t)$, with the exception of the interval end points.  Correspondingly,
$q_{N-2}$ has polynomial degree $N-2$ rather than $N$, as is evident from
\eq{cheb_interp_noend}.  We thus have an alternative approximation for $f(t)$,
which can be computed from the same discretized function values as those needed
for computing $p_N(t)$.  For sufficiently large $N$, one may expect that
$q_{N-2}$ approximates $f(t)$ only slightly less well than $p_N(t)$, given that
its polynomial degree is only smaller by two units.  We may hence use $|\ms
q_{N-2}(t) - p_{N}(t)|$ as a conservative estimate for the interpolation error
$|f(t) - p_{N}(t)|$.%
\footnote{To be precise, since $p_N(t)$ is more accurate than $q_{N-2}(t)$, the
difference $|\ms q_{N-2}(t) - p_{N}(t)|$ is actually an estimate of the accuracy
of $q_{N-2}(t)$ and thus a conservative estimate for the accuracy of $p_N(t)$.}

To evaluate $q_{N-2}(t)$, we can again use a barycentric formula, namely
\begin{align} \label{eq:bary_noend}
q_{N-2}(t) &= \sum_{j=1}^{N-1} f(t_j) \, \tilde{b}_j(t)
\,.\end{align}
The corresponding basis functions
\begin{align} \label{eq:bary_basis_noend}
\tilde{b}_j(t) &= \, \frac{(-1)^{j} \, \sin^2 \theta_j}{t - t_j}
  \Bigg/ \sum_{i=1}^{N-1} \frac{(-1)^{i} \, \sin^2 \theta_i}{t - t_i}
\end{align}
can easily be obtained from \eq{bary_basis} by comparing the general expressions
\eqref{eq:bary_general} and \eqref{eq:bary_basis_general} for the sets of points
$t_j$ with $j=0, \ldots N$ or $j=1, \ldots N-1$.  We note that using the
barycentric formula \eqref{eq:bary_noend} in the full interval $[-1,1] = [t_N,
t_0]$ involves an extrapolation from the interval $[t_{N-1}, t_1]$ on which
$q_{N-2}(t)$ interpolates $f(t)$.  For sufficiently large $N$, the extrapolation
is however very modest, because $t_{N-1} - t_{N} = t_0 - t_1 \approx \pi^2 / (2
N^2)$.

Formulae for differentiation and integration of $f(t)$ that use the approximant
$q_{N-2}$ are easily derived in analogy to the formulae that use $p_N$.  For
differentiation, one obtains
\begin{align} \label{eq:bary_deriv_noend}
q'_{N-2}(t) &= \sum_{j=1}^{N-1} q'_{N-2}(t_j) \, \tilde{b}_j(t)
\end{align}
and
\begin{align}
q'_{N-2}(t_j) &= \sum_{k=1}^{N-1}\, \widetilde{D}_{j k}\, f(t_k)
\end{align}
with
\begin{align} \label{eq:diff_mat_noend}
\widetilde{D}_{j j} &= \dfrac{3 \cos\theta_j}{2 \sin^2\theta_j}
\,, &
\widetilde{D}_{j k} &= \dfrac{\sin^2 \theta_k}{\sin^2 \theta_j}\;
         \dfrac{(-1)^{j+k}}{t_j - t_k}
\quad\text{for}\quad j \neq k
\,.\end{align}
For integration, one uses
\begin{align}
\int_{-1}^1 \dd t\, U_{j-1}(t)
&= \begin{cases}
   2/j & \text{for odd $j$,} \\
   0   & \text{for even $j$,}
   \end{cases}
\end{align}
to obtain an open integration rule
\begin{align} \label{eq:Fejer-rule}
\int_{-1}^1 \dd t \, f(t)
&\approx \sum_{\genfrac{}{}{0pt}{}{k=1}{\text{odd}}}^{N-1}\, \frac{2 \ms d_k}{k}
  = \sum_{j=1}^{N-1}\, \tilde{w}_j\ms f(t_j)\;
\end{align}
with weights
\begin{align} \label{eq:Fejer-weights}
\tilde{w}_j &= \frac{4 \sin \theta_j}{N} \,
   \sum_{\genfrac{}{}{0pt}{}{k=1}{\text{odd}}}^{N-1}\,
   \frac{\sin (k \ms \theta_j)}{k}
\,.\end{align}
This is well known as \emph{Fej{\'e}r's second rule}, see e.g.\
\refcite{Waldvogel:2006bit}.\footnote{This paper is also available at
\url{http://www.sam.math.ethz.ch/~waldvoge/Papers/fejer.html}.}

Using \eq{Fejer-rule} to estimate the error of Clenshaw-Curtis integration
\eqref{eq:CC-rule} is similar to \emph{Gauss-Kronrod} quadrature
\cite{Gauss-Kronrod,Patterson:1968mat}, where for a grid with $2 N+1$ points one
has an integration rule of order $3 N + 1$ and a Gauss rule of order $2 N - 1$,
where the latter uses a subset of the points and is used to estimate the
integration uncertainty.\footnote{An integration rule is of order $p$ if
polynomials up to order $p$ are integrated exactly.  For odd $N$, the order of
the Gauss-Kronrod rule is increased from $3 N + 1$ to $3 N + 2$, because odd
polynomials are correctly integrated to zero for symmetry reasons
\protect\cite{Patterson:1968mat}.  Likewise, the order of the quadrature rules
\protect\eqref{eq:CC-rule} and \protect\eqref{eq:Fejer-rule} for even $N$ is
increased from $N$ to $N+1$ and from $N-2$ to $N-1$, respectively.}
The difference in polynomial orders between the two rules is hence larger in
this case than for the pair of Clenshaw-Curtis and Fej{\'e}r rules, so that one
may expect the error estimate in the latter case to be closer to the actual
error.  We shall come back to this in \sec{error_estimate}.

Given the Chebyshev grid \eqref{eq:cheb_points} with $N+1$ points for even $N$,
one might also think of estimating the integration or interpolation accuracy by
using the subgrid $t_0, t_2, \ldots t_{N-2}, t_{N}$, which has $N/2 + 1$ points
and is again a Chebyshev grid.  For the grids and functions we will consider in
this work, this would however give a gross overestimate of the actual error,
because the interpolant with $N/2 + 1$ points is significantly worse than the
one with $N + 1$ points.  By contrast, using interpolation without the end
points of the original $N + 1$ point grid, we obtain error estimates that are
rather reliable, as will be shown in \sec{error_estimate}.

\section{Chebyshev interpolation of PDFs}
\label{sec:pdf_interpolation}

\subsection{Interpolation strategy}
\label{sec:interpolation_strategy}

To interpolate a parton distribution function $f(x)$ in the momentum fraction
$x$, we interpolate the function $\tilde{f}(x) = x f(x)$ in the variable $u =
\ln x$.  For a given minimum momentum fraction $x_0$, the interval $[x_0, 1]$ is
thus mapped onto the interval $[u_0, 0]$.  For reasons discussed below, we
usually split the $x$ interval into a few subintervals.  On each subinterval, we
perform a linear transformation from $u = \ln x$ to $t \in [-1,1]$ and introduce
a Chebyshev grid in $t$, which is used to interpolate the function as described
in \sec{chebyshev}. To specify the full grid, which is a conjunction of $k$
subgrids, we use the notation
\begin{equation}
[x_0,\ms x_1,\ms \ldots,\ms 1]_{(n_1,\ms n_2,\ms \ldots,\ms n_k)}
\,,\end{equation}
where the $x_i$ are the subinterval boundaries and $n_i = N_i + 1$ is the number
of Chebyshev points for subgrid $i$. We will refer to this as an $(n_1, n_2,
\ldots, n_k)$-point grid. Note that adjacent subgrids share their end points, so
that the total number of grid points is $n_\pts = \sum_i n_i - (k - 1)$.

To be specific, let us consider one subgrid with $N+1$ points $u_0, \ldots,
u_N$, which is mapped by a linear transform onto the Chebyshev grid $t_0,
\ldots, t_N$ given by \eq{cheb_points}.  The corresponding grid points in $x$
are $x_i = e^{u_i}$.  We can then interpolate the PDF using the barycentric
formula
\begin{align} \label{eq:barycentric_pdf}
\tilde{f}(x) &\approx \sum_{j=0}^N \tilde{f}_j \, b_j(\ln x)
&
\text{for~} x_0 \le x \le x_N
\,,\end{align}
where
\begin{align} \label{eq:bary_basis_u}
\tilde{f}_j &= x_j \ms f(x_j)
\,, &
b_j(u) &= \, \frac{\beta_j \, (-1)^{j}}{u - u_j}
  \Bigg/ \sum_{i=0}^{N} \frac{\beta_i \, (-1)^{i}}{u - u_i}
\,.\end{align}
Here we have used that the form of the barycentric basis functions
\eqref{eq:bary_basis} remains unchanged under a linear transform of the
interpolation variable.  Formulae analogous to \eq{barycentric_pdf} can be used
to interpolate functions derived from PDFs, such as their derivatives (see
\sec{differentiation}) or Mellin convolutions of PDFs with an integral kernel (see
\sec{mellin_convolution}).

One reason to use subgrids concerns error propagation.  As seen in
\eq{barycentric_pdf}, the interpolation of
$\tilde{f}$ at a certain value $x$ involves the values of $\tilde{f}$ on
\emph{all} grid points in the interpolation interval, although the weight of
points close to $x$ is higher than the weight of points far away.  In an $x$
region where $\tilde{f}$ is much smaller than its maximum in the
interval, numerical errors from regions of large $\tilde{f}$ can strongly affect
the interpolation accuracy.  This is not much of an issue in our tests below,
where $\tilde{f}$ is computed from an analytic expression, but it does become
important when interpolating a PDF that has been evolved to a higher scale and
is thus affected by numerical errors from the solution of the DGLAP equations.

So on one hand, the accuracy depends on the behavior of the interpolated
function on each subgrid. On the other hand, using multiple subgrids for a fixed
total number $n_\pts$ of points decreases the polynomial degree of the
interpolant on each subgrid, and the accuracy quickly degrades when the
polynomial degree becomes too small. Hence, for given $n_\pts$ and $x_0$, there is a
certain range for the number of subgrids that give the best performance for
typical PDFs. We find that the optimum is to take 2 or 3 subgrids for the values
of $n_\pts$ and $x_0$ used in the following.

To study the accuracy of Chebyshev interpolation for typical PDFs, we consider a
number of representative test functions, covering a broad range of shapes and
analytic forms,
\begin{align} \label{eq:sample_functions}
x f_1 (x) &=
0.0703 \; x^{-0.415 \, (1 + 4.44 \, x) \, (1 + 0.0373 \, \ln x)} \; (1 - x)^{7.75}
\,, \nn \\*
x f_2 (x) &=
17.217 \; x^{-0.33293} \, (1 - x)^{5.3687}
\nn \\*
&\quad \times \bigl[ 1 - 1.664 \; T_1 (1 - 2 \sqrt{x}) + 0.99169 \; T_2 (1 - 2 \sqrt{x})
\nn \\*
&\quad \quad  - 0.42245 \; T_3 (1 - 2 \sqrt{x}) + 0.10176 \; T_4 (1 - 2 \sqrt{x}) \bigr]
\,, \nn \\*
x f_3 (x) &=
4.34 \; x^{-0.015} \, (1-x)^{9.11} - 1.048 \; x^{-0.167} \, (1-x)^{25.0}
\,, \nn \\*
x f_4 (x) &=
7.4 \; x^{0.92} \, (1-x)^{4.6} \, \left( 1 - 2.8 \; \sqrt{x} + 4.5 \; x - 2.0 \; x^2 \right)
\,,\end{align}
where $T_k$ denotes the Chebyshev polynomials defined in \eq{TU-def}.  These
functions correspond to PDFs at the input scales of several common PDF sets.
Specifically, we have
$f_1 = \bar{u}$ at NNLO for ABMP16 \cite{Alekhin:2017kpj},
$f_2 = g$ at LO for MMHT2014 \cite{Harland-Lang:2014zoa},
$f_3 = g$ at NLO for HERAPDF2.0 \cite{Abramowicz:2015mha},
and $f_4 = d_v$ at NLO for JR14 \cite{Jimenez-Delgado:2014twa}.
We have studied several other functions, including $x f(x) \propto x^{-0.7}
\ms (1-x)^{9.2}$, which decreases very steeply and behaves like a typical gluon
density at high scale. The results shown in the following are representative of
this more extended set of functions.

Throughout this work, the relative numerical accuracy of interpolation for a
given quantity is obtained as
\begin{align} \label{eq:accuracy-def}
\text{relative accuracy} =
\bigl| \text{interpolated result} / \text{exact result} - 1 \bigr|
\,,\end{align}
where the exact result is evaluated using the analytic form of the functions in
\eq{sample_functions}. In most plots, we also show the exact result itself as a
thick black line, which is solid (dashed) when the result is positive
(negative). The relative accuracy for other numerical operations is obtained in
full analogy to \eq{accuracy-def}.

\subsection{Interpolation accuracy and comparison with splines}

We now compare Chebyshev interpolation with local interpolation.  As a
prominent example for a method widely used in high-energy physics calculations,
we take the interpolation provided by the LHAPDF
library \cite{Buckley:2014ana}, which has become a standard interface for
accessing parton densities.  LHAPDF offers linear and cubic splines in either
$x$ or $\ln x$.  We use cubic splines in $\ln x$, which is the default in LHAPDF
and gives the most accurate results among these options.\footnote{Technically,
we generate LHAPDF data files for the functions in \eq{sample_functions} and
then run the LHAPDF interpolation routines with the option \texttt{logcubic}.
The LHAPDF data files are generated with double precision to avoid any
artificial loss of numerical accuracy due to the intermediate storage step.} For
brevity, we refer to these as ``L-splines'' in the following. Since spline
interpolants come in a wide variety, we also consider the cubic splines used by
the \texttt{Interpolation} command of Mathematica (versions 11 and 12) and refer
to them as ``M-splines''.  For M-splines, both the first and second derivative
of the interpolant are continuous, whilst for L-splines only the first
derivative is continuous but the second is not.

\begin{table}[ht]
\centering
\begin{tabular} {l | r @{\hspace{2em}} c @{\hspace{2em}} c}
\hline\hline
PDF set & \multicolumn{1}{c}{$x_0$} & $n_\pts$ & $\rho = -n_\pts/\log_{10}(x_0)$
\\\hline
MMHT2014~\cite{Harland-Lang:2014zoa} & $10^{-6}$             &  64 &  10.7 \\
ABMP16~\cite{Alekhin:2017kpj}        & $10^{-7}$             &  99 &  14.1 \\
NNPDF3.1~\cite{Ball:2017nwa}         & $10^{-9}$             & 150 &  16.7 \\
CT18~\cite{Hou:2019efy}              & $0.926\times 10^{-9}$ & 161 &  17.8 \\
JR14~\cite{Jimenez-Delgado:2014twa}  & $10^{-9}$             & 190 &  21.1 \\
MSHT20~\cite{Bailey:2020ooq}         & $10^{-6}$             & 127 &  21.2 \\
NNPDF4.0~\cite{Ball:2021leu}         & $10^{-9}$             & 196 &  21.8 \\
CT14~\cite{Dulat:2015mca}            & $10^{-9}$             & 240 &  26.7 \\
HERAPDF2.0~\cite{Abramowicz:2015mha} & $0.99\times 10^{-6}$  & 199 &  33.1
\\\hline\hline
\end{tabular}
\caption{%
LHAPDF grid parameters for several common PDFs ordered by increasing grid
density.}
\label{tab:LHAPDFgrids}
\end{table}

The interpolation grids used in LHAPDF depend on the PDF set and cover a wide
range in the minimum momentum fraction $x_0$ and in the total number of grid
points $n_\pts$, as shown in \tab{LHAPDFgrids} for several common PDF sets. The
resulting average density of points per decade
in $x$ is  $\rho = - n_\pts / \log_{10} (x_0)$ and essentially determines the
numerical accuracy of the spline interpolation.
For the sake of comparison, we use the grids with the smallest
and the largest density among common PDF sets, which happen to be the MMHT2014
grid ($n_\pts = 64$ with $\rho = 10.7$) and the HERAPDF2.0 grid ($n_\pts = 199$
with $\rho = 33.1$).

\begin{figure*}[t!]
\centering
\includegraphics[width=\WidthTwoSubfigs]{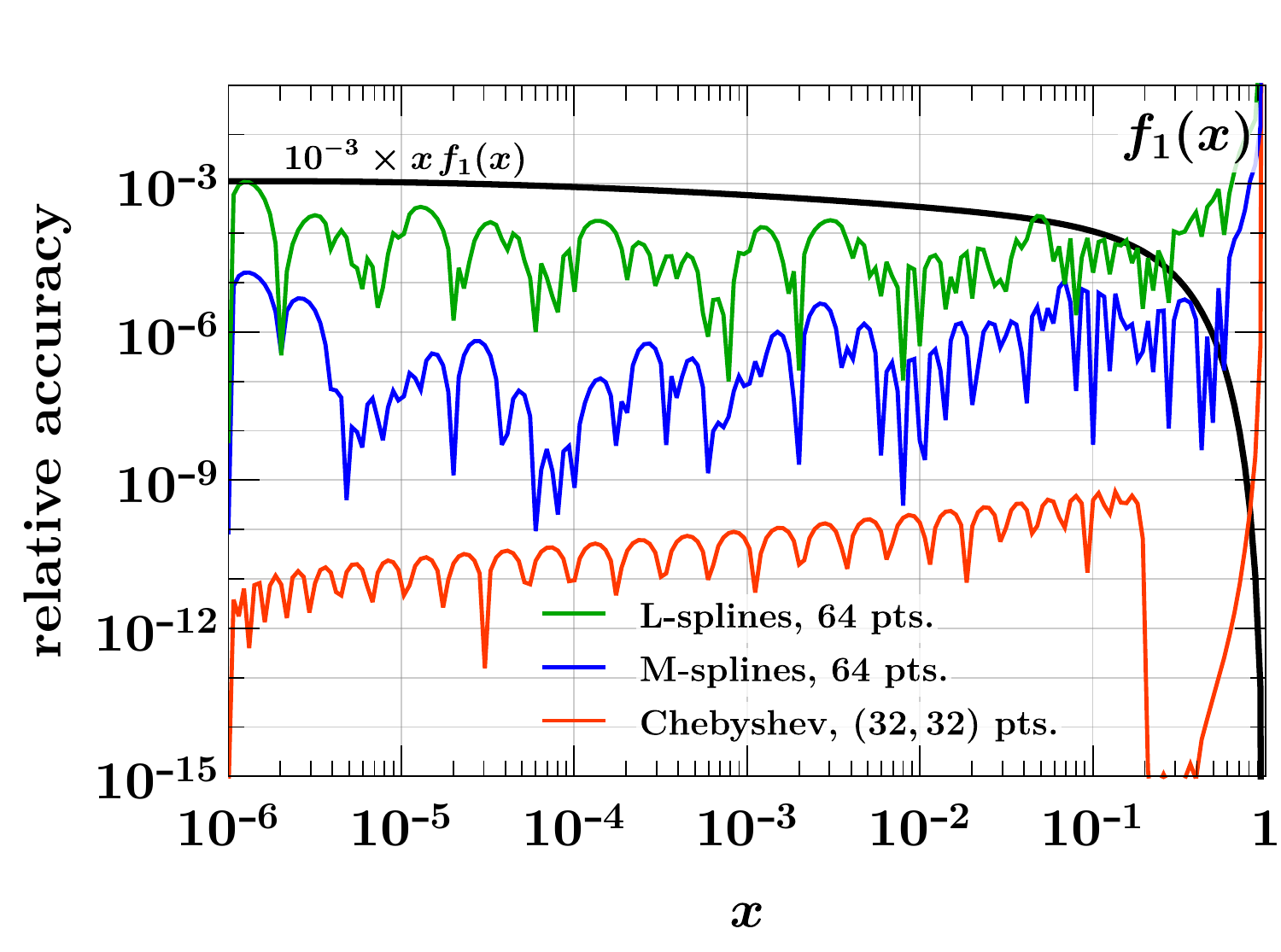}%
\includegraphics[width=\WidthTwoSubfigs]{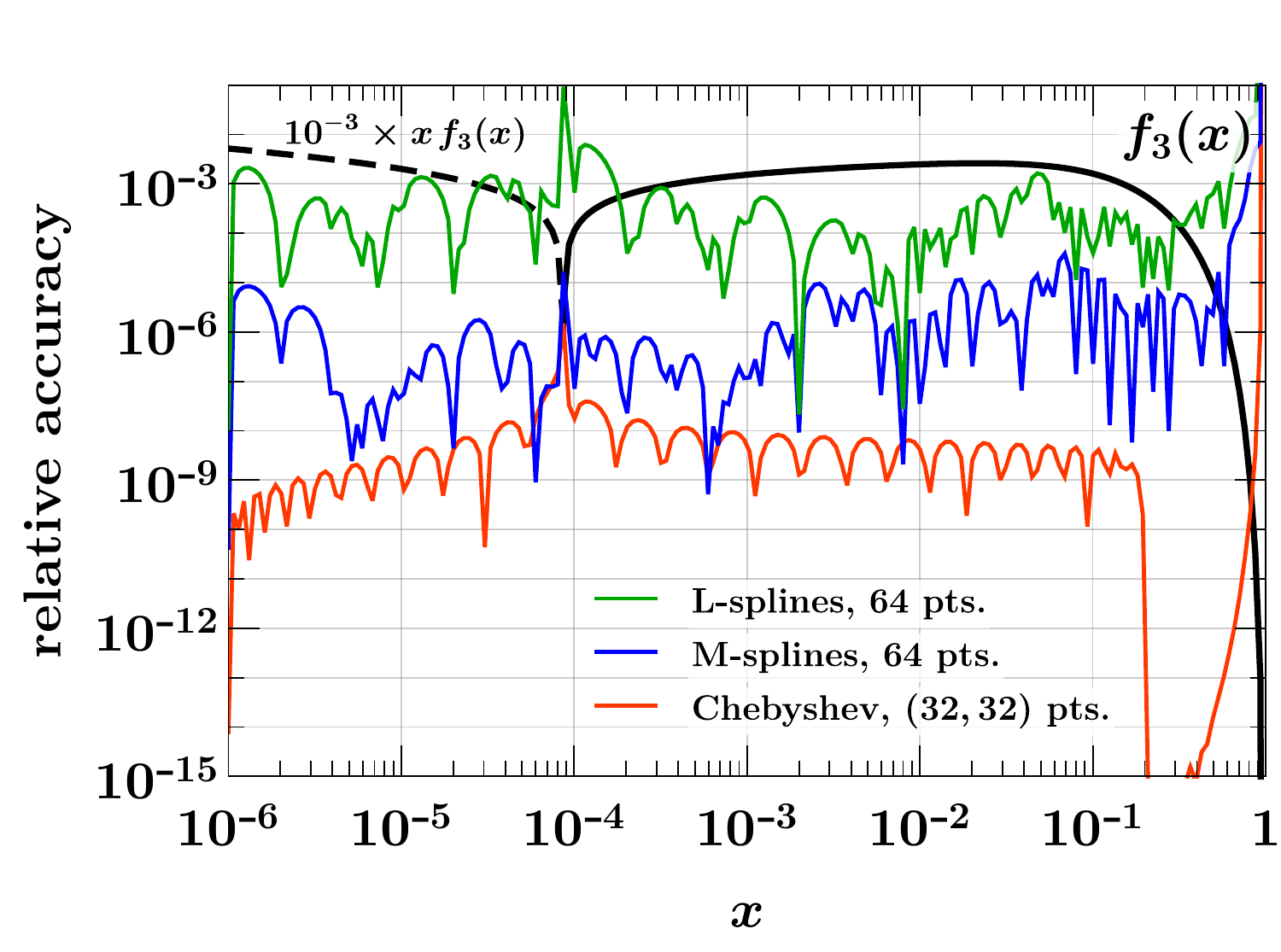}%
\caption{\label{fig:interpolation_comparison_MMHT}%
Relative interpolation accuracy~\eqref{eq:accuracy-def} for the sample PDFs
$f_1(x)$ and $f_3(x)$ in \eq{sample_functions}.  The spline interpolants (green
and blue) use the low-density grid with $n_\pts = 64$. The Chebyshev
interpolation (red) uses the grid $[10^{-6}, 0.2, 1]_{(32,\ms 32)}$, which has
$n_\pts = 63$.
Here and in similar plots, the exact result that is being interpolated is
shown in black (and dashed where the result is negative).
}
\vspace{3ex}
\centering
\includegraphics[width=\WidthTwoSubfigs]{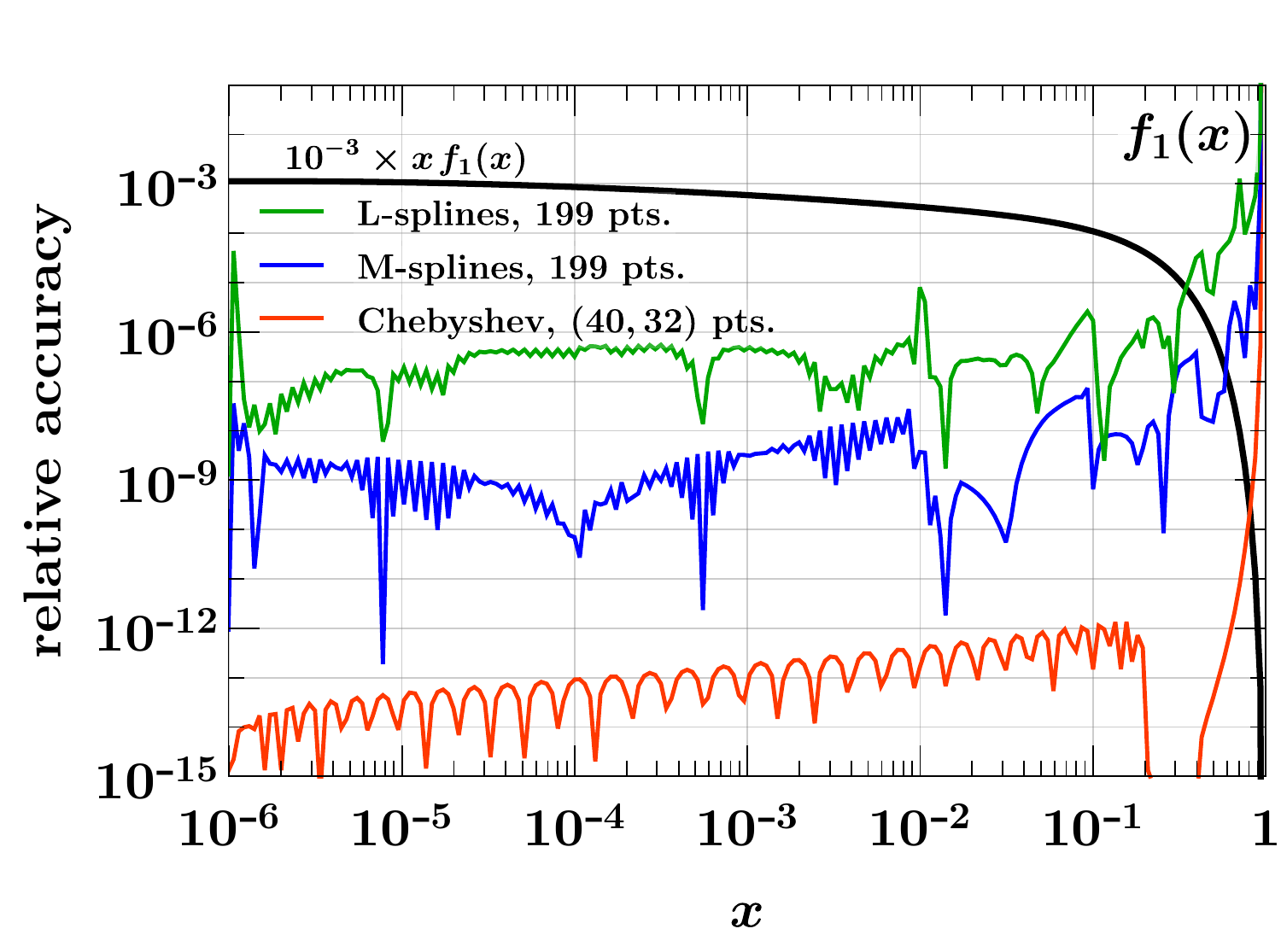}%
\includegraphics[width=\WidthTwoSubfigs]{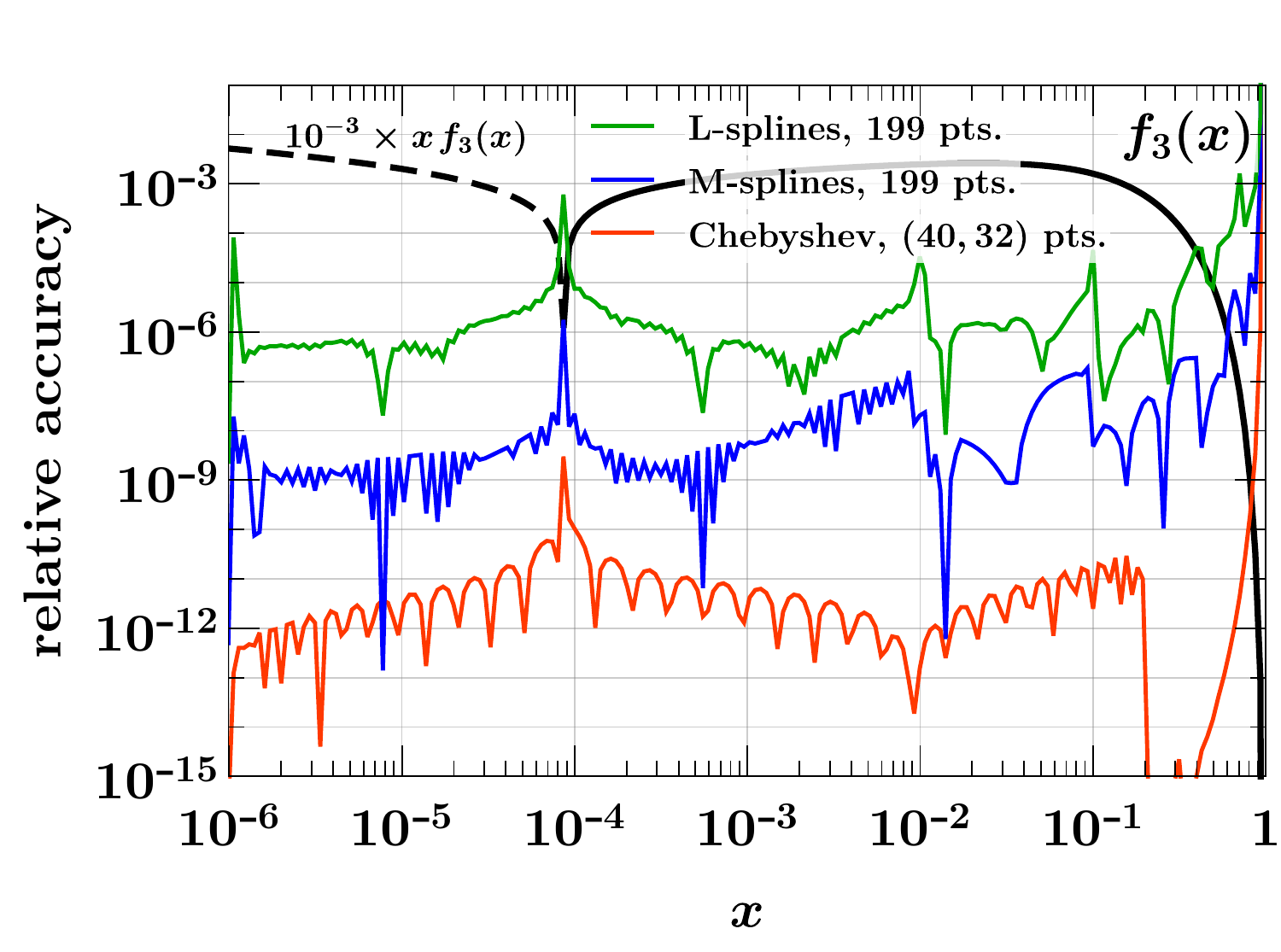}%
\caption{\label{fig:interpolation_comparison_HERA}%
As \fig{interpolation_comparison_MMHT}, but for denser grids.  The spline
interpolants (green and blue) use the high-density grid with $n_\pts = 199$. The
Chebyshev interpolation (red) uses the grid $[10^{-6}, 0.2, 1]_{(40,\ms 32)}$,
which has $n_\pts = 71$.
} \end{figure*}

\begin{figure*}[t!]
\centering
\includegraphics[width=\WidthTwoSubfigs]{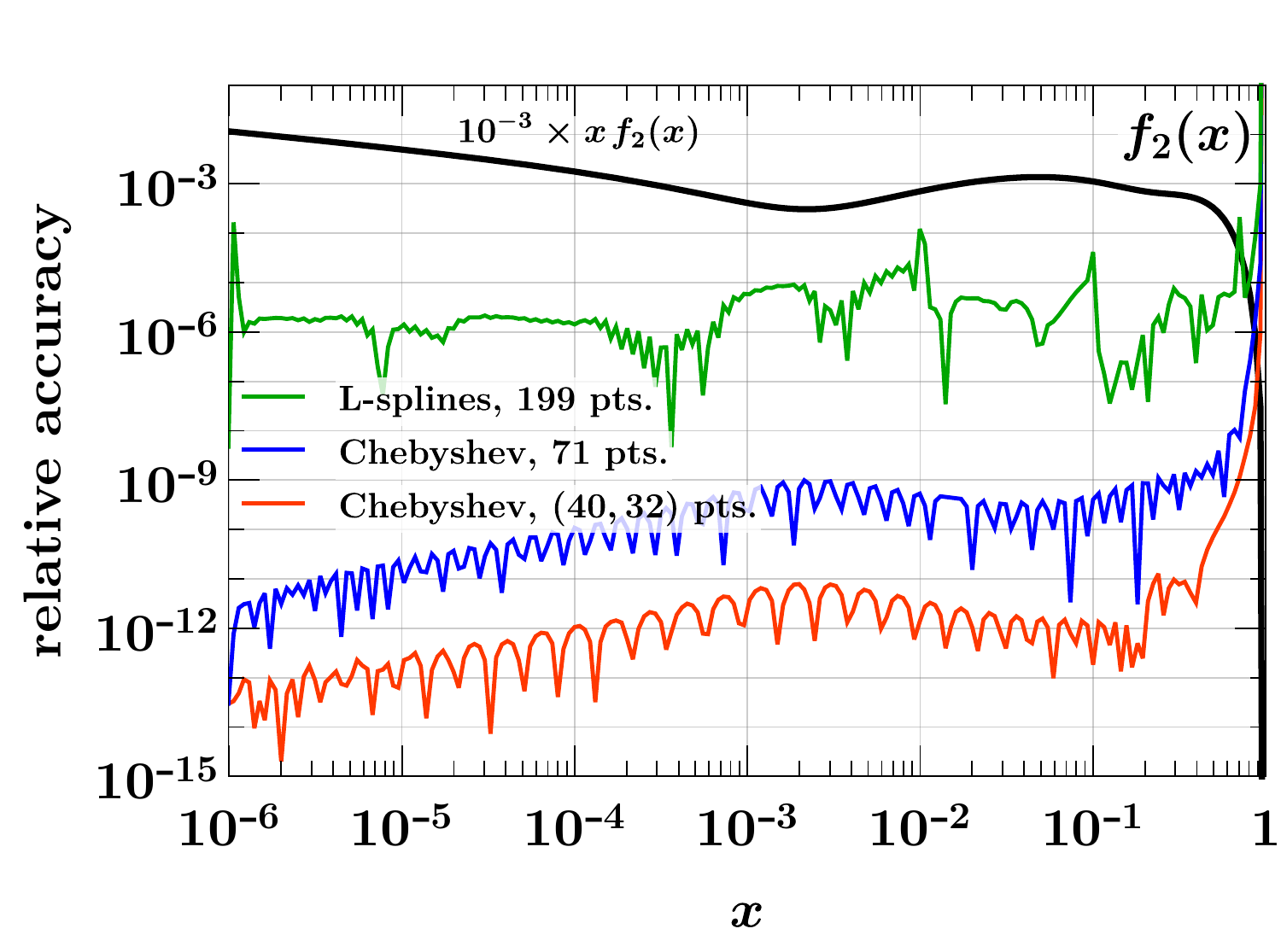}%
\includegraphics[width=\WidthTwoSubfigs]{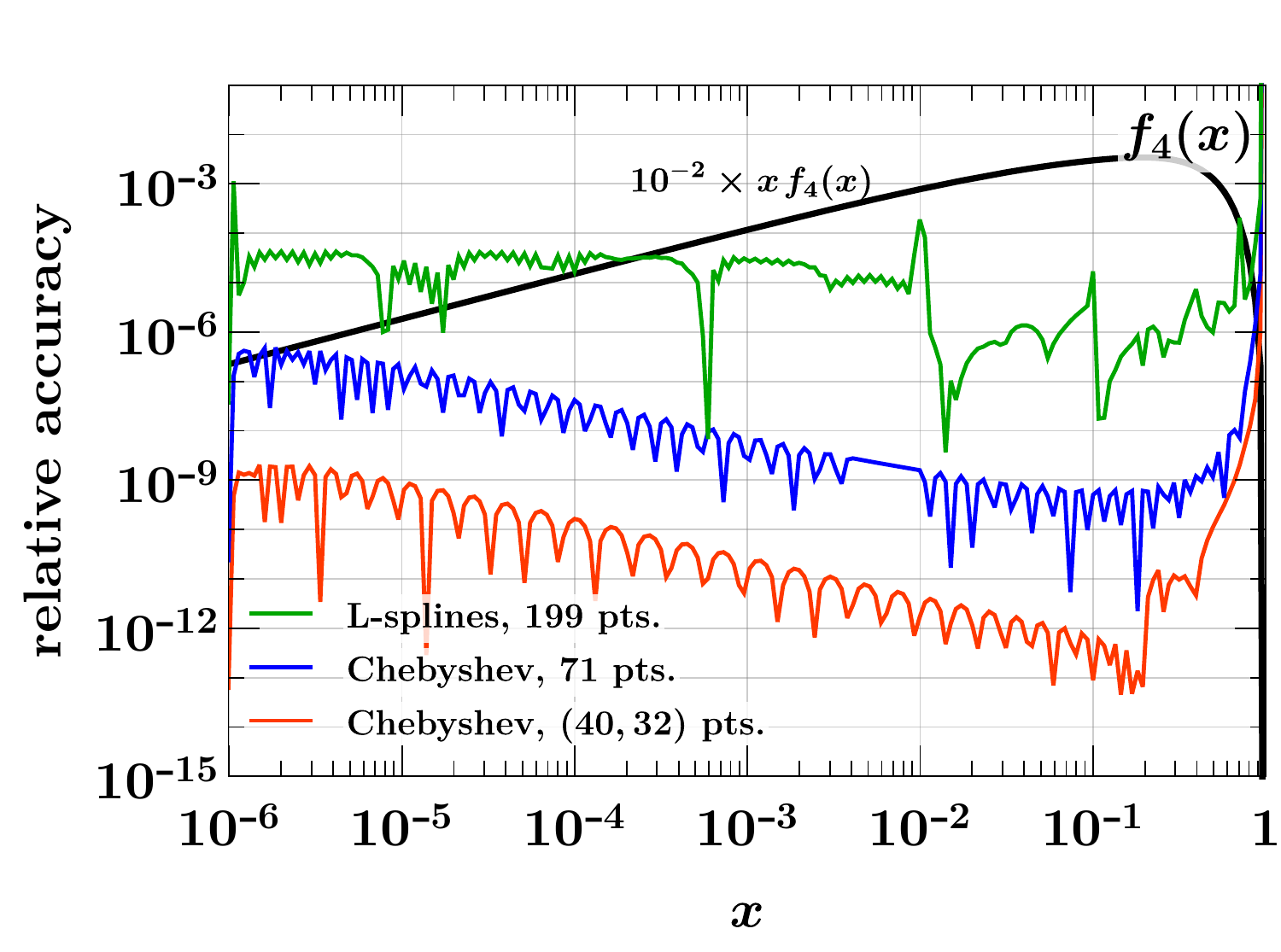}%
\caption{%
Relative interpolation accuracy for the sample PDFs $f_2(x)$ and $f_4(x)$ in
\eq{sample_functions}. The L-splines (green) use the high-density grid with
$n_\pts = 199$. The Chebyshev interpolations use grids with $n_\pts = 71$,
either with a single subgrid (blue) or with two subgrids $[10^{-6}, 0.2,
1]_{(40,\ms 32)}$ (red).
}
\label{fig:interpolation_comparison_subgrids}
\vspace{3ex}
\centering
\includegraphics[width=\WidthTwoSubfigs]{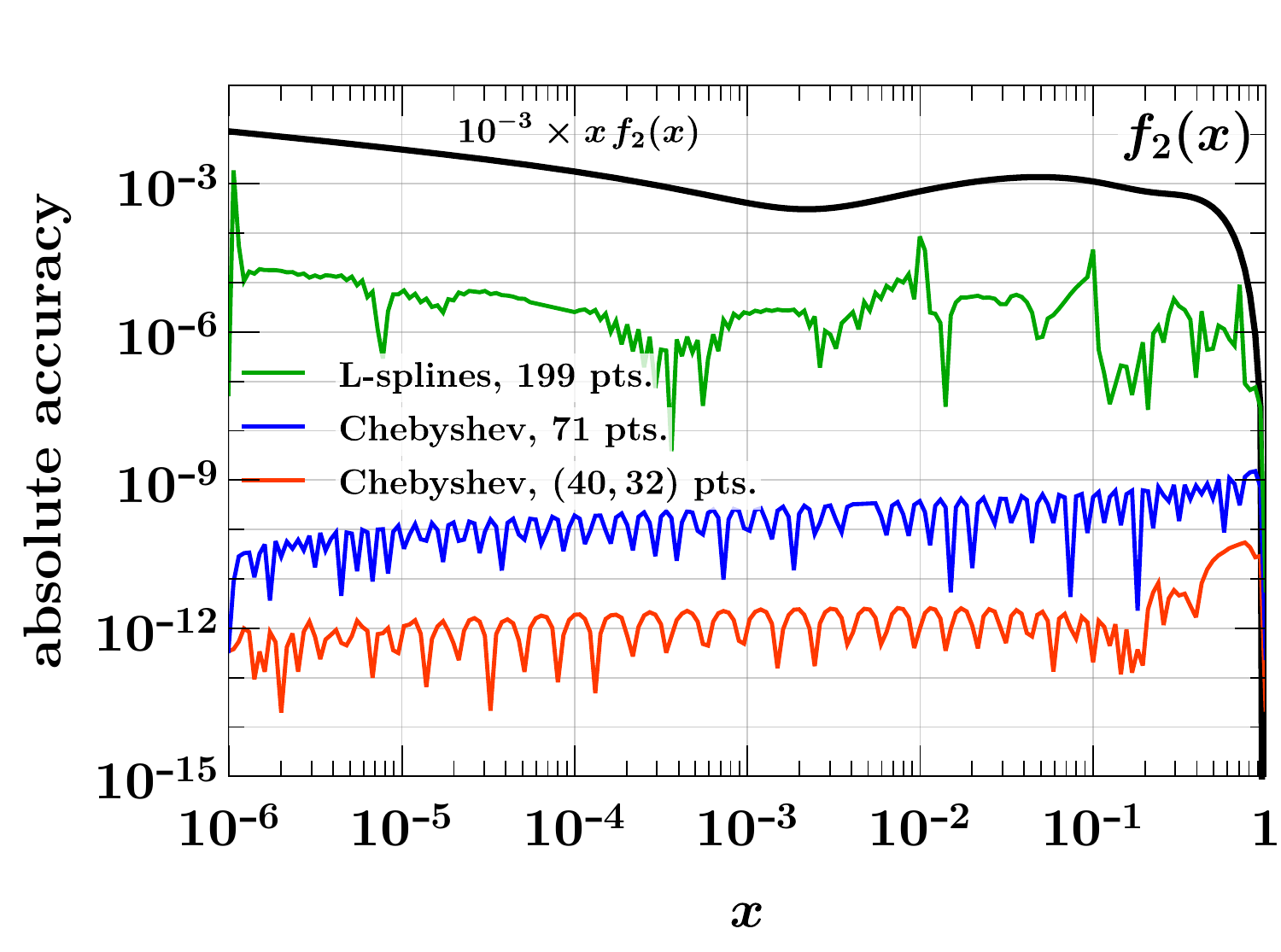}%
\includegraphics[width=\WidthTwoSubfigs]{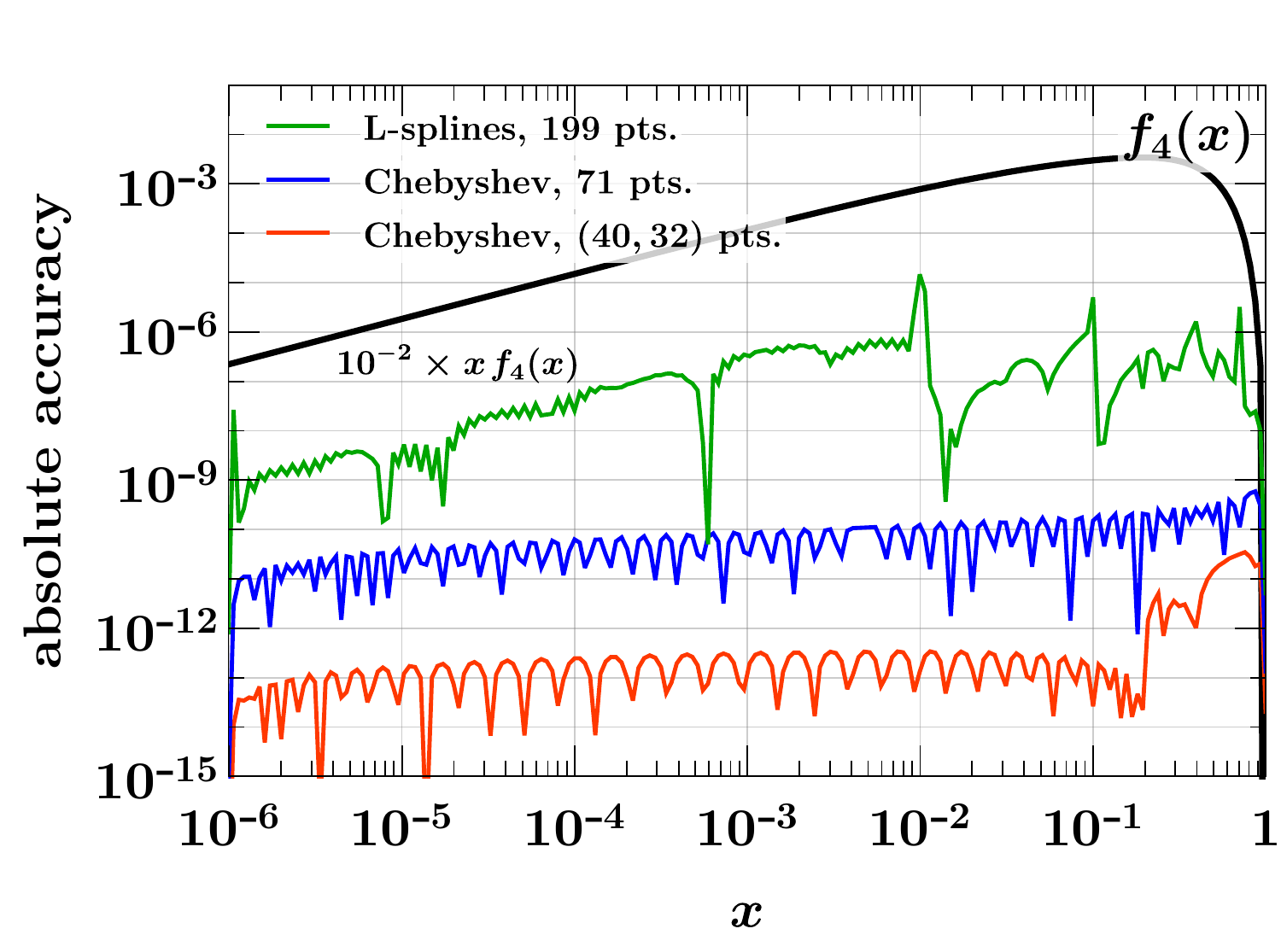}%
\caption{%
As \protect\fig{interpolation_comparison_subgrids}, but showing the absolute
instead of relative interpolation accuracy for $x f(x)$, i.e.\ the absolute
difference between the interpolated and exact results.
}
\label{fig:interpolation_comparison_subgrids_abs}
\end{figure*}


In \fig{interpolation_comparison_MMHT} we compare the spline interpolation on
the low-density grid (MMHT2014 grid with $n_\pts = 64$) with Chebyshev
interpolation on a $(32,32)$-point grid, which has nearly the same total number
of points ($n_\pts = 63$). The M-splines turn out to be more accurate than the
L-splines, but this comes at the expense of them being more complex to
construct. The Chebyshev interpolation achieves a significantly higher accuracy
by several orders of magnitude than either of the splines.  This reflects that,
contrary to splines, Chebyshev interpolation uses polynomials of a high degree.
In \fig{interpolation_comparison_HERA}, we compare splines on the high-density
grid (HERAPDF2.0 grid with $n_\pts = 199$) with Chebyshev interpolation on a
$(40,32)$-point grid with a total of $n_\pts = 71$.  Here, even with less than
half the number of points, the Chebyshev interpolation achieves several orders
of magnitude higher accuracy. This also highlights that the interpolation
accuracy scales much better with the number of points for Chebyshev
interpolation than for splines.

In \fig{interpolation_comparison_subgrids} we compare the accuracy of
interpolation on a single Chebyshev grid with the accuracy obtained with the two
subgrids used in \fig{interpolation_comparison_HERA}.  We see that
for the same total number of points, interpolation on two subgrids is more
accurate, as anticipated above.

We observe in figures \ref{fig:interpolation_comparison_MMHT} to
\ref{fig:interpolation_comparison_subgrids} that the relative accuracy varies
with $x$ for Chebyshev interpolation somewhat more than it does for
splines.  In fact, the \emph{absolute} accuracy of Chebyshev interpolation
varies much less with $x$, as can be seen by comparing
\fig{interpolation_comparison_subgrids} with
\fig{interpolation_comparison_subgrids_abs}.  The opposite holds for splines,
where the relative accuracy shows less variation than the absolute one.  This
reflects that Chebyshev interpolation is ``global'' over the full interpolation
interval, whilst splines are quite ``local'' (although the continuity conditions
for neighboring splines lead to some correlation over larger distances in $x$).

As is seen in figures \ref{fig:interpolation_comparison_MMHT} to
\ref{fig:interpolation_comparison_subgrids}, the relative accuracy degrades in
the limit \mbox{$x \to 1$} for both splines and Chebyshev interpolation.  This
is not surprising, because in this limit the PDFs in \eq{sample_functions}
approach zero. Moreover, they behave like $(1 - x)^\beta$ with noninteger
$\beta$ and are hence nonanalytic at $x=1$. This behavior cannot be accurately
reproduced by interpolating polynomials in the vicinity of $x=1$, whatever their
degree.
We emphasize that this problem concerns the \emph{relative} interpolation
accuracy.  As seen in \fig{interpolation_comparison_subgrids_abs}, the
\emph{absolute} error of interpolation does remain small for $x$ up to $1$, as
is expected from \eq{cheb_convergence}. This is sufficient for many practical
purposes, including the evaluation of convolution integrals that appear in cross
sections or evolution equations.  If high relative accuracy is required at large
$x$, one needs to use subgrids with a sufficient number of points tailored to
the region of interest.

\subsection{Differentiation and integration}
\label{sec:differentiation}

We now turn to Chebyshev interpolation for derivatives of the function
$\tilde{f}(x) = x f(x)$.  We use the barycentric formula \eqref{eq:bary_deriv}
and its analog for the second derivative, multiplying of course with the
Jacobian for the variable transformation from $t$ to $x$.  For comparison, we
also take the numerical derivative $( \tilde{f}_{\text{sp}}(x+h) -
\tilde{f}_{\text{sp}}(x-h) ) / 2h$ of the L-spline interpolant
$\tilde{f}_{\text{sp}}(x)$.  We use a variable step size $h = 10^{-4} \, x$,
having verified that the result remains stable when decreasing $h$ even further.
The second derivative of $\tilde{f}_{\text{sp}}$ is evaluated in an analogous
way.

\begin{figure*}[t!]
\centering
\includegraphics[width=\WidthTwoSubfigs]{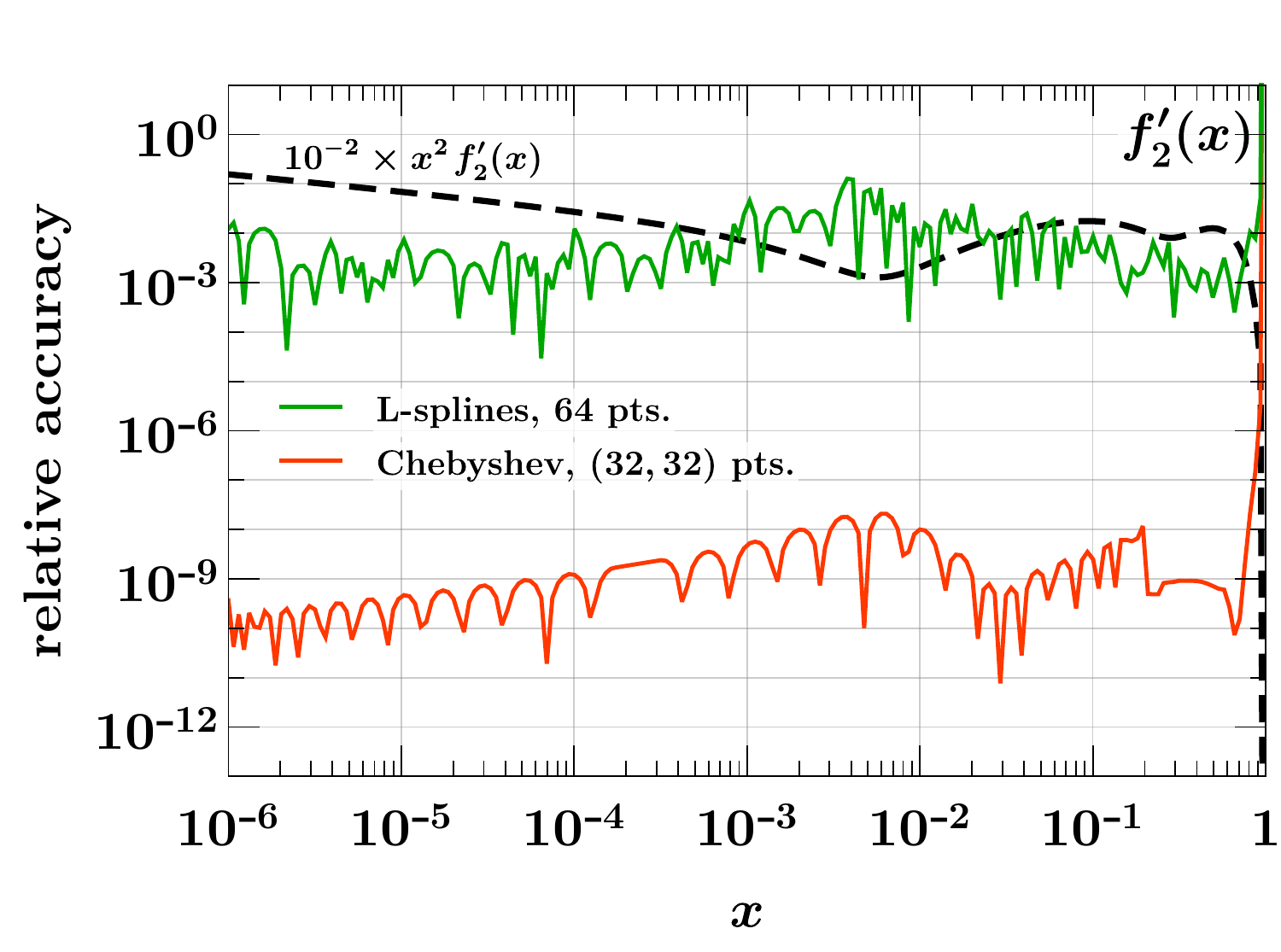}%
\includegraphics[width=\WidthTwoSubfigs]{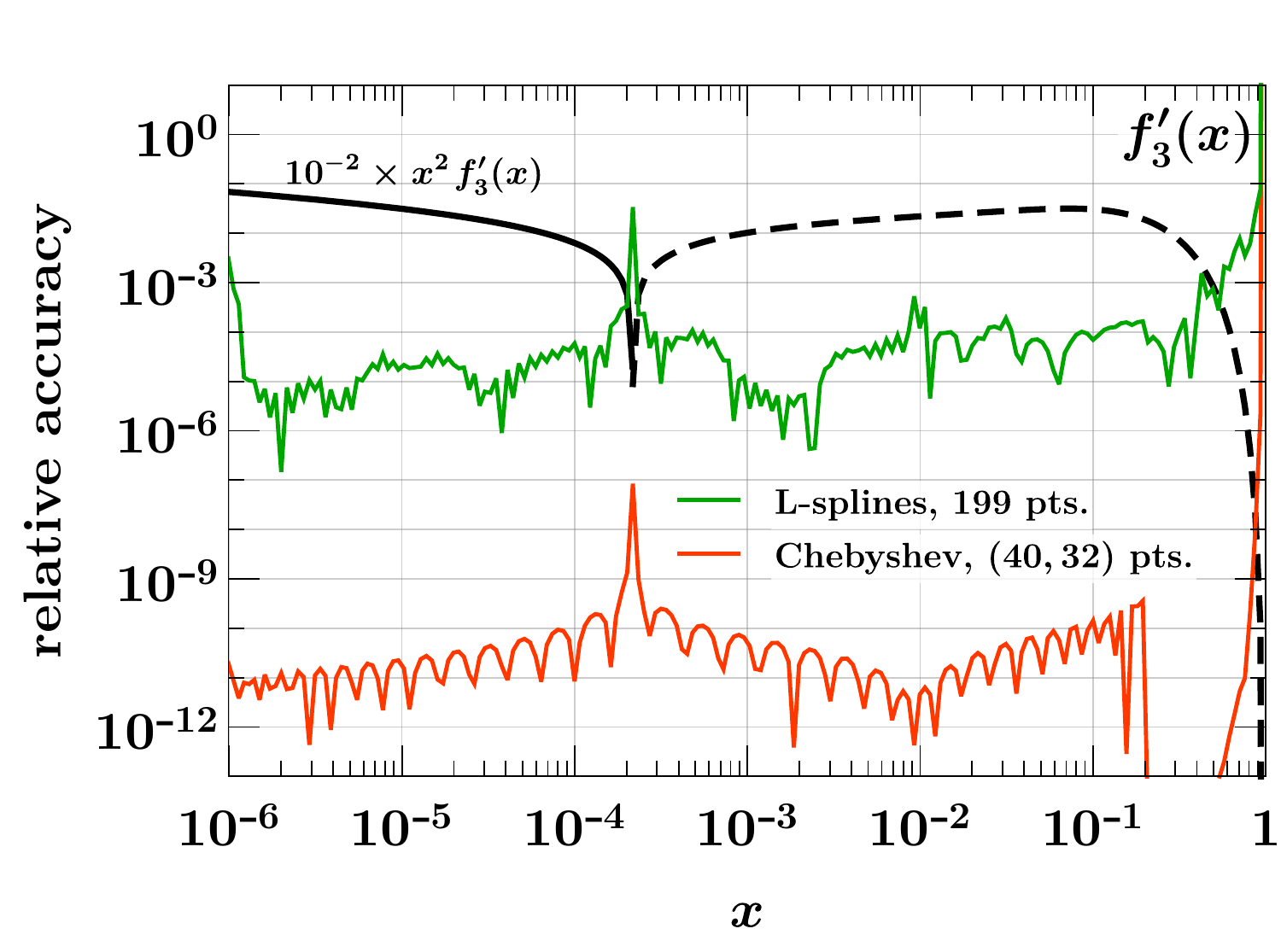}%
\\%
\includegraphics[width=\WidthTwoSubfigs]{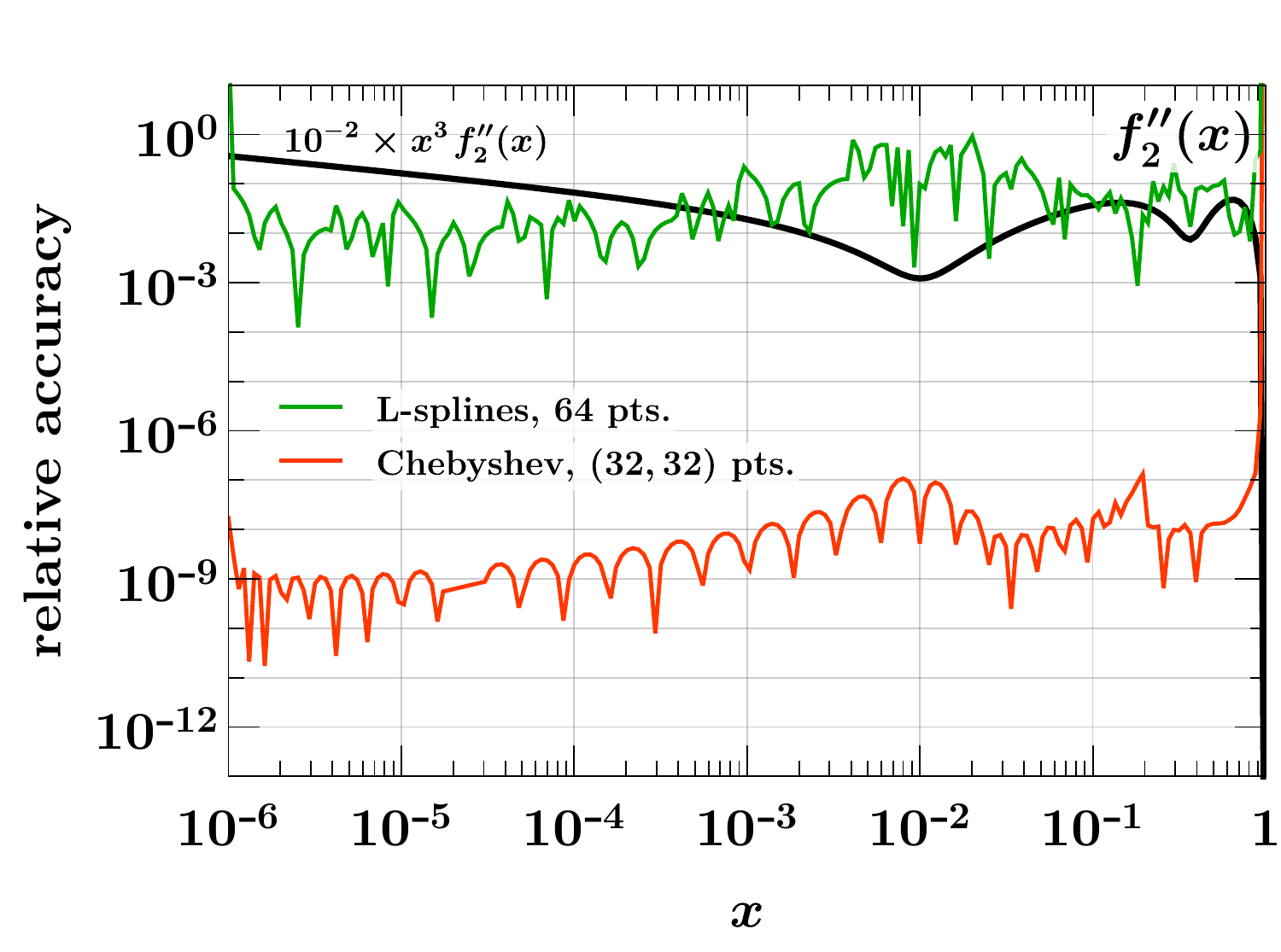}%
\includegraphics[width=\WidthTwoSubfigs]{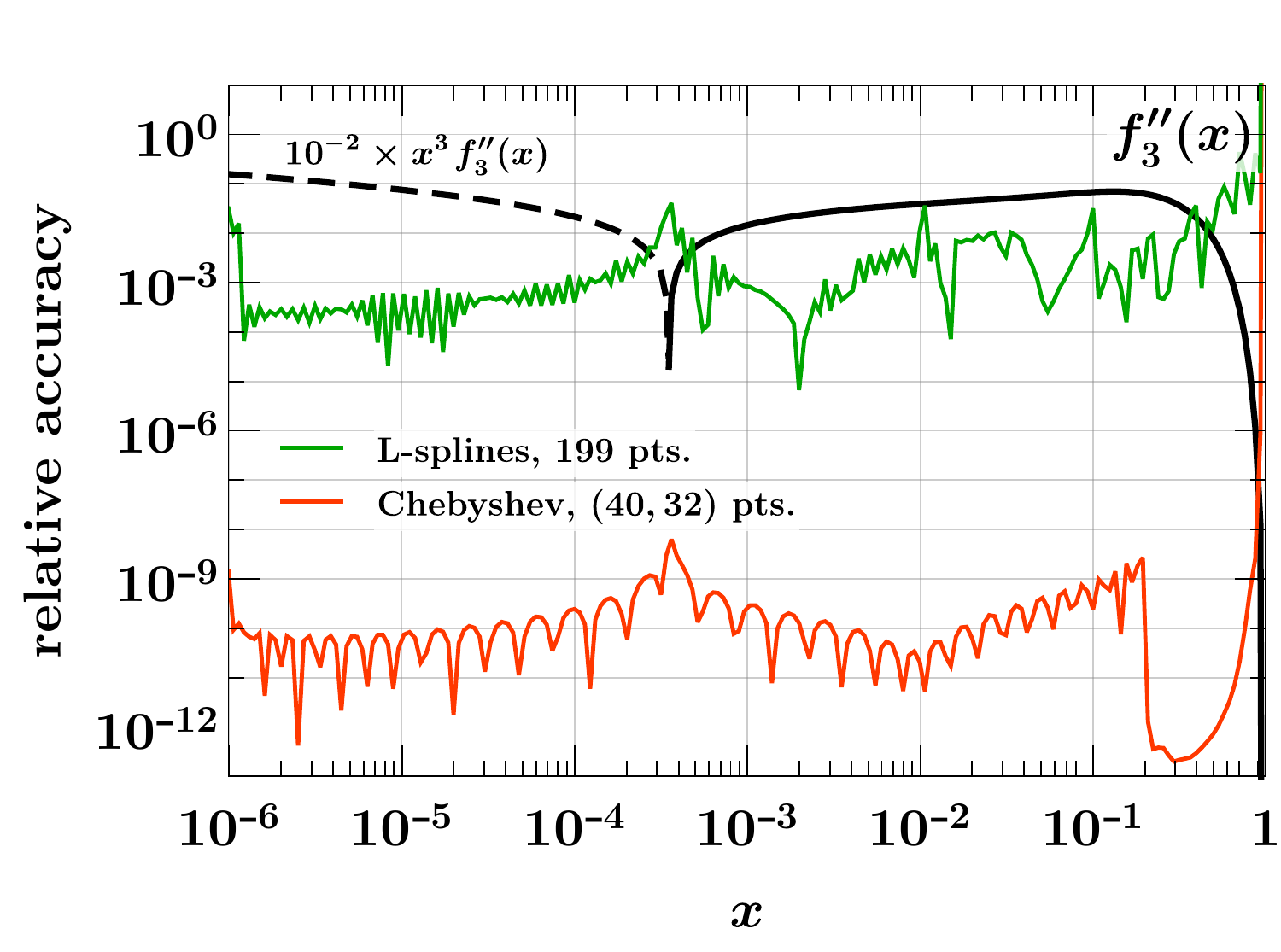}%
\caption{%
Relative accuracy of the first derivative (top) and second derivative (bottom)
for Chebyshev interpolation (red) and numerical differentiation of L-splines
(green).  The grids used on the left have lower $n_\pts$ as in
\fig{interpolation_comparison_MMHT}, and those on the right have higher $n_\pts$
as in \fig{interpolation_comparison_HERA}.
}
\label{fig:derivatives}
\end{figure*}

In \fig{derivatives} we consider the quantities
\begin{align} \label{eq:pdf-derivative}
x^2 f^{\ms\prime}(x) &= x\ms \tilde{f}^{\ms\prime}(x) - \tilde{f}(x)
\,, &
x^3 f^{\ms\prime\prime}(x)
&= x^2\ms \tilde{f}^{\ms\prime\prime}(x)
   - 2 x \ms \tilde{f}^{\ms\prime}(x) + 2 \tilde{f}(x)
\end{align}
and the accuracy of evaluating them using the two methods just described.  We
see that with Chebyshev interpolants, a significantly higher accuracy is
obtained.  This is not surprising, since the polynomials approximating the
derivatives are of a high degree in that case, whereas with cubic splines, one
locally has a quadratic polynomial for the first derivative and a linear
approximation for the second derivative.  For the low-density grid with $n_\pts
= 64$, the L-splines in fact give errors around $10\%$ for the first and $100\%$
for the second derivative in parts of the $x$ range.
We note that on each Chebyshev grid, the absolute accuracy of the derivatives
(not shown here) has a milder variation in $x$ than the relative one,
following the pattern we saw in \fig{interpolation_comparison_subgrids_abs} for
$x f(x)$ itself.

\begin{figure*}[t]
\centering
\includegraphics[width=\WidthTwoSubfigs]{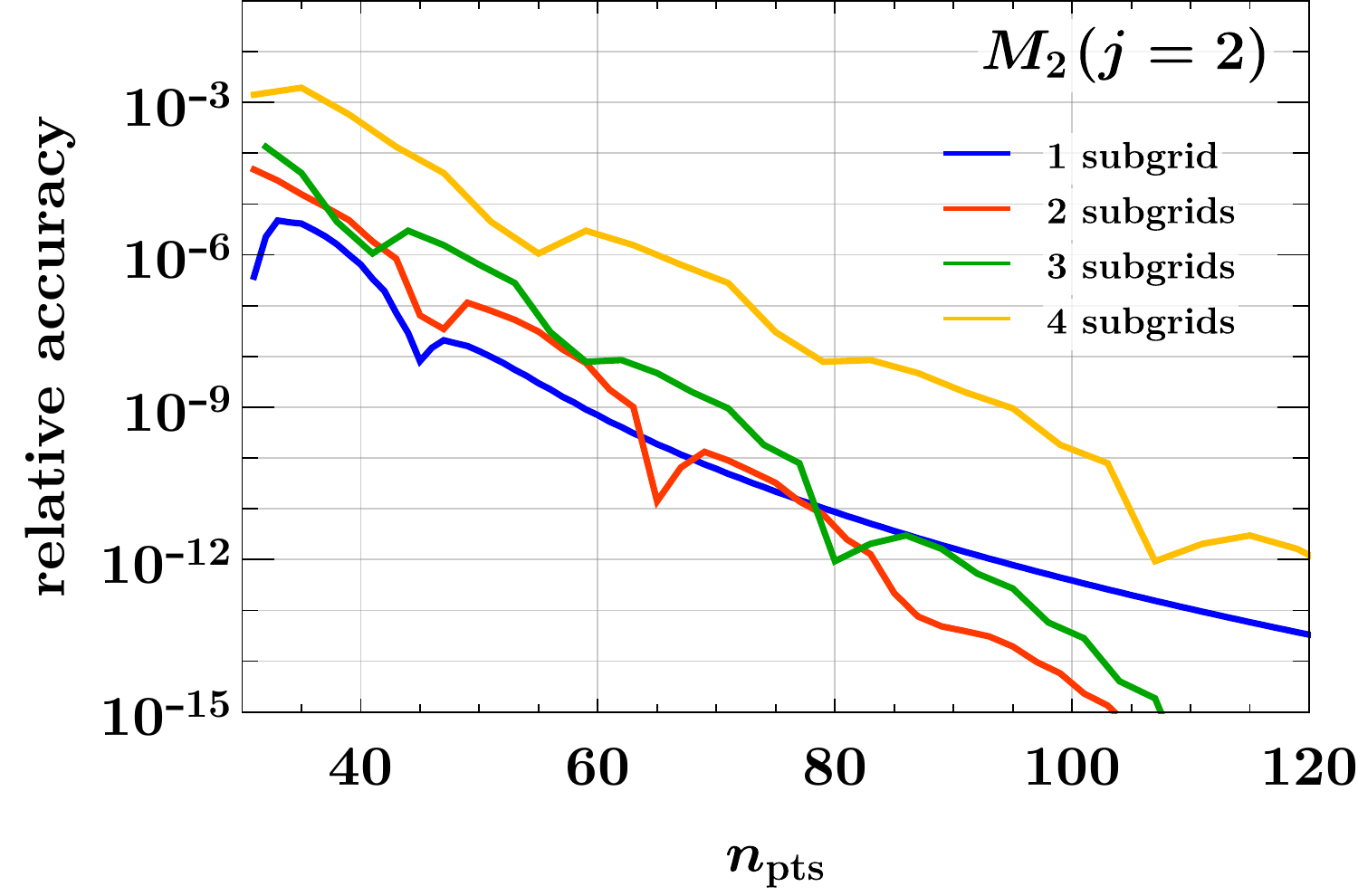}%
\includegraphics[width=\WidthTwoSubfigs]{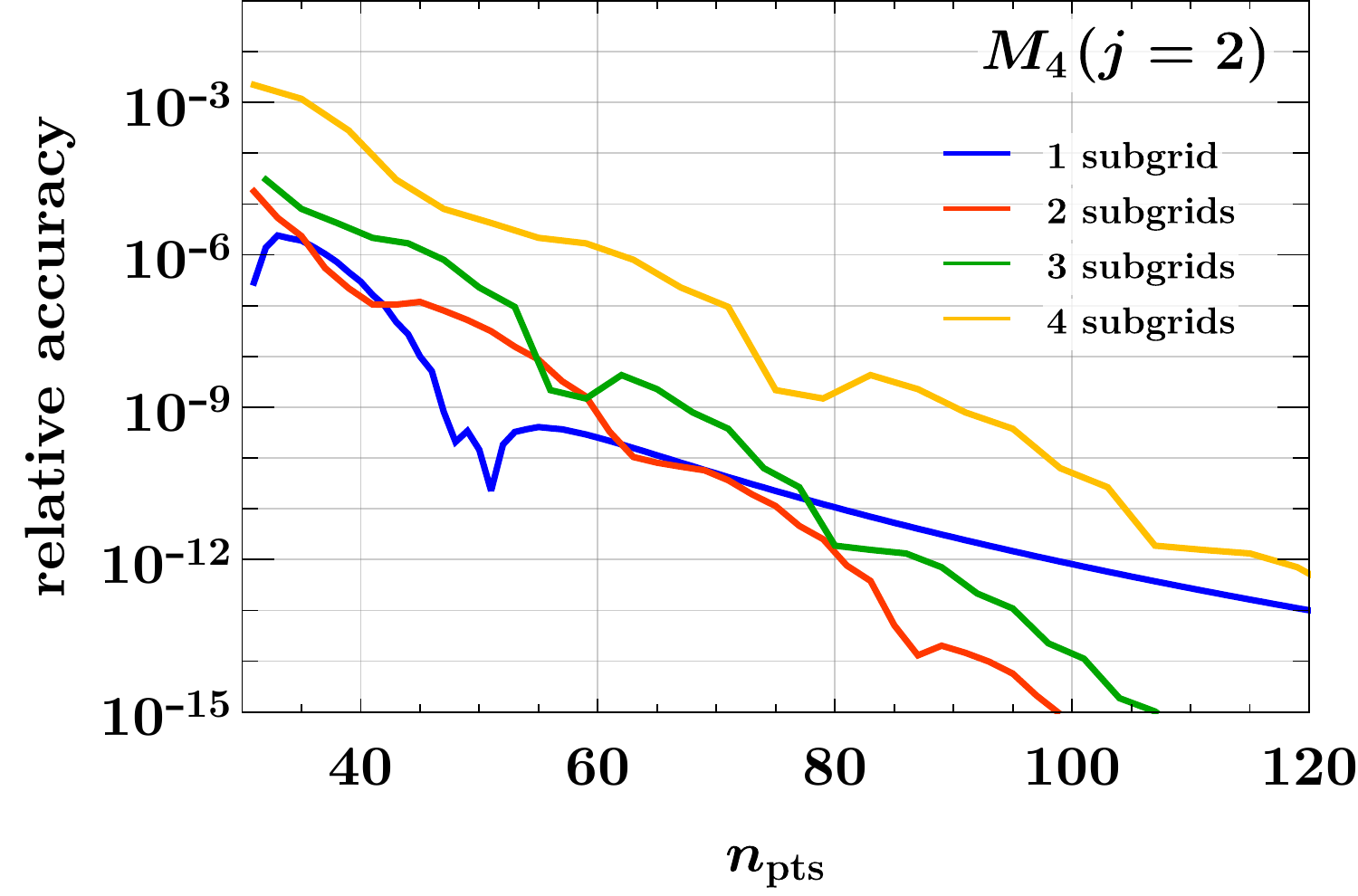}%
\\[15pt]%
\includegraphics[width=\WidthTwoSubfigs]{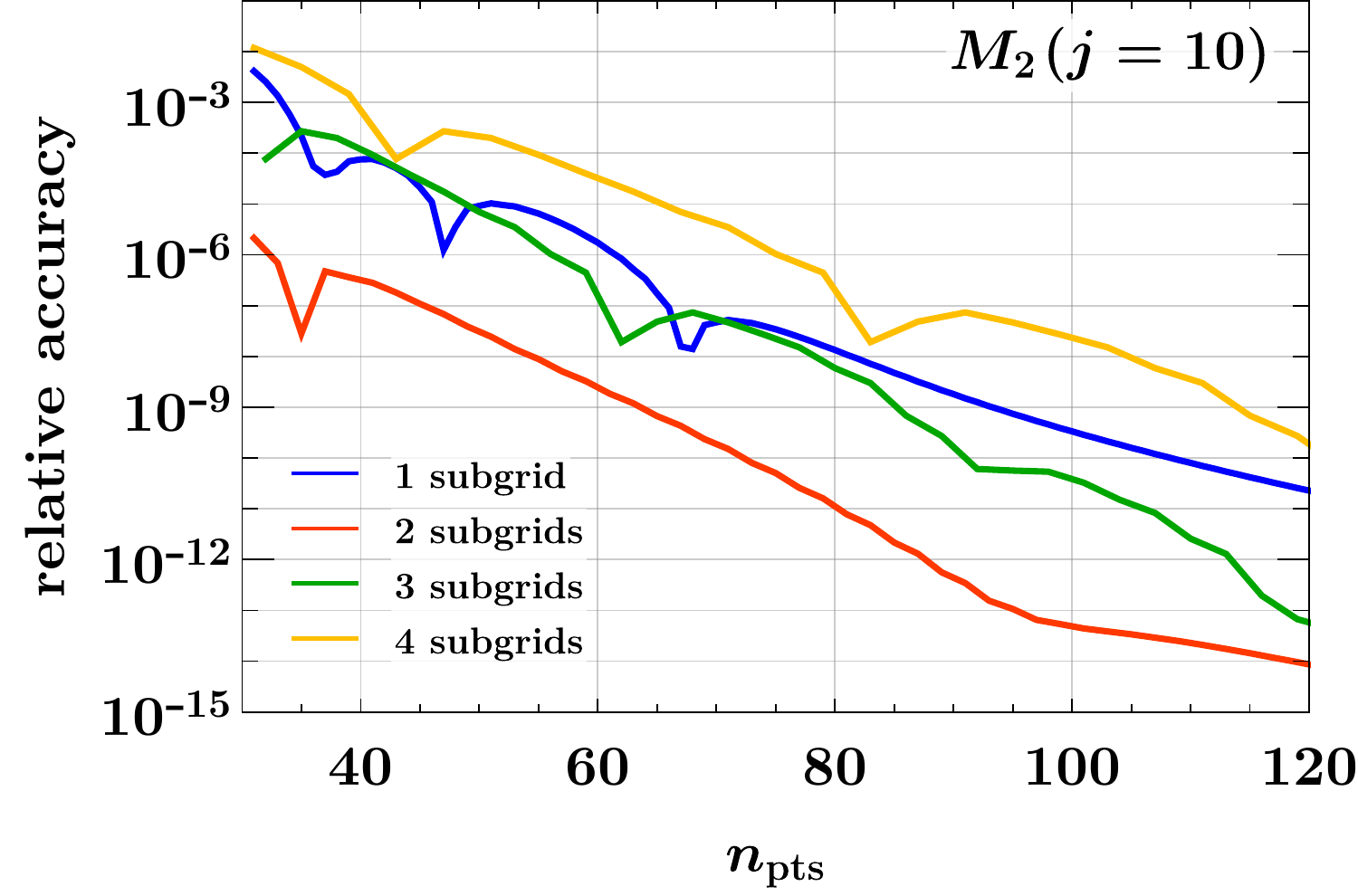}%
\includegraphics[width=\WidthTwoSubfigs]{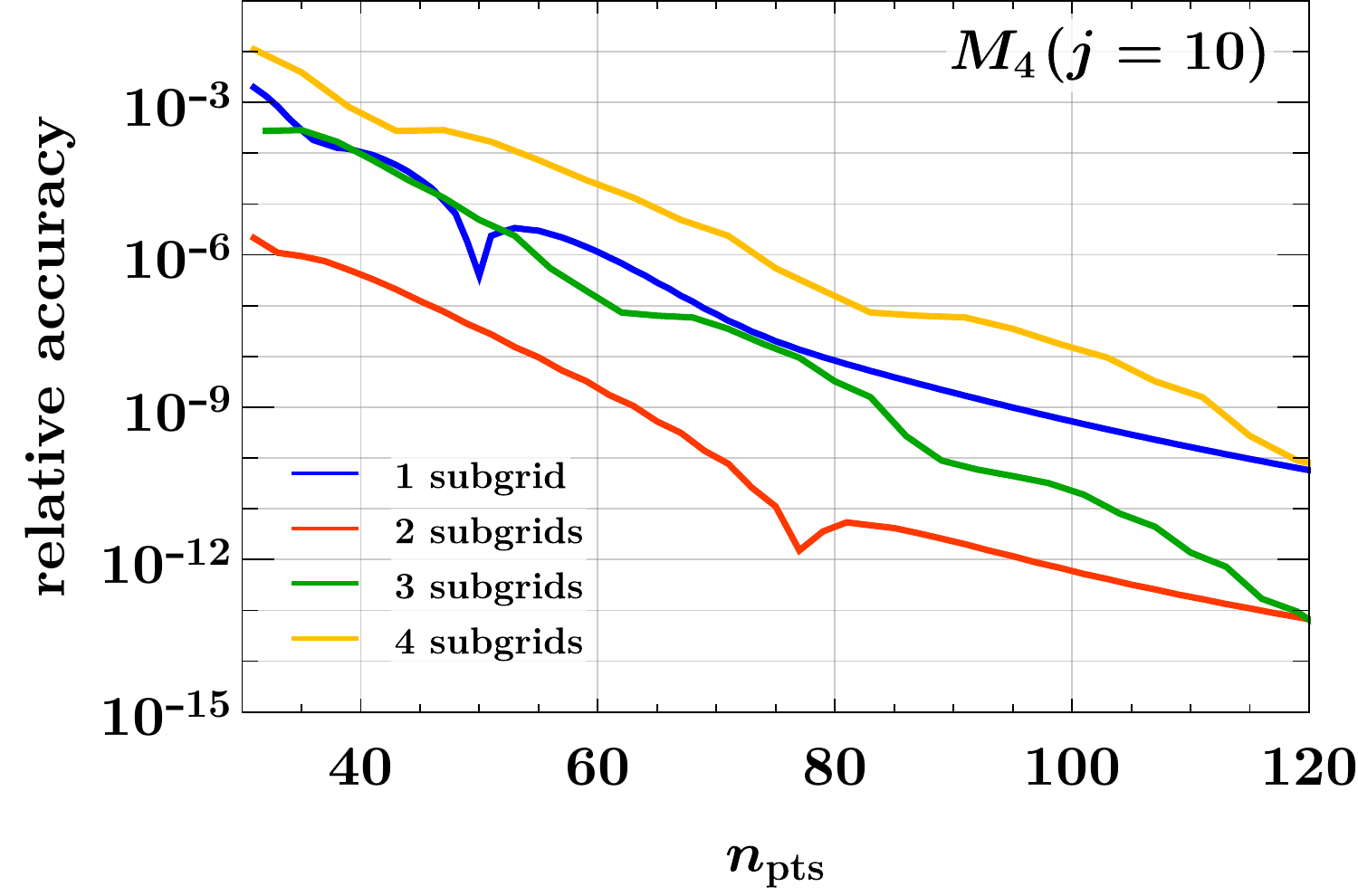}%
\caption{%
Relative accuracy of the truncated Mellin moments \eqref{eq:mellin-moment} with
$x_0 = 10^{-9}$ and with $j=2$ (top) or $j=10$ (bottom), for $f_2(x)$ (left) or
$f_4(x)$ (right). We use Chebyshev grids with $k=1$ to $4$ subgrids and
$n_\pts/k$ points per subgrid, namely
$[x_0, 1]$ (blue),
$[x_0, 0.2, 1]$ (red),
$[x_0, 10^{-3}, 0.5, 1]$ (green), and
$[x_0, 10^{-6}, 10^{-3}, 0.5, 1]$ (yellow).
The results for different $n_\pts$ are connected by lines to guide the eye.
}
\label{fig:mellin_moments}
\end{figure*}

To explore the accuracy of numerical integration, we consider the truncated
Mellin moments
\begin{align} \label{eq:mellin-moment}
M_i(j) &= \int_{x_{0}}^1 \dd x\; x^{j - 1}\, f_i(x)
\,,\end{align}
where the lower integration limit is the lowest point $x_0$ of the grid.  In
\fig{mellin_moments}, we show the dependence of the relative accuracy on the
total number of grid points when using $1$ to $4$ subgrids of equal size. The
advantage of using subgrids is clearly seen, especially for high moment index $j$,
which
emphasizes the large $x$ region of the integrand. We find that taking 2 or 3
subgrids gives best results for a wide range of $j$.  The disadvantage of using
interpolants with lower polynomial degree becomes the dominant limiting factor
with 4 or more subgrids.

\subsection{Estimating the numerical accuracy}
\label{sec:error_estimate}

Let us now take a closer look at methods to estimate the numerical accuracy of
interpolation or integration with Chebyshev polynomials.  An obvious option is
to re-compute the quantity of interest with an increased number of points.
However, the examples in \fig{mellin_moments} show that the accuracy is not a
strictly decreasing function of $n_\pts$, and we see fluctuations with local
minima and maxima as $n_\pts$ varies by about $10$ units.  For a reliable
estimate, one should therefore take a sufficiently large increase in $n_\pts$.
Increasing $n_\pts$ in several steps provides an additional way of ensuring that
the estimate is trustworthy.

The procedure just described can give sound accuracy estimates and is what we
will adopt for assessing the accuracy of DGLAP evolution in \sec{accu}.
However, since the result must be evaluated multiple times on increasingly dense
grids, this procedure can
be computationally expensive, especially when the cost scales more
than linearly with the number of grid points.  This is indeed the case of DGLAP
evolution, where the evaluation of convolution integrals involves $n_\pts \times
n_\pts$ matrices.

\begin{figure*}[t]
\centering
\centering
\includegraphics[width=\WidthTwoSubfigs]{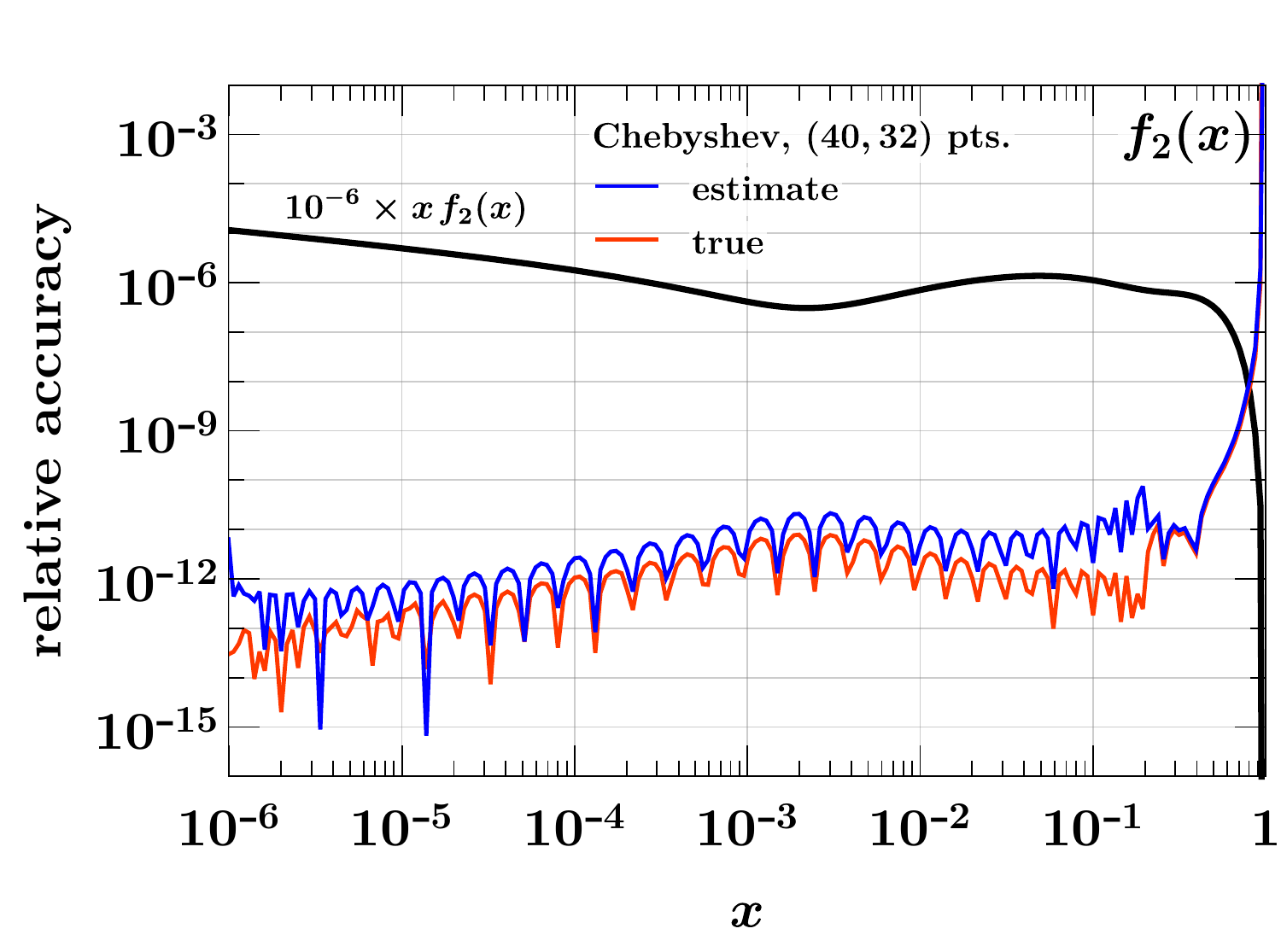}%
\includegraphics[width=\WidthTwoSubfigs]{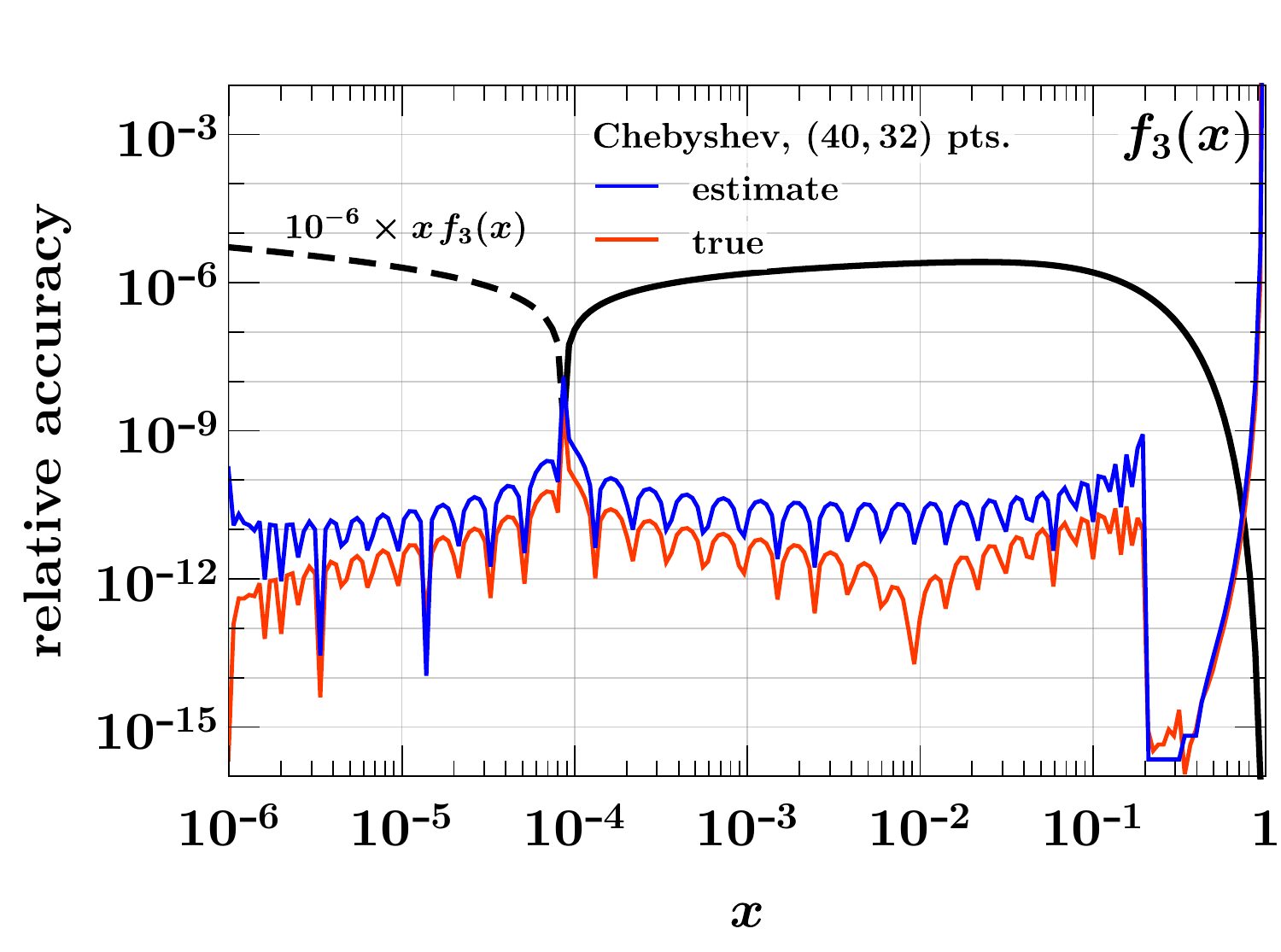}%
\caption{%
Comparison of the true relative interpolation accuracy (red) and its estimate
(blue) obtained from the difference between interpolation on the Chebyshev grid
with and without its end points.  The $(40, 32)$-point grid used is the same as
in \fig{interpolation_comparison_HERA}.
}
\label{fig:interpolation_error_est}
\end{figure*}

\begin{figure*}
\centering
\includegraphics[width=\WidthTwoSubfigs]{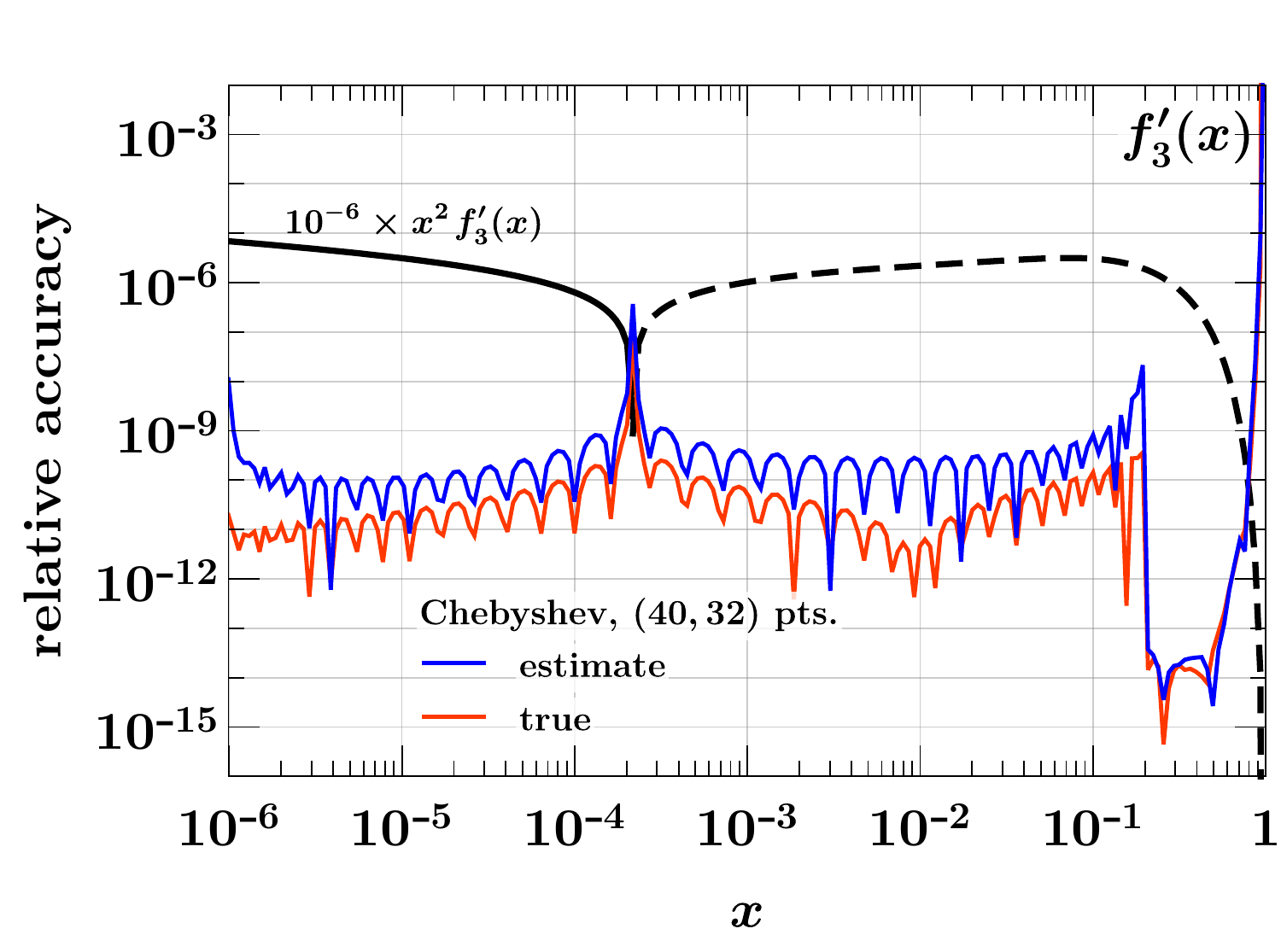}%
\includegraphics[width=\WidthTwoSubfigs]{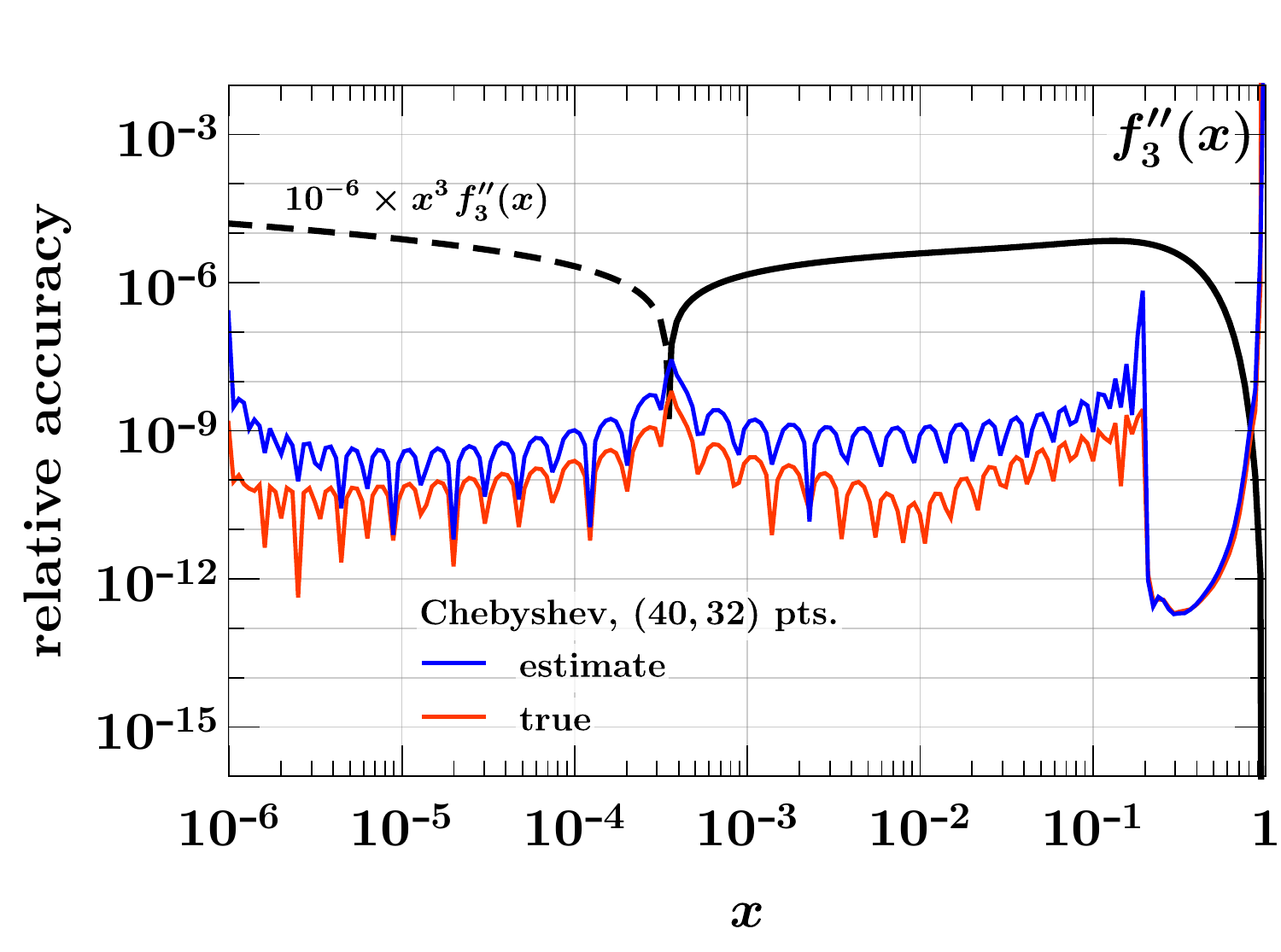}%
\caption{%
Comparison of true (red) and estimated (blue) interpolation accuracy as in
\fig{interpolation_error_est}, but for the first and second derivative of the
sample PDF $f_3(x)$.
}
\label{fig:differentiation_error_est}
\end{figure*}

An alternative with much less computational overhead is to estimate the accuracy
from the difference between Chebyshev interpolation and interpolation on the
same grid without the end points, as described in \sec{chebyshev}.  This does
not require any additional function evaluations. We compare this estimate with
the actual interpolation accuracy in \fig{interpolation_error_est}.   In the
subinterval for large $x$, the estimate is very close to the true error, whilst
in the subinterval for small to intermediate $x$ it somewhat overestimates the
actual numerical error.  The amount of overestimation is highest close to the
interval limits, which is not surprising, because this is where the two
interpolation methods differ most. The corresponding comparison for
differentiation is shown in \fig{differentiation_error_est} and follows the same
pattern. We also note that the quality of the estimate for other sample PDFs is
as good as or even better than in the examples shown here.

\begin{figure*}
\centering
\includegraphics[width=\WidthTwoSubfigs]{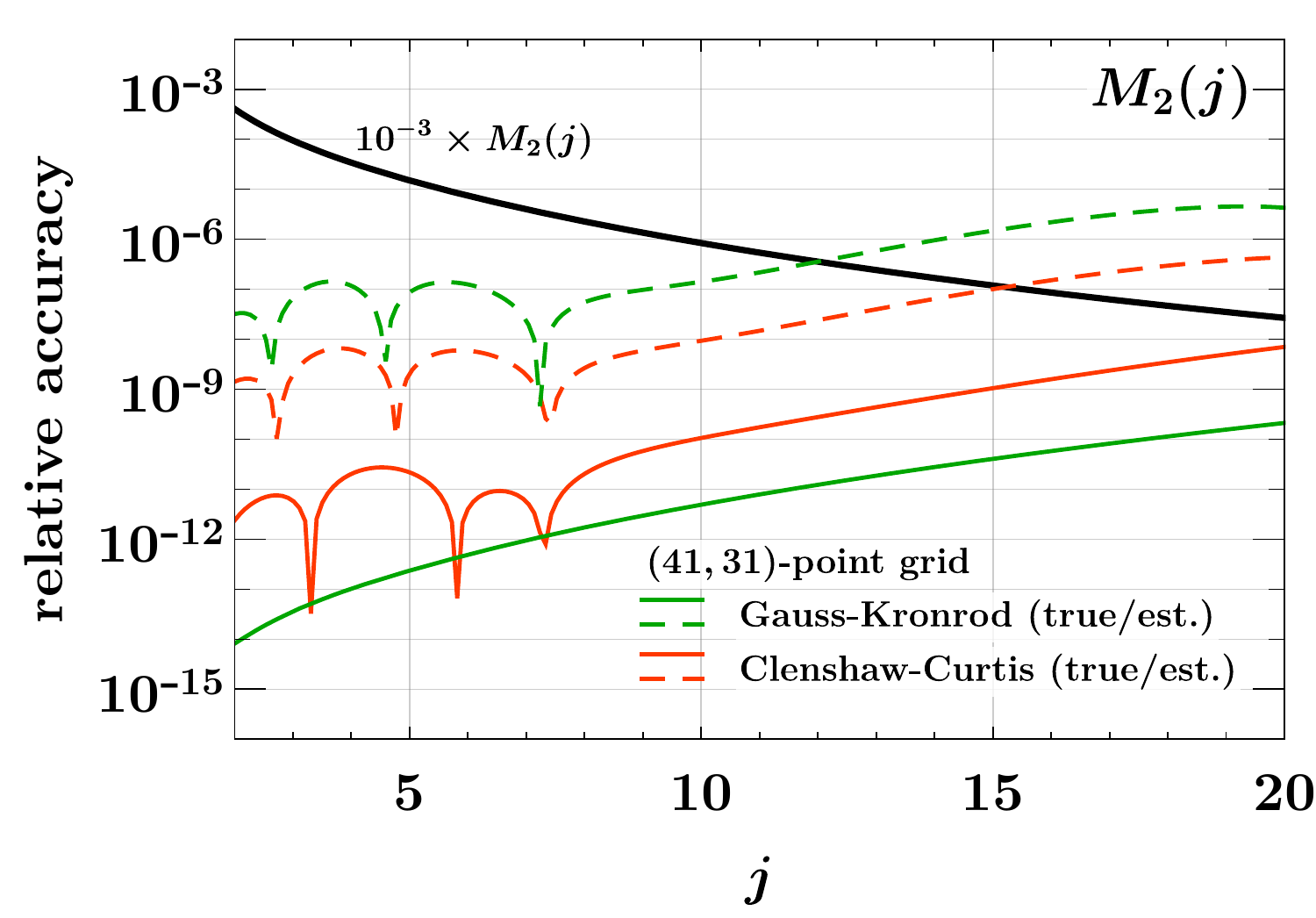}%
\includegraphics[width=\WidthTwoSubfigs]{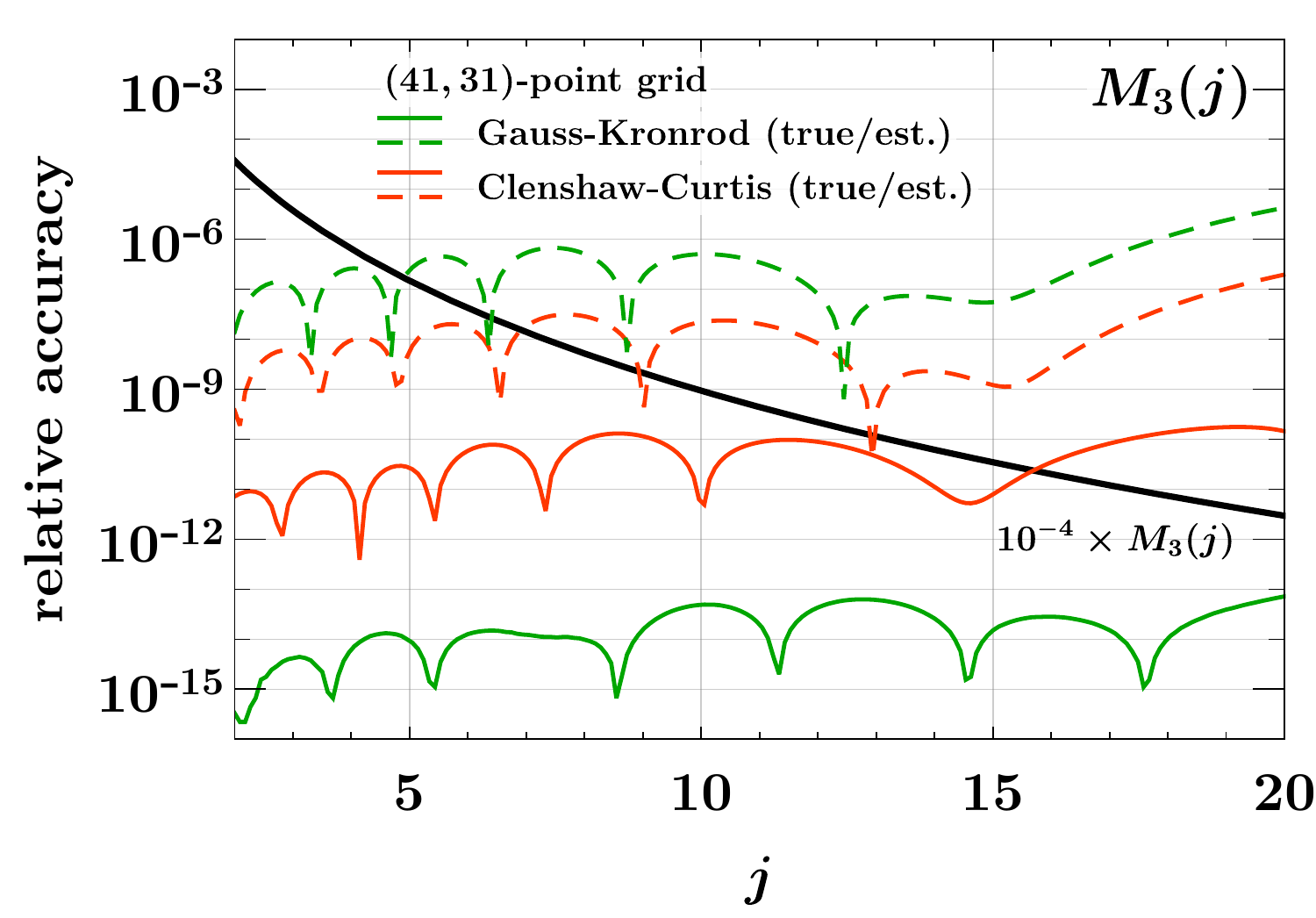}%
\caption{%
Comparison of the relative integration accuracy (solid) and its estimate
(dashed) for Clenshaw-Curtis (red) and Gauss-Kronrod (green) quadrature.  The
considered integrals are the truncated Mellin moments $M_i(j)$ of the test
functions $f_i(x)$ with $i=2,3$. For both methods we use two subgrids $[10^{-9},
0.2, 1]_{(41,\ms 31)}$ with the appropriate Chebyshev or Gauss-Kronrod points for
the integration variable $\ln x$. In
the accuracy estimates, the results for a given integration rule are first added
for the two subintervals, and then the absolute difference between the involved
higher-order and lower-order rules is taken.
}
\label{fig:integration_error_est}
\end{figure*}

The analog for integration of the previous method is to estimate the accuracy
of Clenshaw-Curtis integration from the difference to the Fej{\'er} quadrature
rule.%
\footnote{We always refer to what is known in the literature as
Fej{\'e}r's \emph{second} rule, given in \eq{Fejer-rule}.}
In \fig{integration_error_est}, we compare this estimate with the actual
numerical error for the Mellin moment \eqref{eq:mellin-moment}.
We find that it tends to overestimate the actual error
by two to three orders of magnitude, which is more than what we saw for
interpolation.
For comparison, we also show the relative accuracy obtained with the widely used
Gauss-Kronrod rule, using the same two subintervals and the same number of grid
points per subinterval as for Clenshaw-Curtis. As is commonly done, the accuracy estimate for
Gauss-Kronrod is obtained from the difference between the nominal Gauss-Kronrod
rule and the lower-order Gauss rule on a subset of the grid points. We see that the
Gauss-Kronrod rule has
the highest accuracy.  In practical applications, the true result is however not
known, and an integrator is only as good as its accuracy estimation. Here, the
accuracy estimate for Gauss-Kronrod is much worse than our estimate for
Clenshaw-Curtis.%
\footnote{Expert readers will know that it is in fact not
uncommon for Gauss-Kronrod integrators to have overly conservative accuracy
estimates.}
As discussed at the end of \sec{chebyshev}, this is not unexpected
given the comparison of polynomials orders, but it is interesting to see that
the quantitative effect can be as pronounced as in our examples.  We also note
that the quantitative differences between the integration methods vary
significantly with the shape of the integrands, as is evident from the two
panels in \fig{integration_error_est}.

Overall, as far as numerical accuracy estimates are concerned, using
interpolation, differentiation, or integration without the end points of a
Chebyshev grid can be considered as an inexpensive estimate of reasonable
quality. It is not an alternative to the high-quality estimate one can achieve
by a stepwise increase in the number of grid points. Its biggest advantage is
thus to provide a fast and reliable indicator whether the accuracy of a result
is sufficient or calls for a dedicated, more expensive investigation.

\section{Mellin convolution}
\label{sec:mellin_convolution}

In this section, we consider the Mellin convolution of a PDF with an integral
kernel, such as a DGLAP splitting function, a beam-function matching kernel, or a
hard-scattering coefficient.  In terms of the scaled PDF $\tilde{f}(x) = x
f(x)$, we wish to compute
\begin{align} \label{eq:mellin_convolution}
\bigl( K \otimes \tilde{f} \,\bigr) (x)
&= \int_x^1 \frac{\dd z}{z} \, K(z) \, \tilde{f} \left( \frac{x}{z} \right)
\,.\end{align}
Here, $K(z)$ is the scaled kernel, given e.g.\ by $K(z) = z \ms P(z)$ for a DGLAP splitting function $P(z)$.\footnote{We recall that $h_1 = h_2 \otimes h_3$
implies $\tilde{h}_1 = \tilde{h}_2 \otimes \tilde{h}_3$ with $\tilde{h}_i(z) = z
h_i(z)$.}

Let us see how \eq{mellin_convolution} can be evaluated in a discretized
setting.  For simplicity, we first consider a single Chebyshev grid
for the range $x_0 \le x \le 1$ as described in \sec{interpolation_strategy}.
The convolution $(K\otimes \tilde f)(x)$ is a function of $x$ over the same
domain as $\tilde f(x)$. Hence, we can interpolate it in complete analogy to
$\tilde f(x)$ itself. For this purpose, we only need to evaluate
\eq{mellin_convolution} at the grid points $x_i$,
\begin{align} \label{eq:mellin_convolution_chebpts}
\bigl( K \otimes \tilde{f} \,\bigr) (x_i)
&= \int_{x_i}^1 \! \frac{\dd{z}}{z} \, K(z) \,
   \tilde{f} \left( \frac{x_i}{z} \right)
\approx \int_{x_i}^1 \frac{\dd{z}}{z} \, K(z) \,
   \sum_{j = 0}^N  \tilde{f}_j \, b_j \left( \ln \frac{x_i}{z} \right)
\,,\end{align}
where in the last step we used the barycentric formula
\eqref{eq:barycentric_pdf} to interpolate $\tilde{f} (x_i/z)$ under the
integral.  Introducing the \emph{kernel matrix}
\begin{align} \label{eq:kernel_matrix_def}
K_{i j}
&= \int_{x_i}^1 \frac{\dd{z}}{z} \, K(z) \,
   b_j \left( \ln \frac{x_i}{z} \right)
\,,\end{align}
we can compute the function values of $K\otimes \tilde f$ on the grid by a
simple matrix multiplication,
\begin{align} \label{eq:mellin_product}
\bigl( K \otimes \tilde{f} \,\bigr) (x_i)
&\approx \sum_{j = 0}^N K_{i j} \ms \tilde{f}_j
\,.\end{align}
Importantly, the matrix $K_{i j}$ only depends on the given kernel and grid but
not on $\tilde f$. It can thus be pre-computed once using standard numerical
integration routines.

In general, $K(z)$ is a distribution rather than an ordinary function.
Restricting ourselves to plus and delta distributions, we write
\begin{align}
\label{eq:mellin_kernel_structure}
K(z) &= K_\text{sing} (z) + K_\text{reg} (z) + \delta(1-z) \, K_\delta
\,,\end{align}
where $K_\text{sing}(z)$ is the singular part of the kernel and contains plus
distributions. The function $K_\text{reg}(z)$ contains at most an integrable
singularity at $z=1$, for instance powers of $\ln(1-z)$.  The separation between
$K_{\text{sing}}$ and $K_{\text{reg}}$ is not unique and may be adjusted as is
convenient.  The convolution \eqref{eq:mellin_convolution} can now be written as
\begin{align} \label{eq:mellin_convolution_pieces}
\bigl( K \otimes \tilde{f} \,\bigr) (x)
&= \int_x^1 \frac{\dd{z}}{z} \, K_\text{sing} (z)
   \left[ \tilde{f} \left(\frac{x}{z}\right) - \tilde{f}(x) \right]
   + \int_x^1 \! \frac{\dd{z}}{z} \, K_\text{reg} (z) \,
      \tilde{f} \left(\frac{x}{z}\right)
   + K_d(x) \, \tilde{f}(x)
\,,\end{align}
where
\begin{align} \label{eq:Kd-def}
K_d(x)
&= K_\delta + \int_x^1 \frac{\dd{z}}{z} \, K_\text{sing} (z)
\end{align}
can be evaluated analytically if the separation between $K_{\text{sing}}$ and
$K_{\text{reg}}$ is chosen appropriately.  Due to the explicit subtraction
of $\tilde f(x)$ in the first integral in \eq{mellin_convolution_pieces}, the
plus prescription in $K_{\text{sing}}$ can now be omitted, because
\begin{align}
\int_x^1 \! \dd{z} \, \bigl[ g(z) \bigr]_+ \; h(z)
&= \int_x^1 \! \dd{z} \, g(z) \, h(z)
\qquad\text{for}\quad h(1) = 0
\,.\end{align}
From the definition \eqref{eq:Kd-def} one readily finds that  $K_d(x)$ diverges
like $\ln^{n+1}(1-x)$ for $x\to 1$ if $K(z)$ contains a term
$\mathcal{L}_n(1-z)$, where we write
\begin{equation}
\mathcal{L}_n (y) = \biggl[\frac{\ln^n (y)}{y}\biggr]_+
\end{equation}
for the logarithmic plus distribution of degree $n$.  However, a PDF vanishes
much faster for $x\to 1$ than $K_d(x)$ diverges, such that the product $K_d(x)
\, \tilde{f}(x)$ tends to zero in that limit.

Applying \eq{mellin_convolution_pieces} to the evaluation of the $K_{ij}$ matrix
in \eq{kernel_matrix_def}, we have
\begin{align} \label{eq:kernel_mat_final}
K_{i j}
&= \int_{u_i}^0 \! \dd v \; K_{\text{sing}}(e^v) \,
   \bigl[ b_j(u_i - v) - \delta_{i j} \bigr]
   + \int_{u_i}^0 \! \dd v \; K_{\text{reg}}(e^v) \, b_j(u_i - v)
   + \delta_{i j}\, K_d(x_i)
\,,\end{align}
where $u_i$ and $u_j$ are grid points in $u$, and where we have changed the
integration variable to $v = \ln z$.  The plus prescription in $K_{\text{sing}}$
can again be omitted, because the term in square brackets vanishes at least
linearly in $v$ for $v \to 0$. Hence, all integrals can be evaluated
numerically, for which we use an adaptive Gauss-Kronrod routine.

Given that $x_N = 1$, the matrix element $K_{N \bs N}$ is ill defined if $K$
contains a plus distribution, because $K_d(x)$ diverges for $x\to 1$ in that
case.  This does not present any problem: we already noticed that \rev{$K_d(x) \,
\tilde{f}(x)\to 0$} for $x\to 1$, which translates to $K_{N \bs N} \ms
\tilde{f}_N = 0$ in the discretized version.  This term may hence be omitted in
the matrix multiplication \eqref{eq:mellin_product}, along with all other terms
$K_{i N} \ms \tilde{f}_N$.

It is not difficult to generalize the preceding discussion to the case of
several subgrids in~$u$, each of which is mapped onto a Chebyshev grid.  One
then has a distinct set of barycentric basis functions for each subgrid.  If $u_j$
and $u_i - v$ are not on the same subgrid, then the basis function $b_j(u_i -
v)$ in  \eq{kernel_mat_final} must be set to zero.  As a consequence, the matrix
$K_{i j}$ has blocks in which all elements are zero.  Let us, however, note that
$K_{i j}$ always has nonzero elements below the diagonal $i=j$ and is \emph{not}
an upper triangular matrix.

We now study the numerical accuracy of this method relative to the exact result,
which we obtain by the direct numerical evaluation of the convolution integral
\eqref{eq:mellin_convolution_pieces}. For comparison, we also show the accuracy
of the result obtained by approximating $\tilde{f}(x)$ in
\eq{mellin_convolution_pieces} with its L-splines interpolant and performing the
convolution integral numerically. All explicit numerical integrals in this study
are performed at sufficiently high precision to not influence the results.

\begin{figure*}
\centering
\includegraphics[width=\WidthTwoSubfigs]{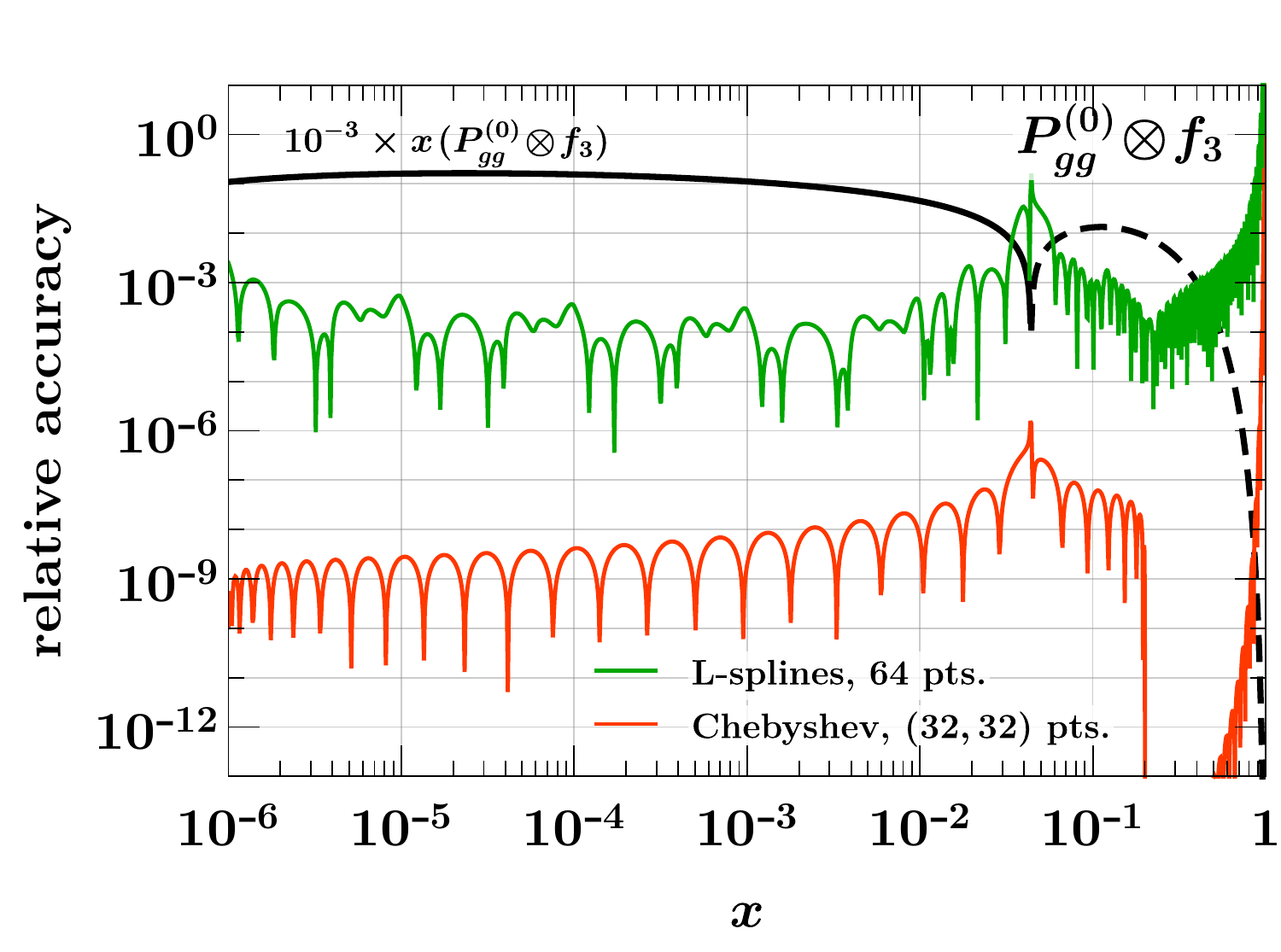}%
\includegraphics[width=\WidthTwoSubfigs]{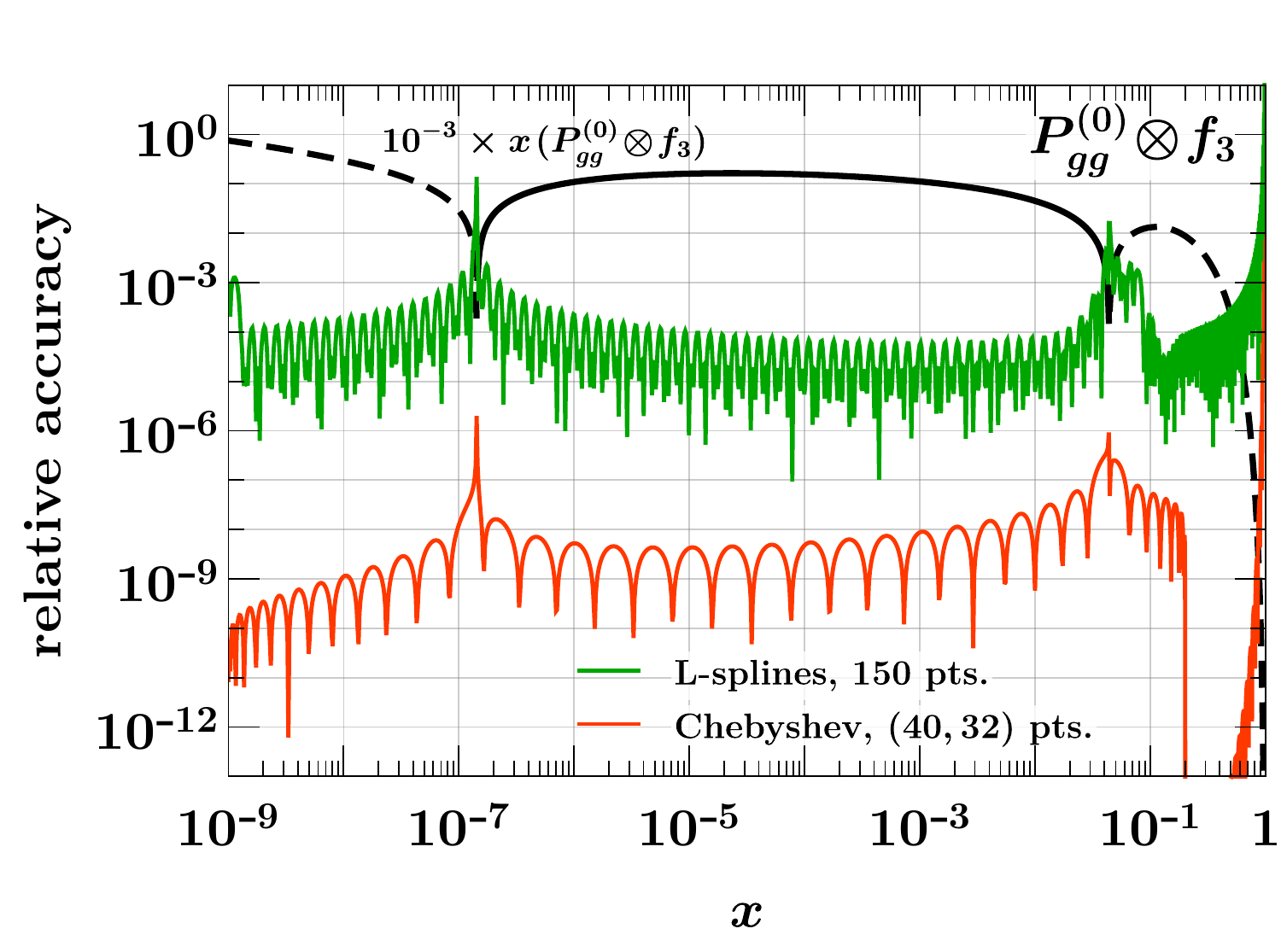}%
\\%
\includegraphics[width=\WidthTwoSubfigs]{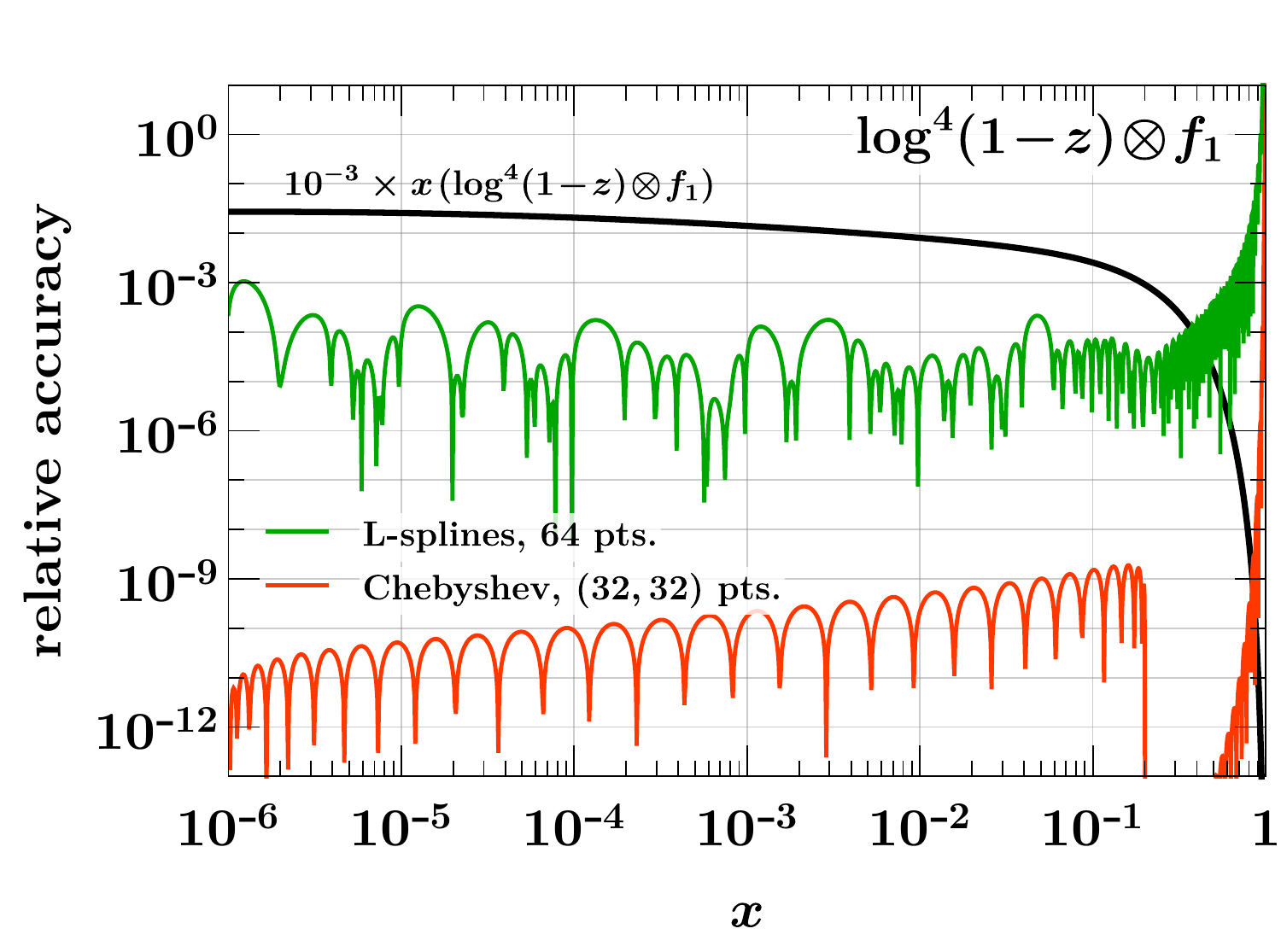}%
\includegraphics[width=\WidthTwoSubfigs]{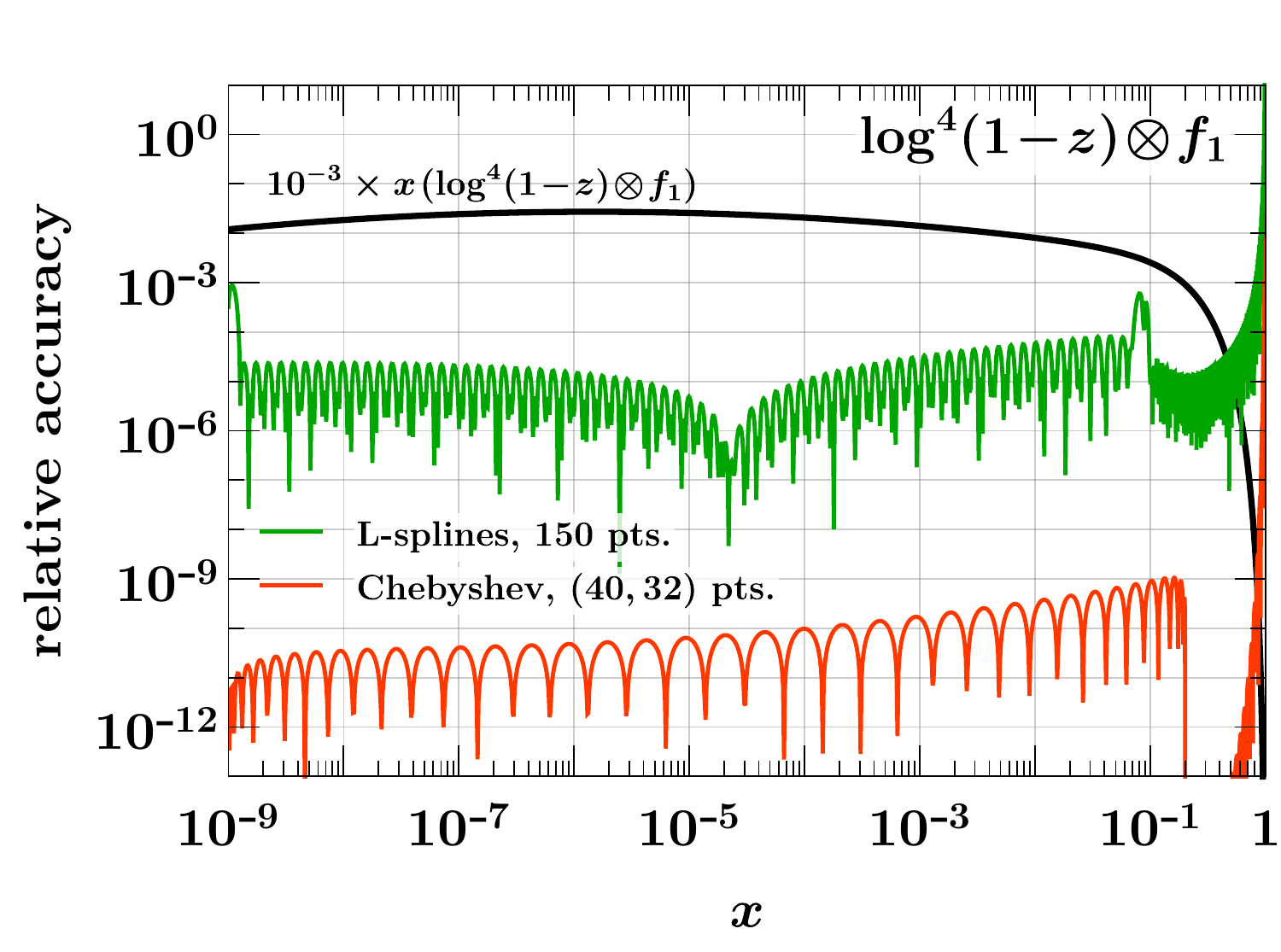}%
\\%
\includegraphics[width=\WidthTwoSubfigs]{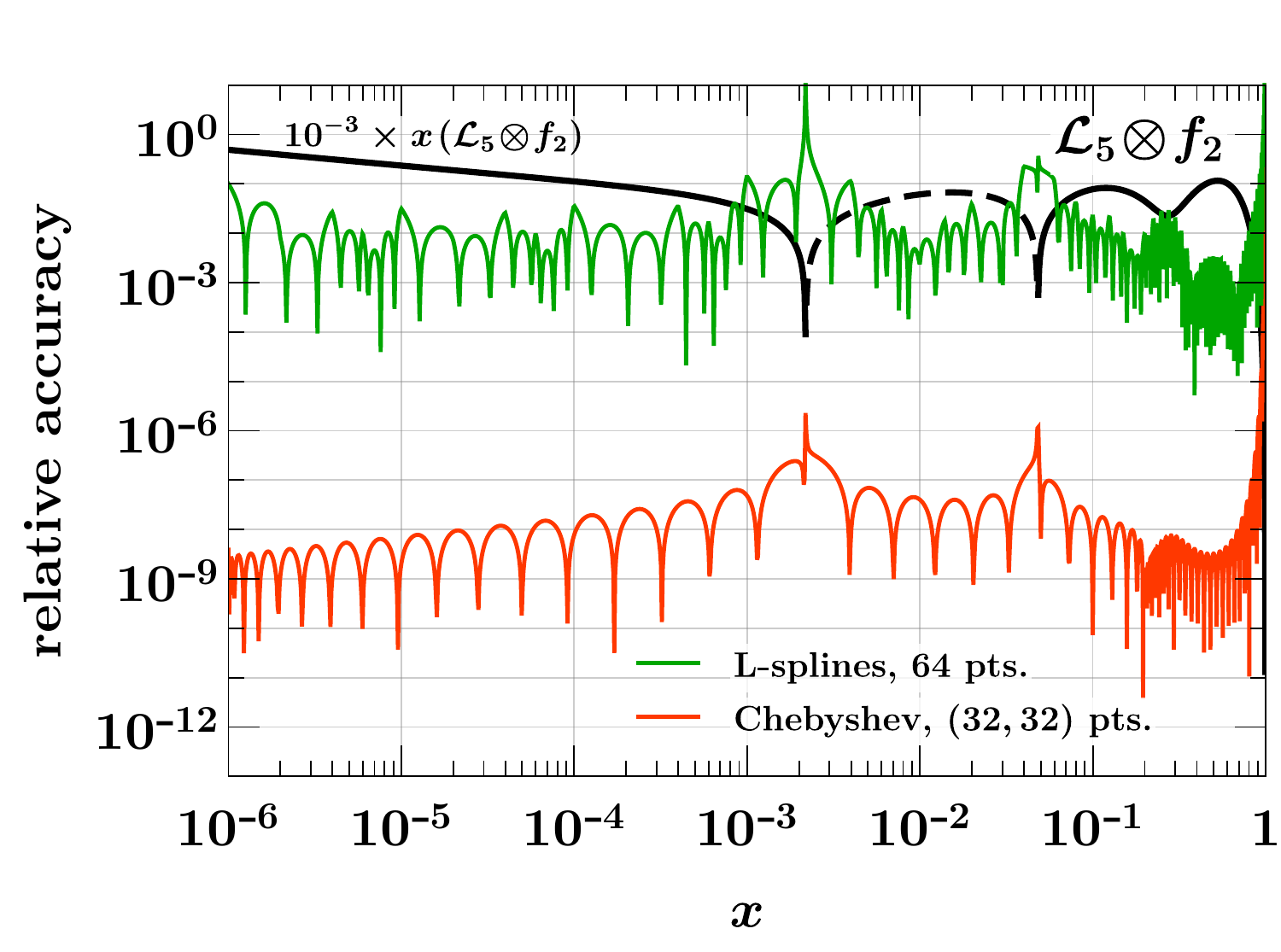}%
\includegraphics[width=\WidthTwoSubfigs]{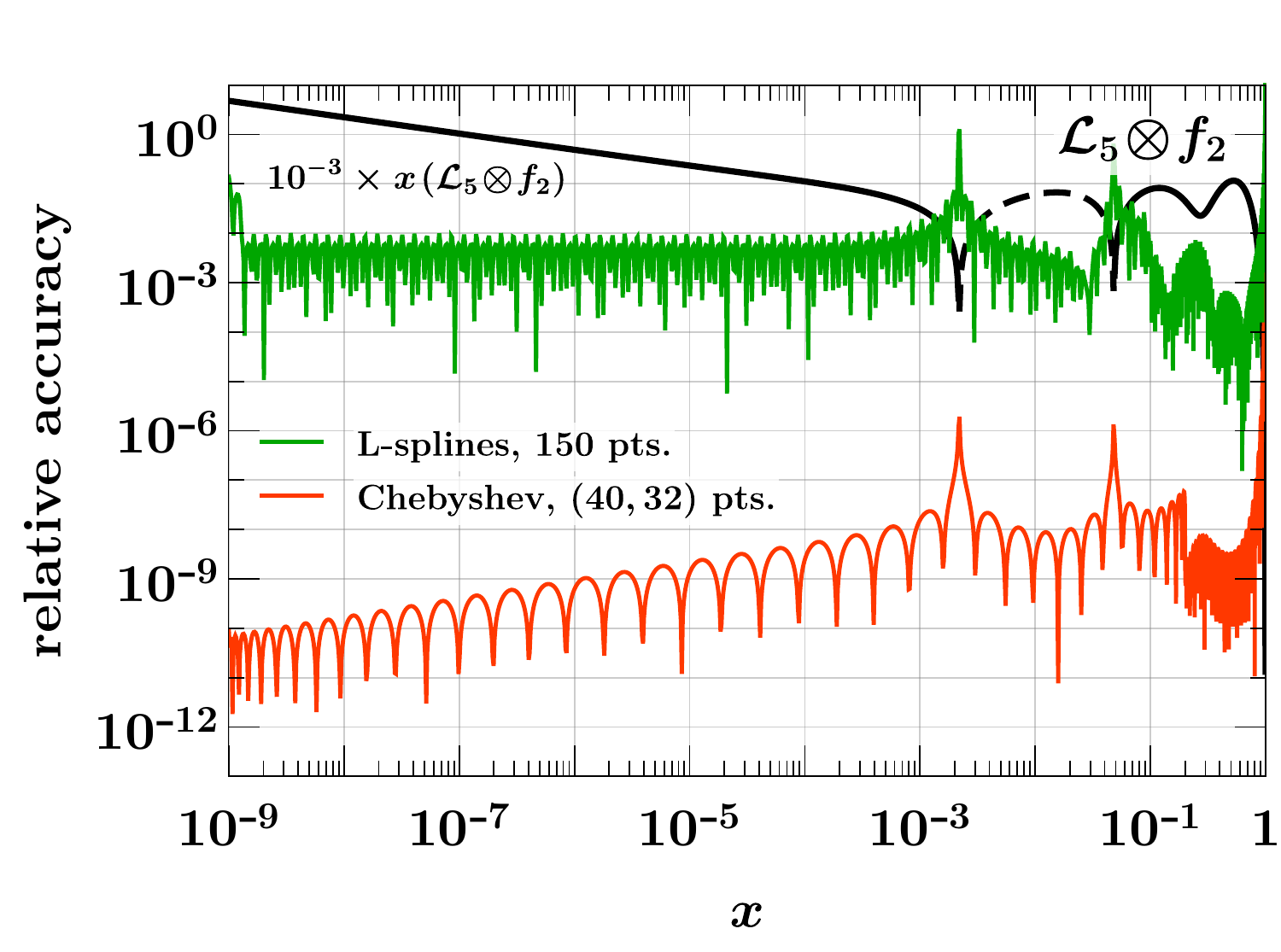}%
\caption{%
Relative accuracy of Mellin convolutions.  The red curves correspond to
Chebyshev interpolation for both the PDF and the convolution result, as
described in the text. The green curves are obtained by performing the numerical
convolution integral for the PDF interpolated with L-splines. Details about the
convolution kernels are given in the text.
On the left, we use the same grids as in \fig{interpolation_comparison_MMHT},
with $n_\pts = 63$ for Chebyshev interpolation and $n_\pts = 64$ for L-splines.
On the right, we use the Chebyshev grid $[10^{-9}, 0.2, 1]_{(40, 32)}$ with
$n_\pts = 71$ and for L-splines a grid with $n_\pts = 150$ (corresponding to the
NNPDF3.1 grid in \tab{LHAPDFgrids}).
}
\label{fig:convolution_accuracy}
\end{figure*}

In \fig{convolution_accuracy}, we show the results of this exercise for the
convolution of our sample PDFs $f_i(x)$ with kernels that have different
singular behavior at the end point $z=1$.  The kernel in the top row is the
leading-order DGLAP splitting function $P_{g g}$, which contains a term
$\mathcal{L}_0 (1-z)$.  The kernel in the second row is $\ln^4 (1-z)$, which is
the most singular term in the three-loop splitting functions $P_{g q}$ and $P_{q
g}$ \cite{Vogt:2004mw,Ablinger:2017tan}.  In the bottom row, we take $\mathcal{L}_5(1-z)$  as
kernel, which appears in the N$^{3}$LO corrections to the rapidity spectrum of
inclusive Higgs production in $pp$ collisions \cite{Dulat:2017prg}.
In all cases, the accuracy of our Chebyshev based method is several
orders of magnitude higher than the one obtained when interpolating PDFs with
L-splines and then performing the convolution integral.

The numerical inaccuracies for the convolution with $\mathcal{L}_5 (1-z)$ using
L-splines are in the percent range over a wide range of $x$ and much higher than
those for the other kernels we looked at.  This is in line with the results of
\refcite{Dulat:2017prg}, which finds that the convolution of
$\mathcal{L}_5(1-z)$ with PDFs from the LHAPDF library can lead to considerable
numerical instabilities.  The PDFs used in that work are those of the NNPDF3.0
analysis \cite{Ball:2014uwa}, and for the sake of comparison we use the
corresponding grid for spline interpolation in the plots on the right of
\fig{convolution_accuracy}.

An alternative method to evaluate the convolution is to use the Chebyshev
interpolant of $\tilde f$ and perform the convolution integral itself
numerically, analogous to what we did with L-splines above. This avoids the
additional interpolation of $(K\otimes \tilde f)$ that happens in the above
kernel matrix method.  The numerical accuracy of this alternative method differs
slightly from the accuracy of the kernel matrix method, but it is
of the same order of magnitude.  Hence, the additional interpolation of
$(K\otimes \tilde f)$ does not introduce a significant penalty in accuracy
beyond that of interpolating $\tilde f$ itself. This makes the kernel matrix
method far more attractive, as it avoids the evaluation of a numerical integral
for each desired $\tilde f$ and $x$.

\section{DGLAP evolution}
\label{sec:dglap_evolution}

\subsection{Numerical solution of DGLAP equations}
\label{sec:dglap_solution}

In this section, we present our approach to the numerical solution of the DGLAP
evolution equations \cite{Dokshitzer:1977sg,Gribov:1972ri,Altarelli:1977zs}.  Up
to order $\alpha_s^{n+1}$, they read
\begin{align} \label{eq:dglap}
\frac{\dd f(x,\mu)}{\dd \ln \mu^2}
&= \sum_{m = 0}^n \, \left[ \frac{\alpha_s(\mu)}{4\pi} \right]^{m+1} \,
   \bigl( P^{(m)} \otimes f(\mu) \bigr)(x)
\,,\end{align}
where $P^{(m)}$ is the splitting function at order $m$.  For notational
simplicity, we suppress the indices for the parton type and the associated sum
on the right-hand side.
To solve \eq{dglap}, we discretize the scaled PDFs $\tilde{f}(x) = x f(x)$ on a
Chebyshev grid and evaluate the Mellin convolutions on the right-hand
side as discussed in \sec{mellin_convolution}. The integro-differential equation
\eqref{eq:dglap} then turns into a coupled system of ordinary differential
equations (ODEs),
\begin{align} \label{eq:dglap_discretized}
\frac{\dd \tilde{f}_{i}(\mu)}{\dd \ln \mu^2}
&= \sum_{j=0}^N \,
   \sum_{m=0}^n \, \left[ \frac{\alpha_s(\mu)}{4\pi} \right]^{m+1} \,
   \widetilde{P}_{i j}^{(m)} \, \tilde{f}_{j}(\mu)
\,,\end{align}
where $\tilde{f}_{i}(\mu) = \tilde{f}(x_i,\mu)$ and
$\widetilde{P}_{i j}^{(m)}$ is the kernel matrix for
$K(z) \equiv z P^{(m)}(z)$ as defined in \eq{kernel_matrix_def}.
This system of ODEs can be solved numerically using the standard Runge-Kutta
algorithm, which is described in more detail in \app{rungekutta}.  This
formulation as a linear system of ODEs by discretization in $x$ is common to
many approaches that solve the DGLAP equations in $x$ space
\cite{Ratcliffe:2000kp, Pascaud:2001bi, Dasgupta:2001eq, Botje:2010ay,
Salam:2008qg, Bertone:2013vaa}.
By using the Chebyshev grid for the discretization in $x$, the resulting
evolved PDF can then be Chebyshev interpolated.

The Runge-Kutta algorithm uses a discretization of the evolution variable $t$
(not to be confused with the argument of Chebyshev polynomials in
\sec{chebyshev}).  It is advantageous if the function multiplying
$\tilde{f}$ on the right-hand side of the evolution equation depends only
weakly on $t$, because this tends to give a uniform numerical accuracy of
evolution with a fixed step size in $t$.  We therefore evolve in the variable
\begin{align} \label{eq:mu-to-t}
t &= - \ln \alpha_s(\mu)
\end{align}
instead of $\ln \mu^2$.  For the running coupling at order $n$, we write
\begin{align} \label{eq:running_rge}
\frac{\dd\ms \alpha_s^{-1}}{\dd \ln\mu^2}
&= - \frac{\beta(\alpha_s)}{\alpha_s^2}
 = \sum_{m=0}^n \frac{\beta_m}{4\pi}\,
   \left( \frac{\alpha_s}{4\pi} \right)^{m}
\end{align}
and use a Runge-Kutta routine (see the end of \app{rungekutta})
to solve for $\alpha_s^{-1}$ as a function of $\ln \mu^2$.
At leading order, the DGLAP equation \eqref{eq:dglap} then becomes
\begin{equation} \label{eq:dglap_t_lo}
\frac{\dd f(t)}{\dd t}
= \frac{1}{\beta_0} \, P^{(0)} \otimes f(t)
\,.\end{equation}
with no explicit $t$ dependence on the right-hand side.
Using $\alpha_s = e^{-t}$, we have at NLO
\begin{equation} \label{eq:dglap_t_nlo}
\frac{\dd f(t)}{\dd t}
= \frac{1}{\beta_0  + (\beta_1 / 4 \pi) \, e^{-t} }
  \biggl[ P^{(0)} + \frac{P^{(1)}}{4\pi} \, e^{-t} \biggr] \otimes f(t)
\,,\end{equation}
where an explicit $t$ dependence appears in the two-loop terms with $\beta_1$
and $P^{(1)}$.  The corresponding equations at NNLO and higher are easily
written down.

The pattern of mixing between quarks, antiquarks, and gluons under DGLAP
evolution is well known.  Its structure at NNLO is given in
\refscite{Moch:2004pa,Vogt:2004mw} and remains the same at higher orders.  To
reduce mixing to a minimum, we work in the basis formed by
\begin{align} \label{eq:evol_diagonal_basis}
& g, &
& \Sigma^{\pm} = \sum_i q_i^\pm , \hspace{-3em}
\nn \\
& u^\pm    - d^\pm   , &
& d^\pm    - s^\pm   , &
& s^\pm    - c^\pm   , &
& c^\pm    - b^\pm   , &
& b^\pm    - t^\pm
\,,\end{align}
where $q^\pm = q \pm \bar{q}$.  Contrary to the often-used flavor nonsinglet
combinations $u + d - 2 s$, $u + d + s - 3 c$, etc., the differences between
consecutive flavors in \eq{evol_diagonal_basis} are less prone to a loss of
numerical accuracy due to rounding effects in regions where the strange and
heavy-quark distributions are much smaller than their counterparts for $u$ and
$d$ quarks. Note that at NNLO, the combination $\Sigma^-$ has a different
evolution kernel than the flavor differences in the second line of
\eq{evol_diagonal_basis}. This is due to graphs that have a $t$ channel cut
involving only gluons.

We implement the unpolarized splitting functions at LO and NLO in their exact
analytic forms.  For the unpolarized NNLO splitting functions, we use the
approximate expressions given in \refscite{Moch:2004pa,Vogt:2004mw}, which are
constructed from a functional basis containing only the distributions
$\mathcal{L}_0(1-z)$, $\delta(1-z)$ and polynomials in $z$, $1/z$, $\ln z$, and
$\ln (1-z)$.  With these parameterizations, the numerical evaluation of the
kernel matrices in all channels takes about as long for $P^{(2)}$ as it does for
$P^{(1)}$.%
\footnote{Using the approximate, parameterized NNLO splitting functions
is a matter of convenience and for compatibility with the benchmark
results below. It is also possible to use the exact
expressions, since the kernel matrices are evaluated only once.}

Both $\alpha_s$ and the PDFs depend on the number $n_F$ of quark flavors that
are treated as light and included in the $\overline{\text{MS}}$ renormalization
of these quantities.  For the conversion between $\alpha_s$ for different $n_F$,
we use the matching conditions in \refcite{Chetyrkin:1997sg} at the appropriate
order.  For the transition between PDFs with different $n_F$, we implement the
matching kernels given in \refcite{Buza:1996wv}, which go up to order
$\alpha_s^2$. We have verified that they agree with the independent calculation
in \refcite{Behring:2014eya}.  The Mellin convolutions of matching kernels with
PDFs are also evaluated as described in \sec{mellin_convolution}.

\subsection{Validation}
\label{sec:validation}

To validate our DGLAP evolution algorithm, we compare our
results with the benchmark tables in section 1.33 of \refcite{Giele:2002hx} and
section 4.4 of \refcite{Dittmar:2005ed}.  These tables contain PDFs evolved to the
scale $\mu = 100 \GeV$ from initial conditions given in analytic form at $\mu_0
= \sqrt{2} \ms \GeV$.  Evolution is performed at LO, NLO and NNLO, either at
fixed $n_F = 4$ or with a variable number of flavors $n_F = 3 \dots 5$ and
flavor transitions at scales $\mu_{c}$ and $\mu_{b}$.  The default choice is
$\mu_c = m_c = \sqrt{2} \GeV$ and $\mu_b = m_b = 4.5 \GeV$.  We compare with the
LO and NLO results in \refcite{Giele:2002hx} and with the NNLO results in
\refcite{Dittmar:2005ed}.   The latter are obtained with the same parameterized
NNLO splitting functions that we use.  Furthermore, the parameterization
in eq.~(3.5) of \refcite{Vogt:2004ns} is used for the two-loop matching kernel for
the transition from a gluon to a heavy quark.  We also use this parameterization
for the sake of the present comparison, noting that visible differences with
the benchmark results appear when we use the exact analytic form for this kernel
instead.

The benchmark tables were obtained using two programs, \textsc{HOPPET}
\cite{Salam:2008qg} and \textsc{QCD-Pegasus} \cite{Vogt:2004ns}, both run with
high-precision settings.  The former solves the evolution equations in $x$
space, whilst the latter uses the Mellin moment technique.  The tables give
results for the evolved PDFs at 11 values of $x$ between $10^{-7}$ and $0.9$.
The results are given with five significant digits, with the exception of
several sea-quark combinations at $x=0.9$, which are very small and are given to
only four significant digits.

As specified in \refcite{Giele:2002hx}, evolution with \textsc{HOPPET} was
performed by using seventh-order polynomials for the interpolation on multiple
$x$ grids spanning the interval $[10^{-8}, 1]$ with a total of $n_\pts = 1,\!170$.
 A uniform grid in $\ln \mu^2$ was used, with 220 points between $\mu = \sqrt{2}
\ms \GeV$ and $1000 \GeV$.  The results were verified to be stable within a
$10^{-5}$ relative error for $x<0.9$ by comparing them with the results of the
same program with half the number of points on both the $x$ and the $\mu$ grids.
Furthermore, they were compared with the results obtained with
\textsc{QCD-Pegasus}.

For the comparison with our approach, we use a Chebyshev grid with 3 subgrids
$[10^{-8}, 10^{-3}, 0.5, 1]_{(24,\ms 24,\ms 24)}$ which has $n_\pts = 70$. The DGLAP
equations are solved using a Runge-Kutta algorithm with the DOPRI8 method (see
\app{rungekutta}) and a maximum step size in $t$ of $h = 0.1$.  This amounts to
about 12 Runge-Kutta steps from the starting scale to $\mu = 100 \GeV$.  Our
results for the benchmark numbers (rounded as stated above) remain the same if
we take $40$ instead of $24$ points for each subgrid in $x$.  They also remain
unchanged if we use the same Runge-Kutta method with maximum step size $h = 0.3$
or $h = 0.02$.

We compare our results with the numbers reported in tables 2, 3, 4 of
\refcite{Giele:2002hx} and in tables 14 and 15  of \refcite{Dittmar:2005ed},
which cover unpolarized
evolution both with fixed $n_F=4$ and with variable $n_F=3 \ldots 5$.  Whilst
the tables also give results for evolution with different scales
$\mu_{\text{r}}$ in $\alpha_s$ and $\mu_{\text{f}}$ in the PDFs, we always set
$\mu_{\text{r}} = \mu_{\text{f}}$.  We agree with all benchmark numbers,%
\footnote{Note that the value of $\alpha_s(100 \GeV)$, evolved at LO with
$n_F = 4$, is mistyped in table 2 of \refcite{Giele:2002hx} and corrected in
table 16 of \refcite{Dittmar:2005ed}.}
except for those given in our \tabs{benchmark_FFN}{benchmark_VFN}. In all cases
where we differ, the differences can be attributed to the benchmark results and
fall into two categories: 
\begin{enumerate}
\item The benchmark tables contain a number of entries, marked by an asterisk,
for which the results of the two used codes differ in the sixth digit and give
different numbers when rounded to five digits.  The number with the smaller
modulus is then given in the tables.  In several cases, our result agrees with
number with the larger modulus and in this sense agree with the benchmark
results.

\item We differ from the benchmark numbers in five more cases.  For the two
cases at NNLO, the benchmark results have typographical errors in the exponent.
In the other three cases, we differ by one unit in the fifth digit.  We
contacted the authors of the benchmark tables, who confirmed that indeed their
respective codes agree with our numbers \cite{Salam:2019pri}.
\end{enumerate}

\begin{table}[!ht]
\centering
\begin{tabular}{c|c| cccc ccc}
\hline\hline
order & $x$ & $x u_v$ & $x d_v$ & $x L_{-}$ & $2 x L_{+}$
            & $x s_{+}$ & $x c_{+}$ & $g$ \\
\hline
LO    & $0.5$ & & & & & \textcolor{blue}{$7.3137^{-4}$} & & \\
      & $0.9$ & & & & & & $4.8894^{-9}_{*}$ & \\
\hline
NLO   & $10^{-7}$ & & & & & $6.6914^{+1}_{*}$ & & $1.1484^{+3}_{*}$ \\
      & $10^{-4}$ & & & & & & & $9.2873^{+1}_{*}$ \\
      & $10^{-3}$ & & & & $6.1649^{+0}_{*}$ & & & \\
      & $10^{-2}$ & & & & & $8.4221^{-1}_{*}$ & & \\
\hline\hline
\end{tabular}
\caption{%
Results for evolution with $n_F = 4$ for which we differ from the numbers in the
benchmark tables in \protect\refcite{Giele:2002hx}.  The notation for PDF
combinations is $L_\pm = \bar{d} \pm \bar{u}$ and $q_{+} = q + \bar{q}$ for
$q=s,c$.  The number format $a^b$ is shorthand for $a \times 10^b$. As discussed
in the text, all differences can be attributed to the benchmark results, where
entries with an asterisk correspond to category 1 and the entry in blue
corresponds to category 2.  We note that in the table headers of
\protect\refcite{Giele:2002hx} it should read $2 x L_{+}$ instead of $x L_{+}$.
}
\label{tab:benchmark_FFN}
%
\vspace{2em}
%
\centering
\begin{tabular}{c|c| cccc cccc}
\hline\hline
order & $x$ & $x u_v$ & $x d_v$ & $x L_{-}$ & $2 x L_{+}$
            & $x s_{+}$ & $x c_{+}$ & $x b_{+}$ & $g$ \\
\hline
LO    & $10^{-7}$ & & & & & & & \textcolor{blue}{$4.6071^{+1}$} & \\
\hline
NLO  & $10^{-6}$ & & & & & & & & $5.2290^{+2}_{*}$ \\
     & $0.7$     & \textcolor{blue}{$2.0102^{-2}$} & & & & & & & \\
\hline
NNLO & $10^{-7}$ & & \textcolor{blue}{$1.0699^{-4}$} & &
     & & & & \textcolor{blue}{$9.9694^{+2}$} \\
\hline\hline
\end{tabular}
\caption{%
As \tab{benchmark_FFN}, but for evolution and flavor matching from $n_F = 3
\ldots 5$.  The corresponding benchmark tables are in
\protect\refcite{Giele:2002hx} for LO and NLO and in \protect\refcite{Dittmar:2005ed}
for NNLO.
}
\label{tab:benchmark_VFN}
\end{table}

\subsection{Numerical accuracy}
\label{sec:accu}

We now study the numerical accuracy of our method in some detail.  We use
$x_0 = 10^{-7}$ and 3 subgrids. The evolution equations are solved with the
DOPRI8 method (see \app{rungekutta}).  To assess the accuracy, we compare three
settings with different numbers of grid points and different Runge-Kutta steps
$h$:
\begin{enumerate}
\item $[10^{-7}, 10^{-2}, 0.5, 1]_{(24,\ms 24,\ms 24)}$ ($n_\pts = 70$) and $h=0.1$,
\item $[10^{-7}, 10^{-2}, 0.5, 1]_{(40,\ms 40,\ms 40)}$ ($n_\pts = 118$) and $h=0.1$,
\item $[10^{-7}, 10^{-2}, 0.5, 1]_{(40,\ms 40,\ms 40)}$ ($n_\pts = 118$) and $h=0.004$.
\end{enumerate}
To estimate the numerical error due to the discretization in $x$, we take the
difference between settings 1 and 2, whereas the error due to the Runge-Kutta
algorithm is estimated from the difference between settings 2 and 3.  For the
combined error, we take the difference between settings 1 and 3. Note that these
estimates correspond to the accuracy of setting~1. We use the same initial
conditions at $\mu_0 = \sqrt{2} \ms \GeV$ as in the benchmark comparison
described in the previous subsection.  Starting at $n_F=3$, heavy flavors are
added at $m_c = \sqrt{2} \ms \GeV$, $m_b = 4.5 \GeV$, and $m_t = 175 \GeV$.  In
the remainder of this section, we always evolve and match at NNLO.

\begin{figure}[!t]
\centering
\includegraphics[width=\WidthTwoSubfigs]{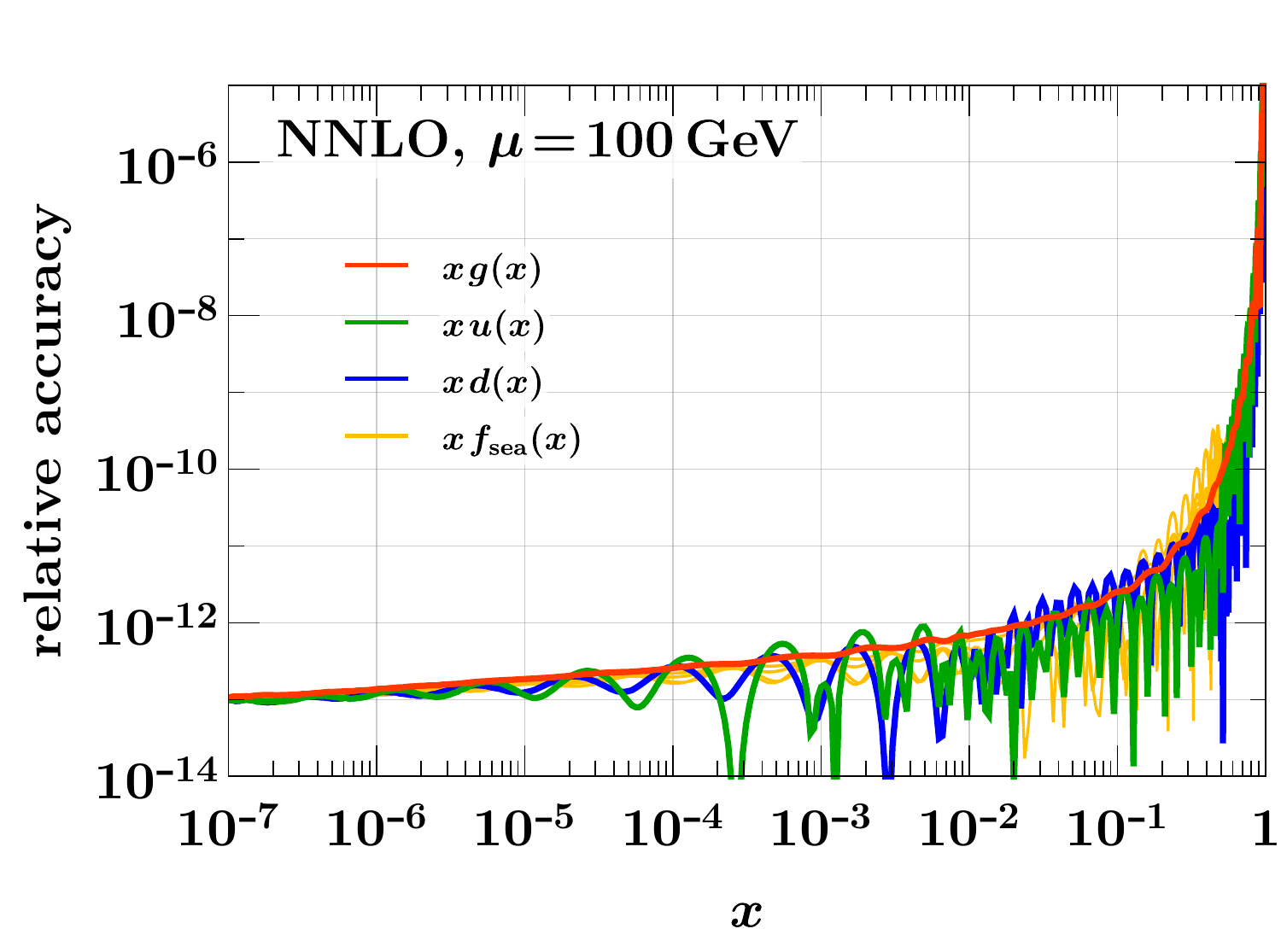}%
\includegraphics[width=\WidthTwoSubfigs]{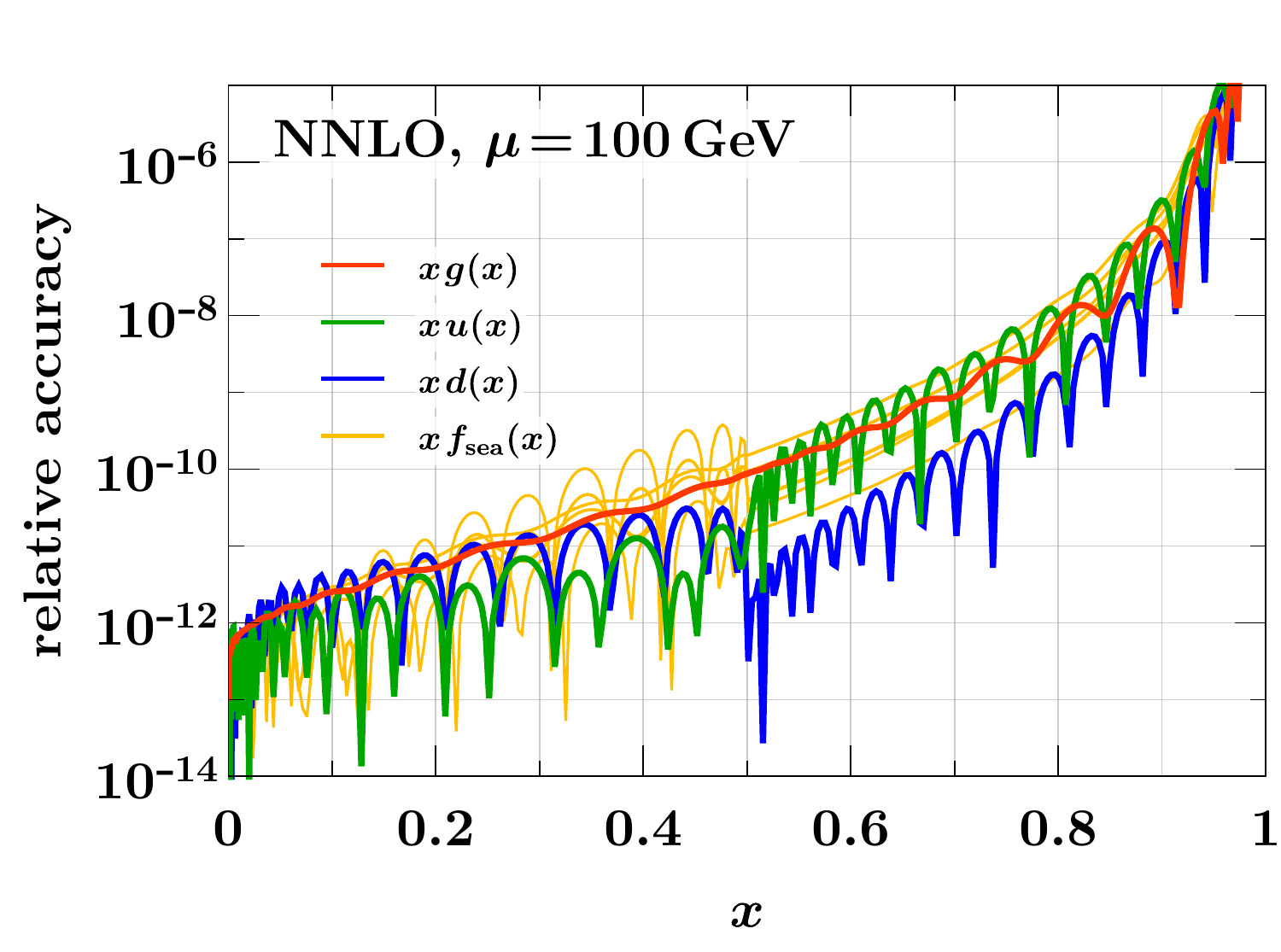}%
\\%
\includegraphics[width=\WidthTwoSubfigs]{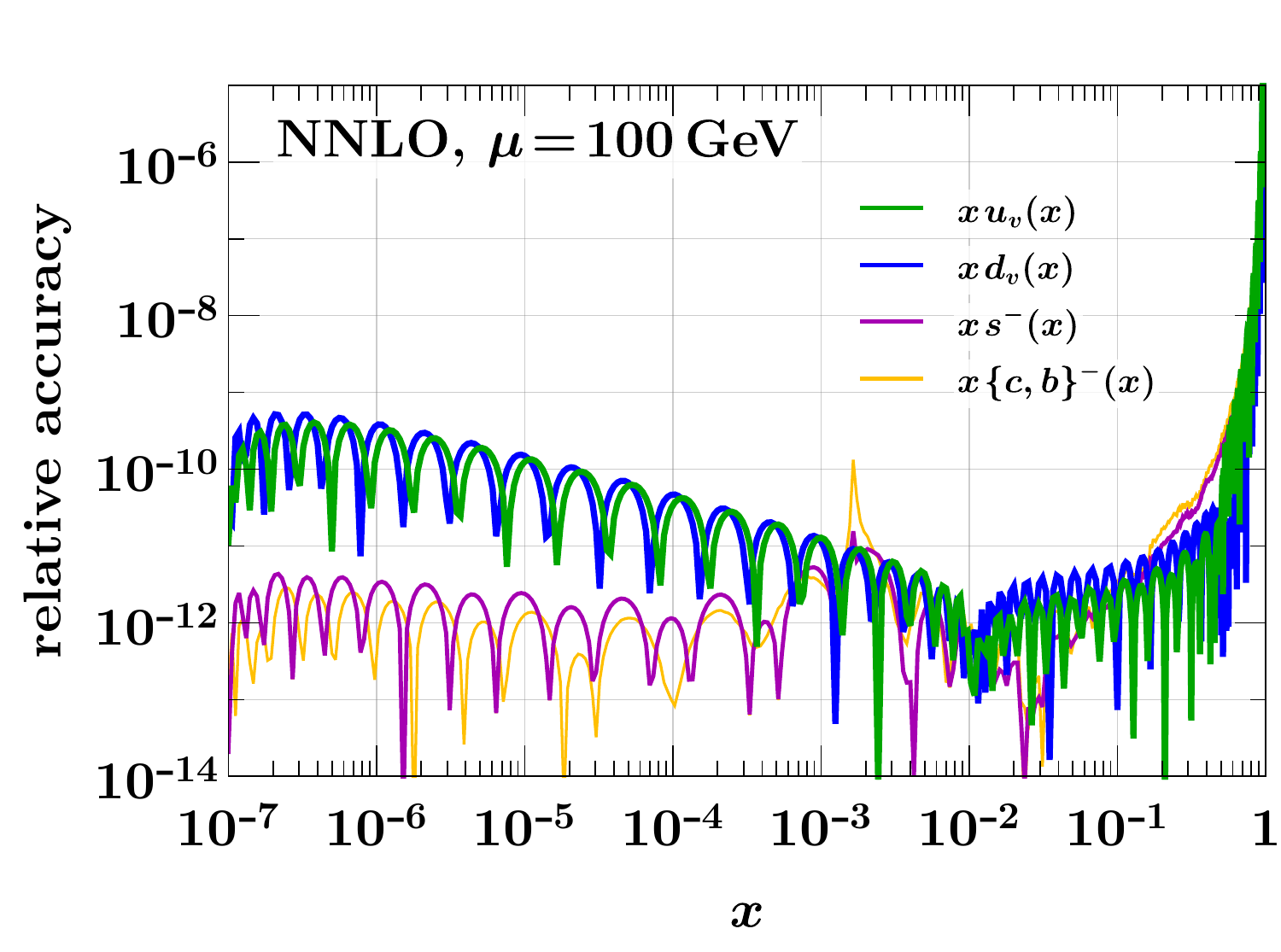}%
\includegraphics[width=\WidthTwoSubfigs]{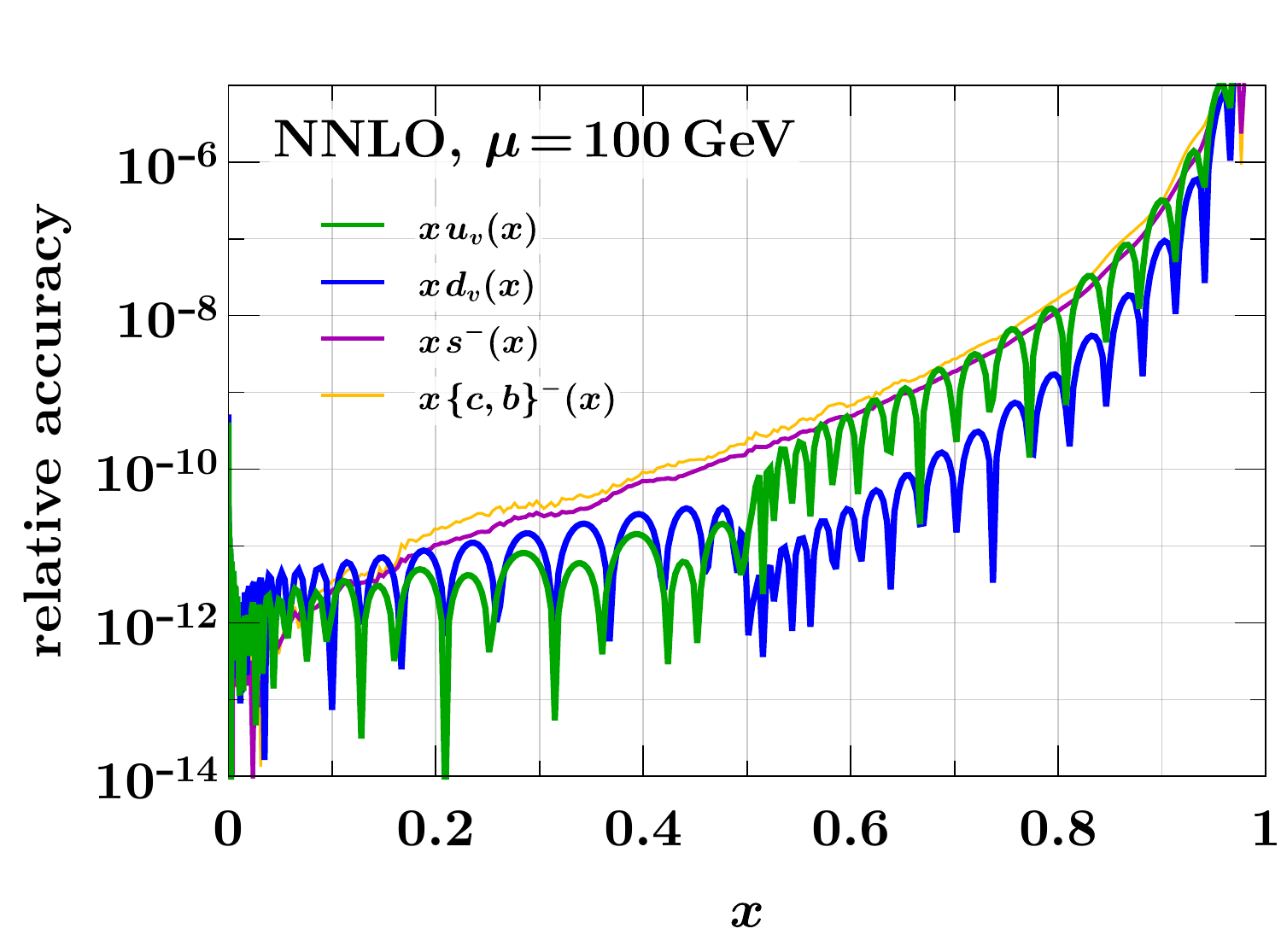}%
\caption{%
Relative numerical accuracy of NNLO DGLAP evolution and flavor matching from
$\mu_0 = \sqrt{2} \ms \GeV$ and $n_F = 3$ to $\mu = 100 \GeV$ and $n_F = 5$.
We use the PDFs defined in \refscite{Giele:2002hx, Dittmar:2005ed}. The top row
shows quark, antiquark, and gluon distributions (with $f_{\text{sea}} \in
\{ \bar{u}, \bar{d}, s, \bar{s}, c, \bar{c}, b, \bar{b} \}$), and the bottom row
shows the differences $q - \bar{q}$.
}
\label{fig:pdf_evolution_compare_100gev}
\end{figure}

\begin{figure}
\centering
\includegraphics[width=\WidthTwoSubfigs]{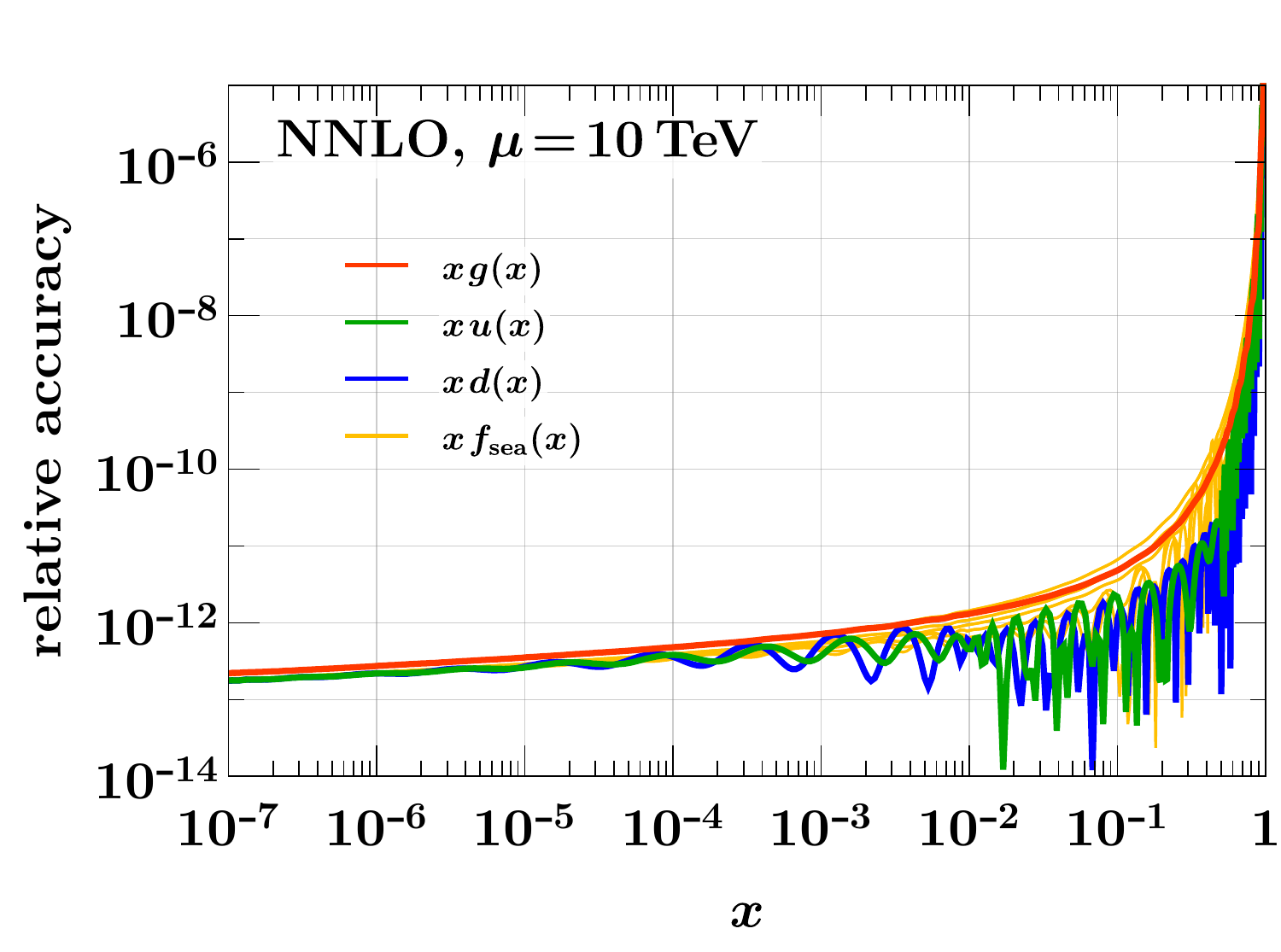}%
\includegraphics[width=\WidthTwoSubfigs]{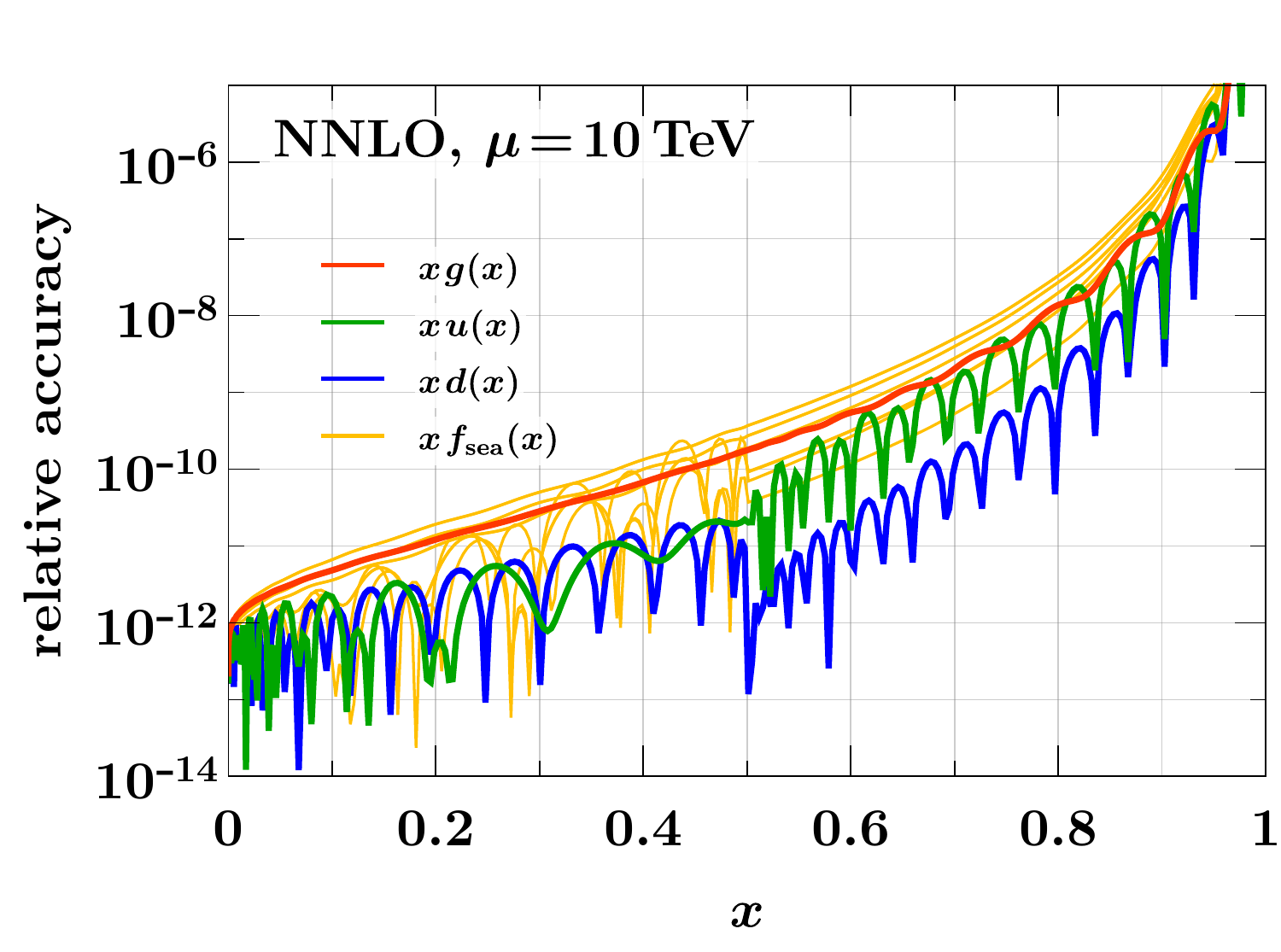}%
\\%
\includegraphics[width=\WidthTwoSubfigs]{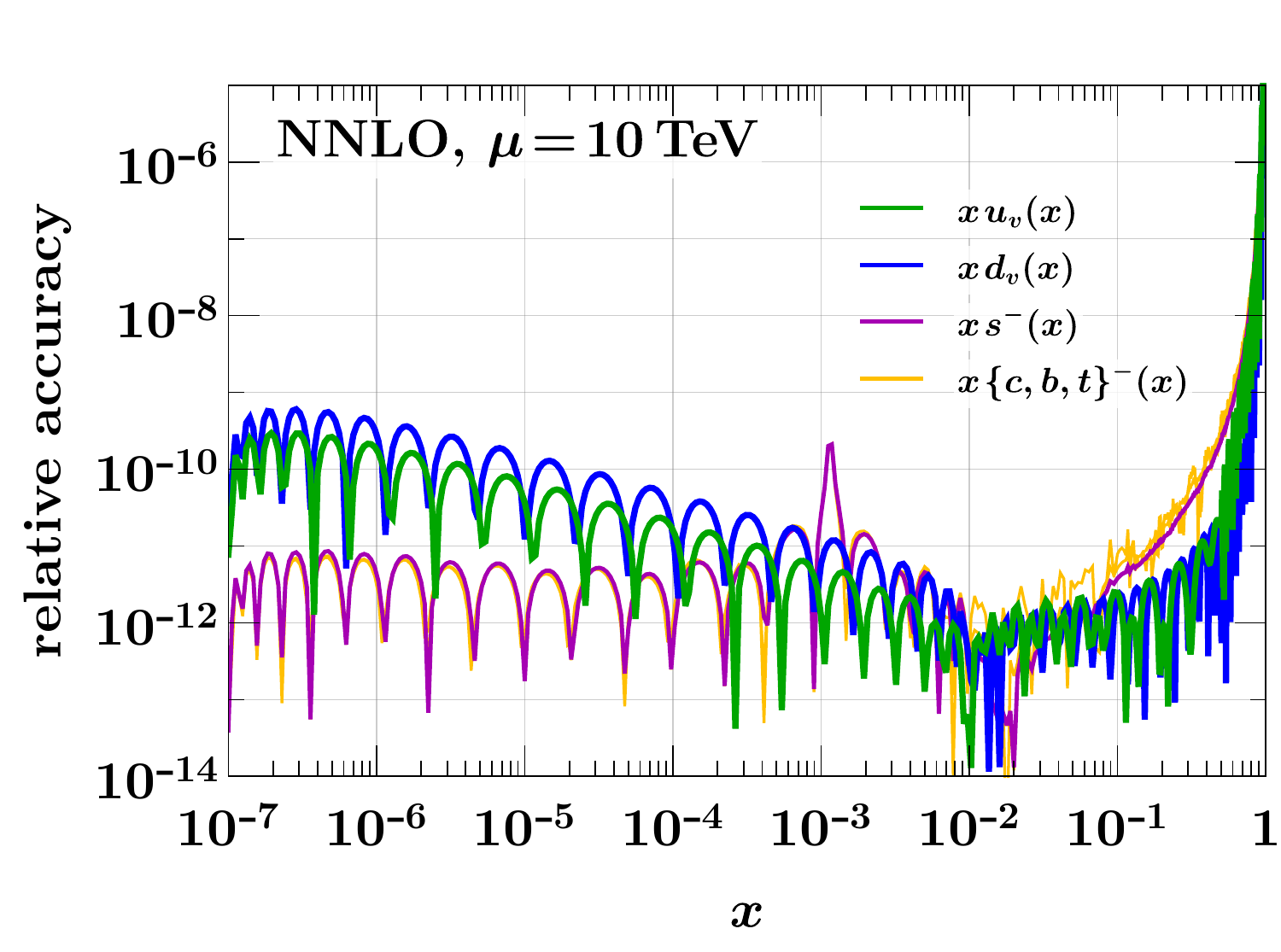}%
\includegraphics[width=\WidthTwoSubfigs]{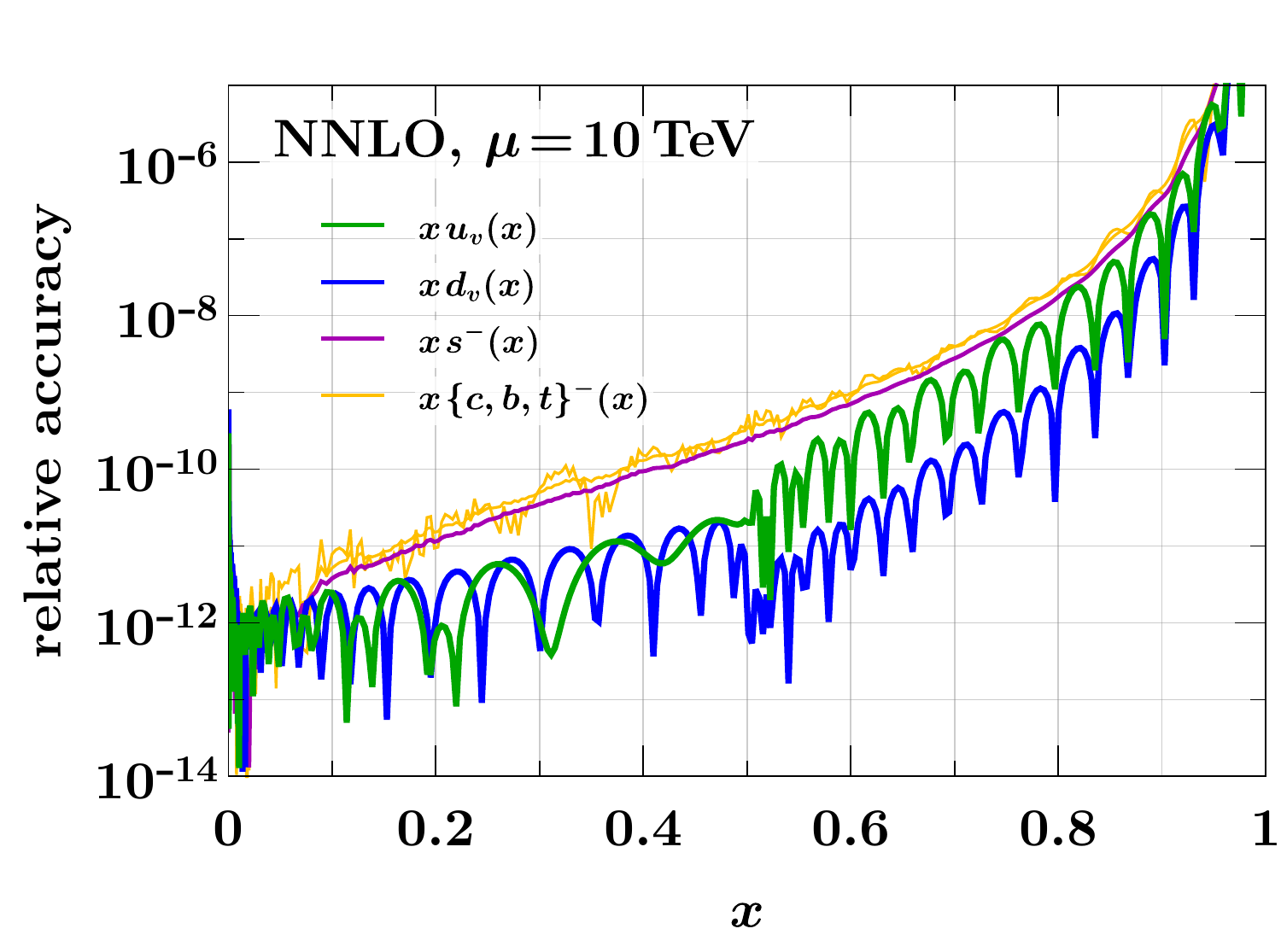}%
\caption{%
As \fig{pdf_evolution_compare_100gev}, but for evolution and flavor matching up
to $\mu = 10 \TeV$ and $n_F = 6$.
}
\label{fig:pdf_evolution_compare_10tev}
\end{figure}

The combined discretization and Runge-Kutta accuracy for evolution to $\mu = 100
\GeV$ and $\mu = 10 \TeV$ is shown in
\figs{pdf_evolution_compare_100gev}{pdf_evolution_compare_10tev}, respectively,
both for the individual parton flavors $g$, $q$, $\bar{q}$, and for the valence
combinations $q^- = q - \bar{q}$.  The relative accuracy is better than
$10^{-7}$ up to $x \le 0.8$, and much better than that for smaller $x$. The same
holds when we evolve to $\mu = 1.01 \ms m_c$ or $\mu = 1.01 \ms m_b$, where the
charm or bottom distributions are very small. The relative accuracy of $u_v$ and
$d_v$ increases towards small $x$.  This reflects the strong decrease of these
distributions in the small-$x$ limit, as we already observed and explained in
\sec{pdf_interpolation}.  The combinations $s^-$, $c^-$, $b^-$, and $t^-$ show
less variation at small~$x$, and so does their relative accuracy.

\begin{figure}
\centering
\includegraphics[width=\WidthTwoSubfigs]{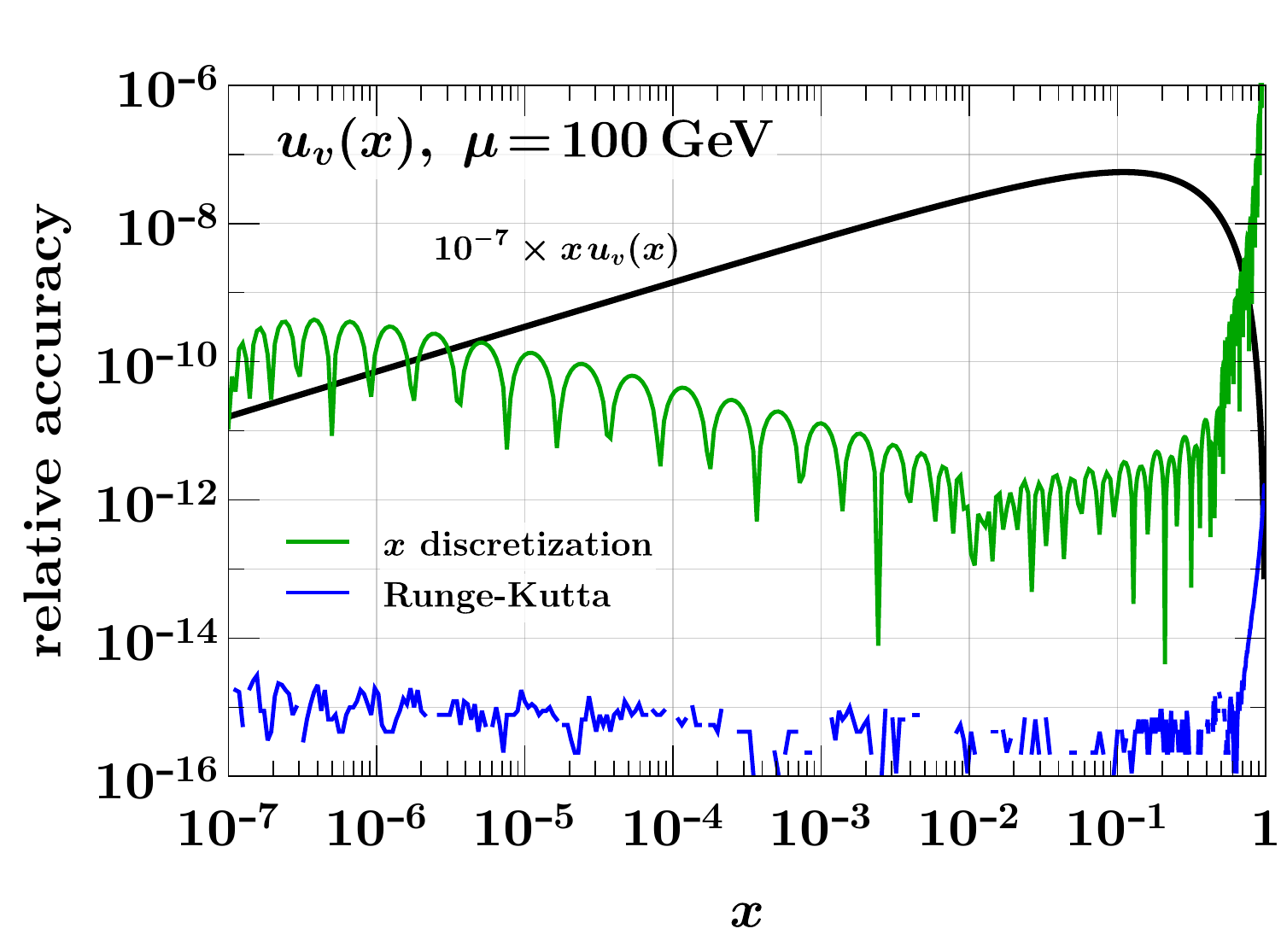}%
\includegraphics[width=\WidthTwoSubfigs]{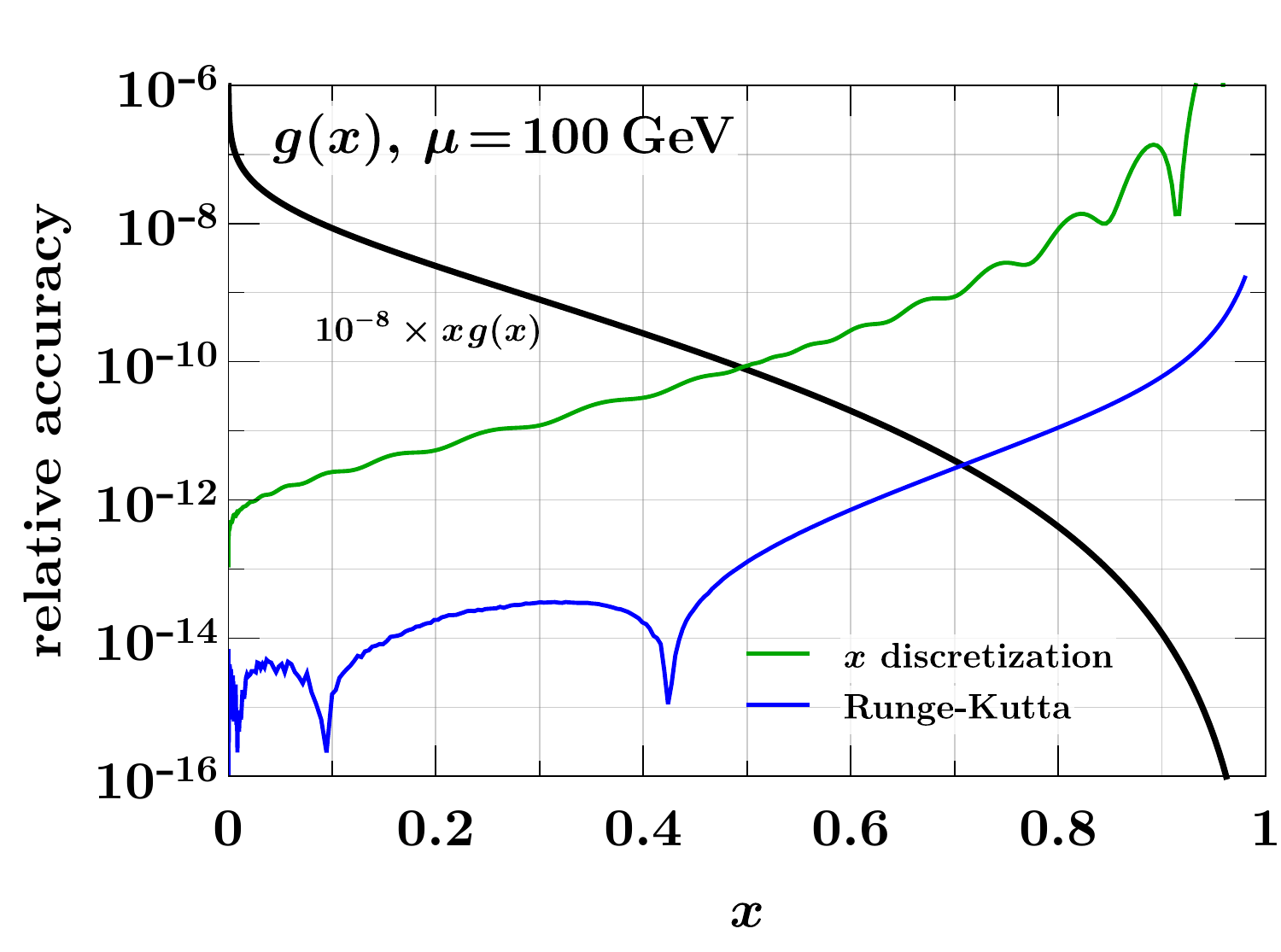}%
\caption{%
Relative accuracy of NNLO DGLAP evolution and flavor matching, distinguishing
the contributions due to $x$-space discretization (green) and the Runge-Kutta
algorithm (blue). The plot on the left shows $x\ms u_v (x)$ and the plot on the
right $x\ms g(x)$.  Notice the different scales in $x$.
}
\label{fig:pdf_evolution_compare_differr}
\end{figure}

In \fig{pdf_evolution_compare_differr} we show examples for the separate errors
due to discretization and the Runge-Kutta algorithm.  With our settings, the
overall numerical accuracy is entirely determined by the discretization in $x$,
whilst the inaccuracy due to the Runge-Kutta algorithm can be neglected.  The
Runge-Kutta accuracy for $u_v$ is in fact determined by the machine precision
except for very large $x$, as signaled by the noisy behavior of the error curve.

It is often found that backward evolution, i.e.\ evolution from a higher to a
lower scale, is numerically unstable.%
\footnote{See, however, \refcite{RuizArriola:1998er} for an early study that
concluded the contrary.}
The structure of the DGLAP equations is such that the relative uncertainties
(physical or numerical) of PDFs become larger when one evolves from a scale
$\mu_1$ to a lower scale $\mu_0$.  This property is shared by other
renormalization group equations in QCD, including the one for $\alpha_s(\mu)$.
To which extent it leads to numerically unreliable results is, however, a
separate question.

\begin{figure}
\centering
\includegraphics[width=\WidthTwoSubfigs]{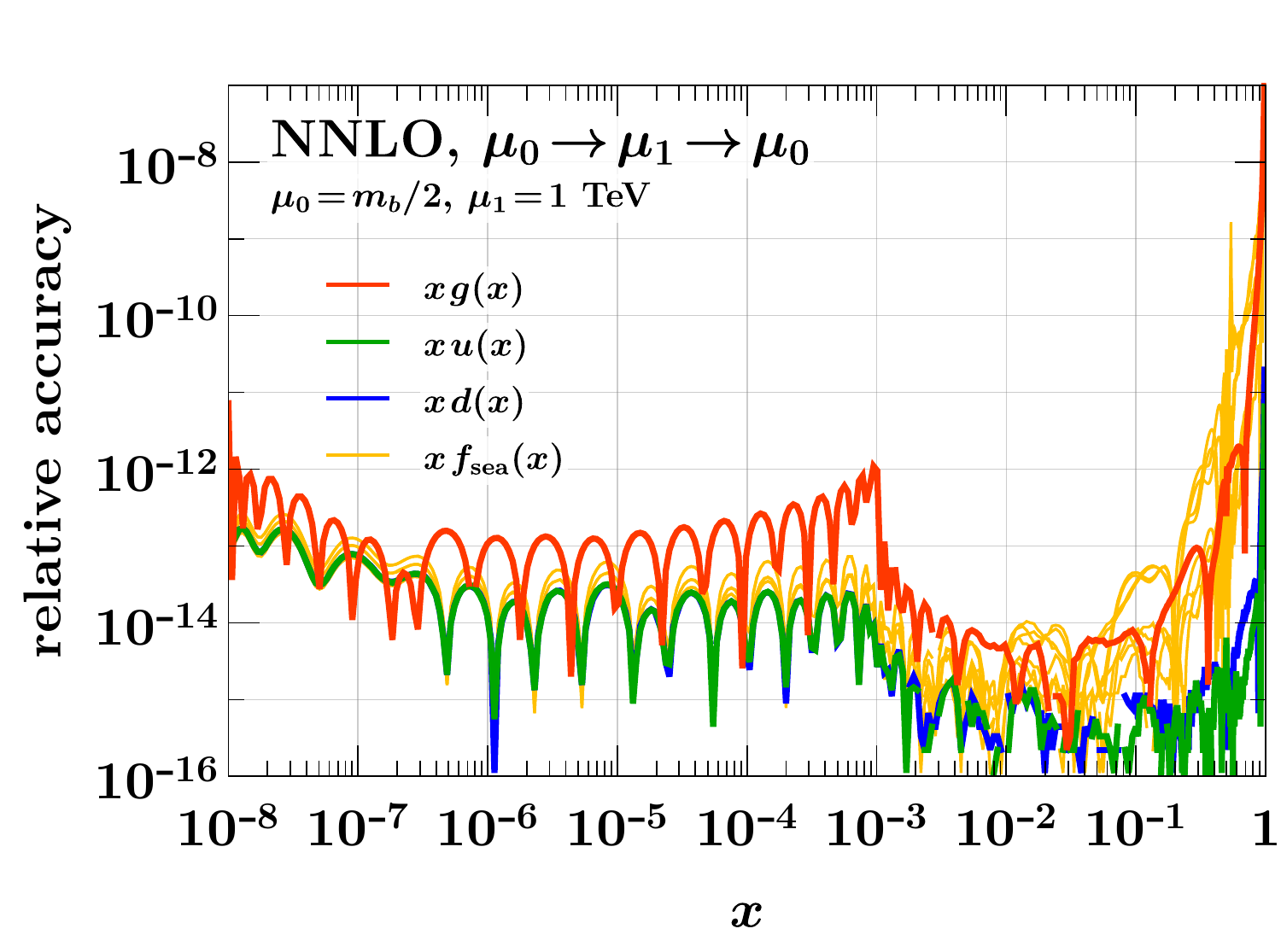}%
\includegraphics[width=\WidthTwoSubfigs]{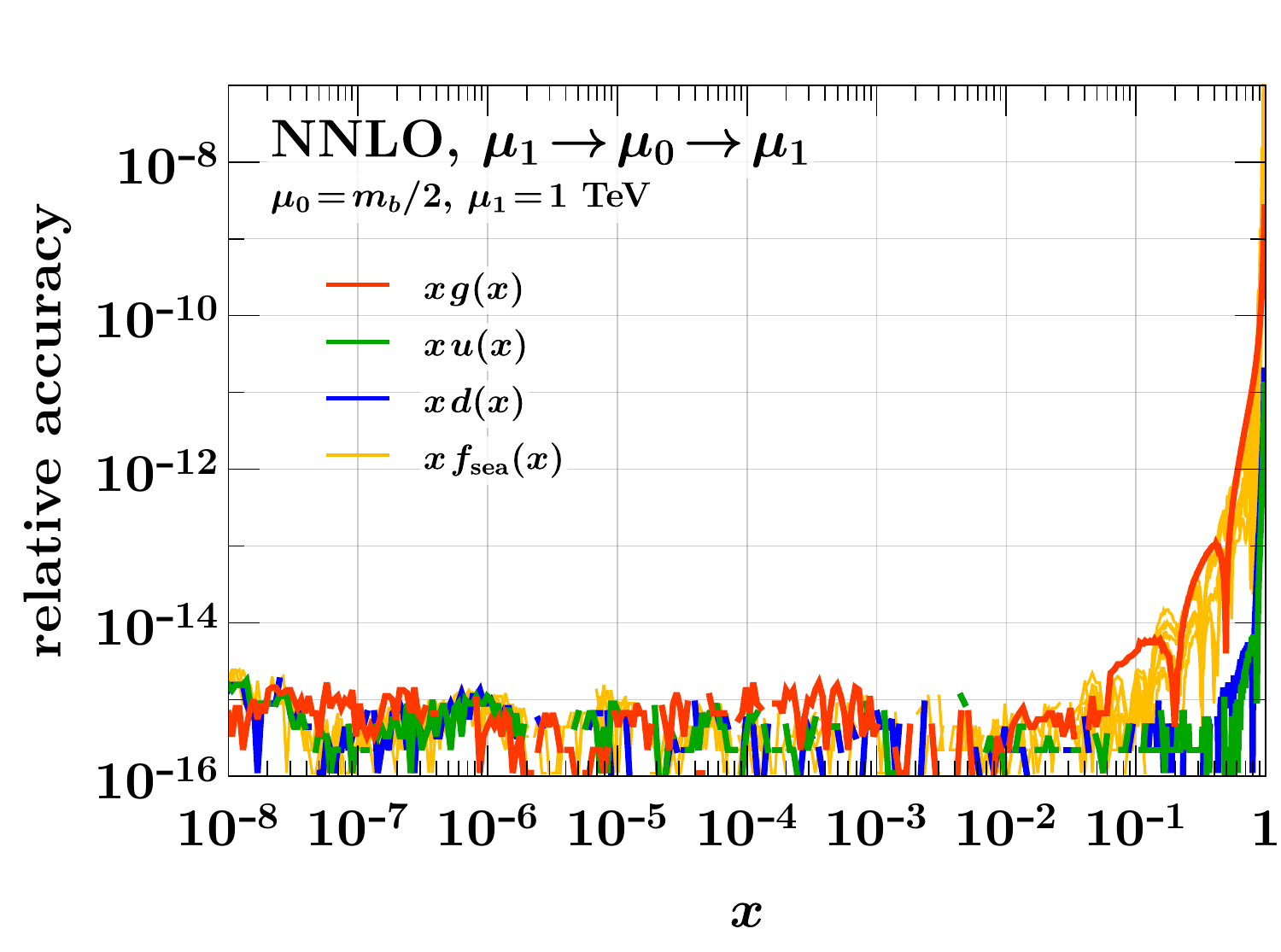}%
\\
\includegraphics[width=\WidthTwoSubfigs]{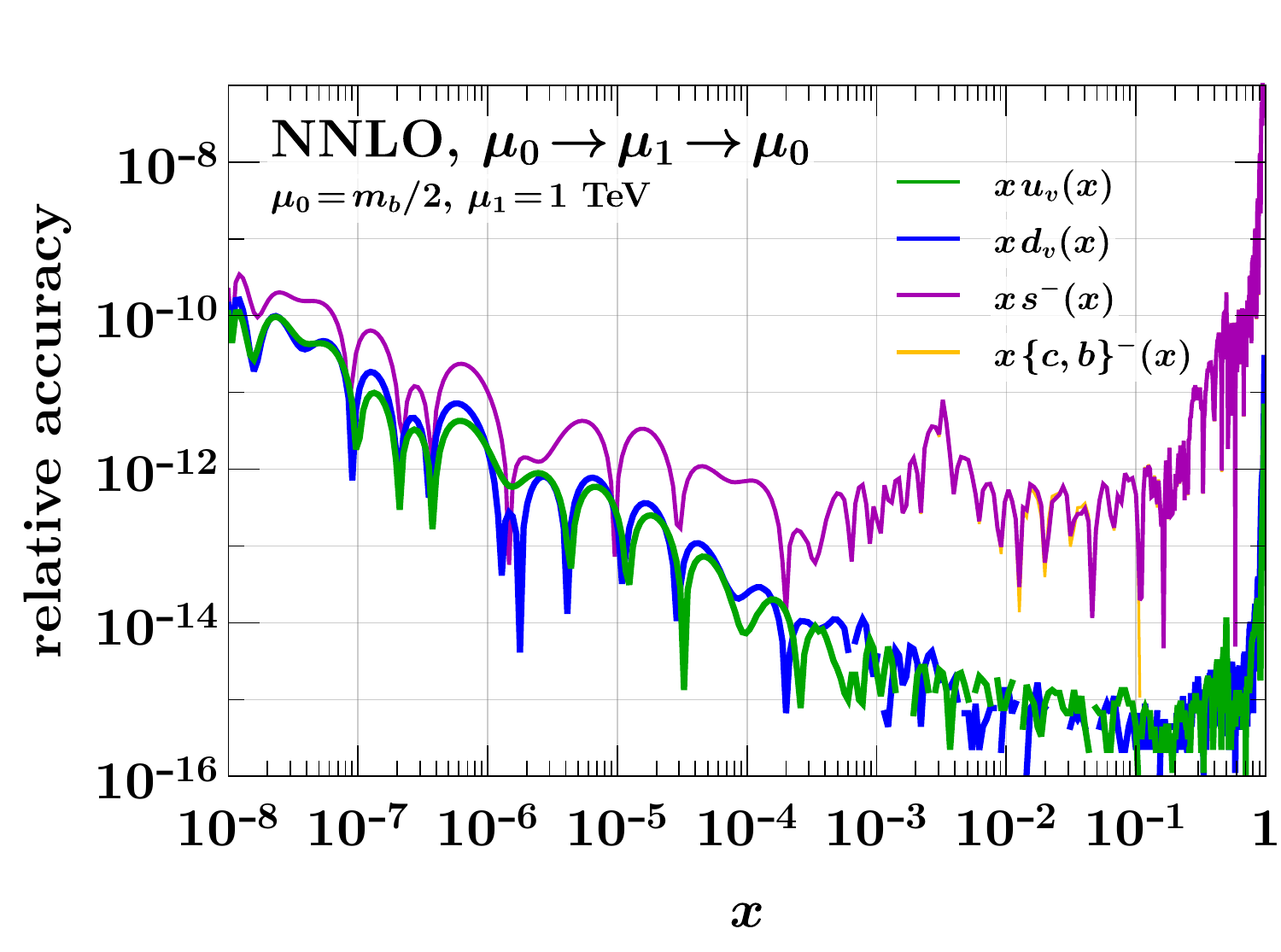}%
\includegraphics[width=\WidthTwoSubfigs]{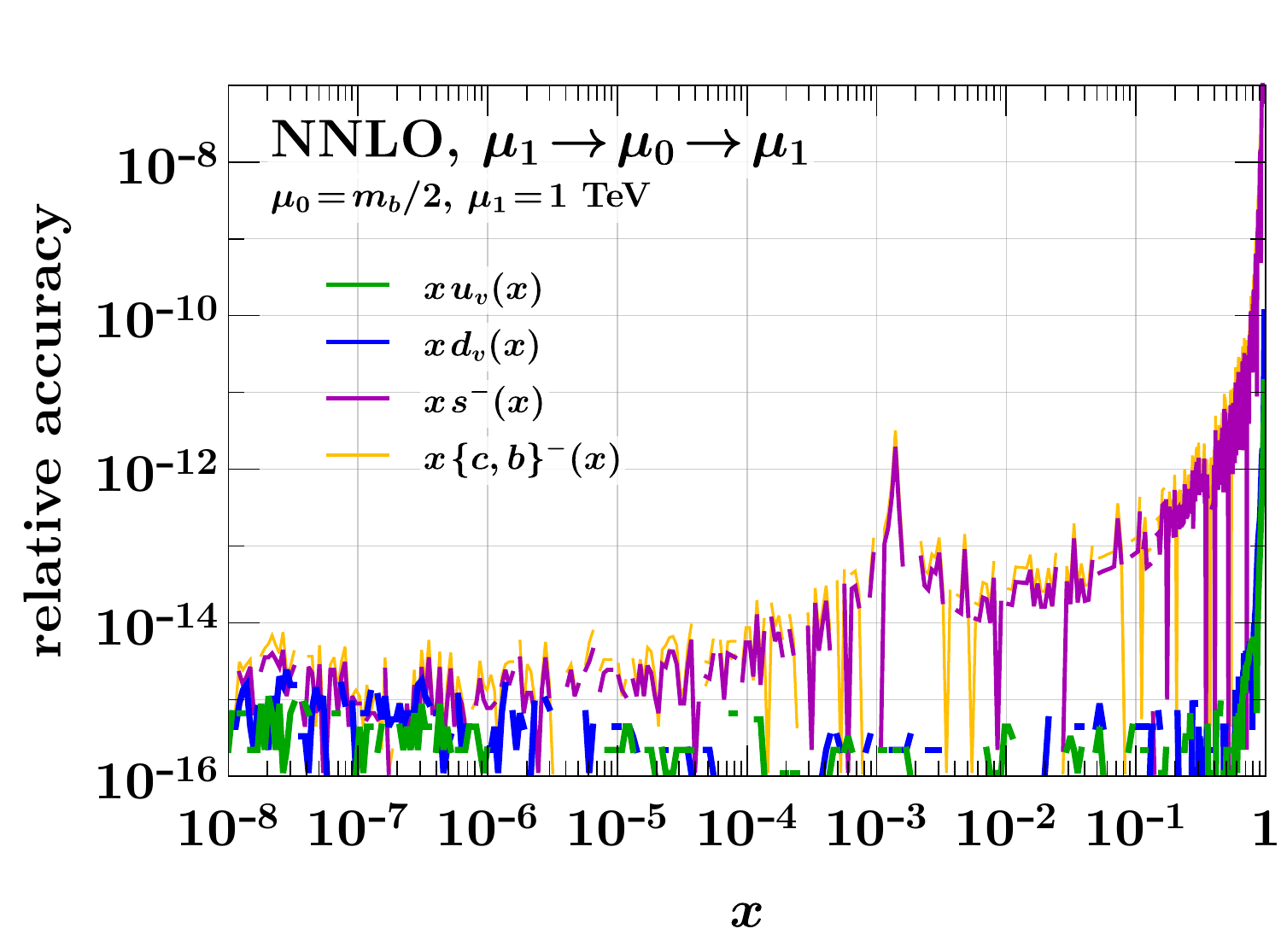}%
\caption{%
Relative accuracy of NNLO DGLAP evolution at fixed $n_F = 5$ from $\mu_0 \to
\mu_1 \to \mu_0$ (left) and from $\mu_1 \to \mu_0 \to \mu_1$ (right).  The
scales are $\mu_0 = 2.25 \GeV$ and $\mu_1 = 1 \TeV$, and the DOPRI8 method with
maximum step size $h = 0.1$ is used to solve the evolution equations.  The same
grid in $x$ is used as for the benchmark comparison in \sec{validation}.
}
\label{fig:backward-evo}
\end{figure}

To study backward evolution within our method, we perform the following
exercise.  We start with the input PDFs of the benchmark comparison, evolve them
with $n_F = 4$ from $\sqrt{2} \ms \GeV$ to $m_b / 2 = 2.25 \GeV$ and match to
$n_F = 5$ at that point.%
\footnote{Note that by matching at $\mu = m_b / 2$, we obtain nonzero values for
the $b$ and $\bar{b}$ distributions.}
The result is taken as initial condition for $n_F = 5$ evolution from $\mu_0 =
2.25 \GeV$ to $\mu_1 = 1 \TeV$ (step~1).  The result of step 1 is evolved down
to $\mu_0$ (step 2), and the result of step 2 is again evolved up to $\mu_1$
(step 3).  We thus have two quantities that are sensitive to the accuracy of
backward evolution:
\begin{itemize}
\item the difference between the output of step 2 and the input to step 1, which
corresponds to the evolution path $\mu_0 \to \mu_1 \to \mu_0$,
\item the difference between the output of step 3 and the input to step 2,
corresponding to the evolution path $\mu_1 \to \mu_0 \to \mu_1$.
\end{itemize}
The relative accuracy for the two evolution paths is shown in \fig{backward-evo}
for the DOPRI8 method with maximum step size $h = 0.1$.  In both cases, we find
very high accuracy, in some cases near machine precision as signaled by the
noisy behavior of the curves.  Relative errors tend to be larger for the
evolution path $\mu_0 \to \mu_1 \to \mu_0$, but they are all below $10^{-8}$ for
$x \le 0.9$ (except in the vicinity of zero crossings).  Note that a high
accuracy is also obtained for the small combinations $s^-$, $c^-$, and $b^-$,
which are induced by evolution at order $\alpha_s^3$.

We have also repeated our exercise with the widely-used RK4 method and $h =
0.03$, which requires approximately the same number of function calls as DOPRI8
with $h = 0.1$ (see the \hyperref[app:rungekutta]{appendix} for more detail).
This yields relative errors that are orders of magnitude larger for both
evolution paths, but still below $10^{-5}$ for $x \le 0.9$ and away from zero
crossings.  This shows the significant benefit of using Runge-Kutta methods with
high order for DGLAP evolution.  Nevertheless, even with the standard RK4
method, we find no indication for numerical instabilities of backward evolution
in our setup.

Finally, we note that a Runge-Kutta method with less than the 13 stages of
DOPRI8 can
be useful if one needs to perform evolution in many small steps, for instance
when computing jet cross sections with $\mu \sim p_T$ for a dense grid in the
transverse jet momentum $p_T$.  Setting $h \sim 0.03$, we find the DOPRI6 method
with its 8 stages to be well suited for such situations.

\section{Conclusions}
\label{sec:conclusions}

We have shown that a global interpolation using high-order Chebyshev polynomials
allows for an efficient and highly accurate numerical representation of PDFs.
Compared with local interpolation methods on equispaced grids, such as splines,
our method can reach much higher numerical accuracies whilst keeping the
computational cost at a comparable or reduced level.

Not only interpolation, but also differentiation and integration of PDFs are
numerically accurate and computationally simple in our approach.  The same holds
for the Mellin convolution of a PDF with an integral kernel, which can be
implemented as a simple multiplication with a pre-computed matrix. In
particular, high accuracy is retained if the kernel has a strong singularity at
the integration end point, such as \mbox{$\bigl[ \ln^5 (1-z) / (1-z) \bigr]_+$}
or $\ln^4(1-z)$. Even for such demanding applications, it is possible to achieve
a numerical accuracy below $10^{-6}$ with only about $60$ to $70$ grid points.
Hence, with our method, numerical inaccuracies become safely negligible for
practical physics calculations involving PDFs. If desired, the inaccuracy
caused by the interpolation can be estimated with modest additional
computational effort by using interpolation
on the same grid without the end points. This yields an estimate that is
appropriately conservative, but not as overly conservative as the estimate we
obtain with Gauss-Kronrod quadrature in the case of integration.

By combining our Chebyshev-based interpolation with a Runge-Kutta method of high
order, we obtain a very accurate implementation of
DGLAP evolution.  Using 70 grid points in $x$ and evolving from $\mu = 1.41
\GeV$ to $\mu = 10 \TeV$, we find relative errors below $10^{-7}$ for $x$
between $10^{-7}$ and $0.8$.  We successfully tested our implementation against
the benchmark evolution tables in \refscite{Giele:2002hx,Dittmar:2005ed}.
Performing backward evolution with our method, we find no indications for any
numerical instabilities.

\rev{Let us briefly point out differences between our implementation of
Chebyshev interpolation and the use of Chebyshev polynomials in parameterizations
of PDFs. Perhaps most striking is that the order of the highest polynomial used
is much larger in our case. There are several reasons for this.  First, our
approach aims at keeping uncertainties due to interpolation small compared to
physics uncertainties.  By contrast, as argued in \cite{Martin:2012da}, a PDF
parameterization need not be much more accurate (relative to the unknown true
form) than the uncertainty resulting from fitting the PDFs to data.  Second, as
we have seen, polynomials up to order 20 or 40 can be handled with ease in our
method, whereas PDF determinations naturally tend to limit the number of
parameters to be fitted to data.  Third, the detailed functional forms
used in the PDF parameterizations \refscite{Pumplin:2009bb, Glazov:2010bw,
Martin:2012da, Harland-Lang:2014zoa} at a fixed scale~$\mu$ differ from each
other and from the form \eqref{eq:barycentric_pdf} we use for interpolation at
any scale $\mu$. It may be interesting to further investigate the impact of
these differences on the number of polynomials required for a satisfactory
description of PDFs.}

\rev{Our approach is} implemented in the C++ library \chili, which is still under
development and which we plan to eventually make
public.  To give a sense of its current performance, we note that the
evaluation of the
interpolated PDF takes less than a microsecond. With the settings in
\sec{validation}, NNLO DGLAP evolution from $\mu = 1.41 \GeV$ to
$\mu = 100 \GeV$ at $n_F = 4$ takes on the order of 10 to 20 milliseconds.%
\footnote{These timings are on a recent desktop or laptop computer with
an Intel\textregistered\ Core\texttrademark\ i5 or i7 processor.
We expect that with a dedicated performance tuning, the code can still
be made faster.}
If significantly faster access to evolved PDFs is needed, it may be preferable
to pre-compute the $\mu$ dependence and  use interpolation in both $x$ and $\mu$.
This can easily be done using the techniques we have described.

To conclude, let us mention additional features that are already available in
\chili\ but not discussed here:
$(i)$ polarized evolution up to NNLO for longitudinal and up to
NLO for transverse and linear polarizations, with flavor matching up to NLO for
all cases,
$(ii)$ combined QCD and QED evolution of PDFs with kernels up to
$\mathcal{O}(\alpha_s \ms \alpha)$ and $\mathcal{O}(\alpha^2)$, with QED flavor
matching up to $\mathcal{O}(\alpha)$,
and
$(iii)$ evolution and flavor matching of double parton distributions $F(x_1,
x_2, \vec{y}\ms; \mu_1, \mu_2)$.  For the latter, accuracy goals are typically
lower than for computations involving PDFs, but the possibility to work with a
low number of grid points is crucial for keeping memory requirements manageable.
This will be described in a future paper.

\acknowledgments
We thank Florian Fabry, Mees van Kampen, and Peter Pl{\"o}{\ss}l for their
contributions to \chili.  It is a pleasure to thank  Arnd Behring, Johannes
Bl\"umlein, Johannes Michel, Sven Moch, and Lorenzo Zoppi for valuable
discussions and input.  Special thanks are due to Gavin Salam and Andreas Vogt
for clarifying the discrepancies between our results and the tables in
\refscite{Giele:2002hx, Dittmar:2005ed}.

This work was in part funded by the Deutsche Forschungsgemeinschaft (DFG, German
Research Foundation) -- Research Unit FOR 2926, grant number 409651613.
The work of RN is supported by the ERC Starting Grant REINVENT-714788. RN also
acknowledges the CINECA award under the ISCRA initiative for the availability of
the performance computing resources needed for this work.

\appendix

\section{Runge-Kutta methods}
\label{app:rungekutta}

In this appendix, we take a closer look at the Runge-Kutta (RK) algorithm, which
we use to solve the discretized DGLAP equation \eqref{eq:dglap_discretized}, as
well as the renormalization group equation \eqref{eq:running_rge} for the
running coupling.  We pay special attention to different Runge-Kutta methods and
their accuracy.

The RK algorithm solves the initial value problem
\begin{align} \label{eq:initial_value_problem}
\dd y(t) / \dd t &= F(t, y)
& \text{with~~}
y (t_0) &= y_0
\,. \end{align}
It approximates the solution $y(t)$ by the set of values $y_i \approx y(t_i)$ at
equispaced points~$t_i$.  The distance between two consecutive points is called
the step size $h = t_{i+1} - t_i$, and the algorithm computes $y_{i+1}$ from the
approximate solution $y_i$ at the previous point.
The solution at a given $t_f$ is obtained by dividing the interval $[t_0, t_f]$
into $N$ steps, such that $t_f = t_N$.  When using the algorithm in our work, we
fix a maximum step size $h_{\text{max}}$.  For any given $t_0$ and $t_f$, the
number of steps is then taken as the smallest integer $N$ that satisfies $|t_f -
t_0| / N \le h_{\text{max}}$.  For brevity, $h_{\text{max}}$ is denoted by $h$
in \sec{dglap_evolution}.

In so-called explicit RK methods, the algorithm for a step is of the form
\begin{equation} \label{eq:rk_step}
y_{i+1} = y_i + h \sum_{j = 1}^s b_j \ms k_j
\end{equation}
with
\begin{align} \label{eq:rk_stages}
k_1 &= F(t_i, \ms y_i)
\,, \nn \\
k_2 &= F(t_i + h \ms c_2 , \ms y_i + h \ms a_{21} \ms k_1 )
\,, \nn \\
k_3 &= F\bigl(t_i + h \ms c_3, \ms y_i + h \ms [ a_{31} \ms k_1
       + a_{32} \ms k_2 ] \ms \bigr)
\,, \nn \\
\vdots
\nn \\
k_s &= F\bigl(t_i + h \ms c_s, \ms y_i + h \ms [ a_{s1} \ms k_1 + \dots
       + a_{s,s-1} \ms k_{s-1} ] \ms \bigr)
\,.\end{align}
In standard nomenclature, one calls $s$ the number of stages, $a_{ij}$ the RK
matrix, $b_i$ the weights and $c_i$ the nodes.  Different RK methods are
characterized by different values of these parameters.

A method is of order $p$ if a single step has a numerical accuracy of
$\mathcal{O}(h^{p+1})$.  For a given set of RK parameters, it is easy to
determine $p$ from \eqs{rk_step}{rk_stages} by a Taylor expansion in~$h$. Since
for given $t_0$ and $t_f$ the number of required steps scales like $1/h$, the
cumulated numerical error of $N$ steps in the solution of the initial value
problem
\eqref{eq:initial_value_problem} is
\begin{align}  \label{eq:rk_error}
y_N = y(t_f) + \mathcal{O}(h^p)
\,.\end{align}
A broad range of RK methods with different orders and numbers of stages is known
\cite{Hairer:1993}.  Widely used is the RK4 method, also called the ``classic RK
method'' or simply ``the RK method'', which has four stages and order $p=4$.  We
also investigate several methods with $p>4$, one by Cash and Karp
\cite{Cash:1990} and three by Dormand and Prince \cite{DORMAND198019,
PRINCE198167}.  According to their order, we denote the latter by DOPRI5,
DOPRI6, and DOPRI8.  The main parameters of these methods are given in
\tab{runge-kutta}.  As noted in \refcite{PRINCE198167}, the order of the DOPRI6
method is 7 rather than 6 if the function $F(t,y)$ in \eq{initial_value_problem}
has no explicit $t$ dependence.  The numerical study described below suggests
that the DOPRI8 method is of order 10 if $F(t,y)$ is $t$ independent.  We do not
attempt to provide a general proof for this conjecture and hence put a question
mark next to the corresponding entry in our table.

\begin{table}[t]
\centering
\begin{tabular}{c|ccccc}
\hline\hline
method & RK4 & Cash-Karp & DOPRI5 & DOPRI6 & DOPRI8 \\
\hline
order $(p)$ &  4  &    5      &    5   &  6(7)  &  8(10?) \\
stages $(s)$ &  4  &    6      &    7   &   8    &  13 \\
\hline\hline
\end{tabular}
\caption{%
Order and number of stages of the different Runge-Kutta methods investigated in
this work.  The entries in parentheses are explained in the text.
}
\label{tab:runge-kutta}
\end{table}

To quantify the performance of the different methods for DGLAP evolution, we
evolve the sample PDF $f_1(x)$ given in \eq{sample_functions} with the DGLAP
kernel for the flavor non-singlet combination $u^+ - d^{\ms+}$, either at LO or
at NLO.  As can be seen in \eqs{dglap_t_lo}{dglap_t_nlo}, LO evolution
corresponds to a $t$ independent function $F(t,y)$, whereas NLO evolution
implies an explicit $t$ dependence in $F(t,y)$.  We evolve from $t_0 = 0.7$ to
$t_f = 1.7$.  According to \eq{mu-to-t}, this respectively gives $\alpha_s
\approx 0.5$ and $\alpha_s \approx 0.18$, so that the $t$ dependence of $F(t,y)$
for NLO evolution is quite strong at the lower end of the evolution interval.

\begin{figure}
\centering
\includegraphics[width=\WidthTwoSubfigs]{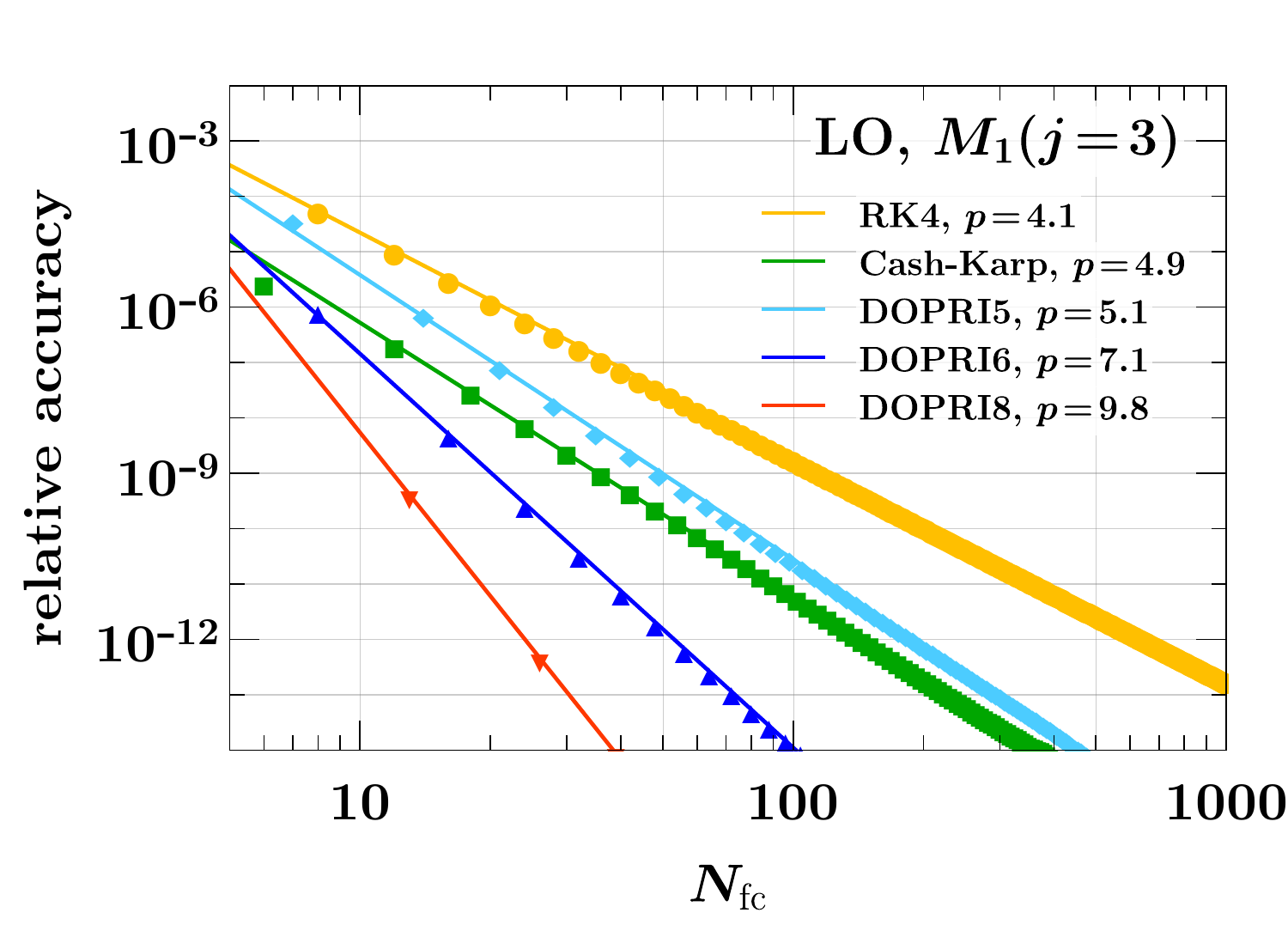}%
\includegraphics[width=\WidthTwoSubfigs]{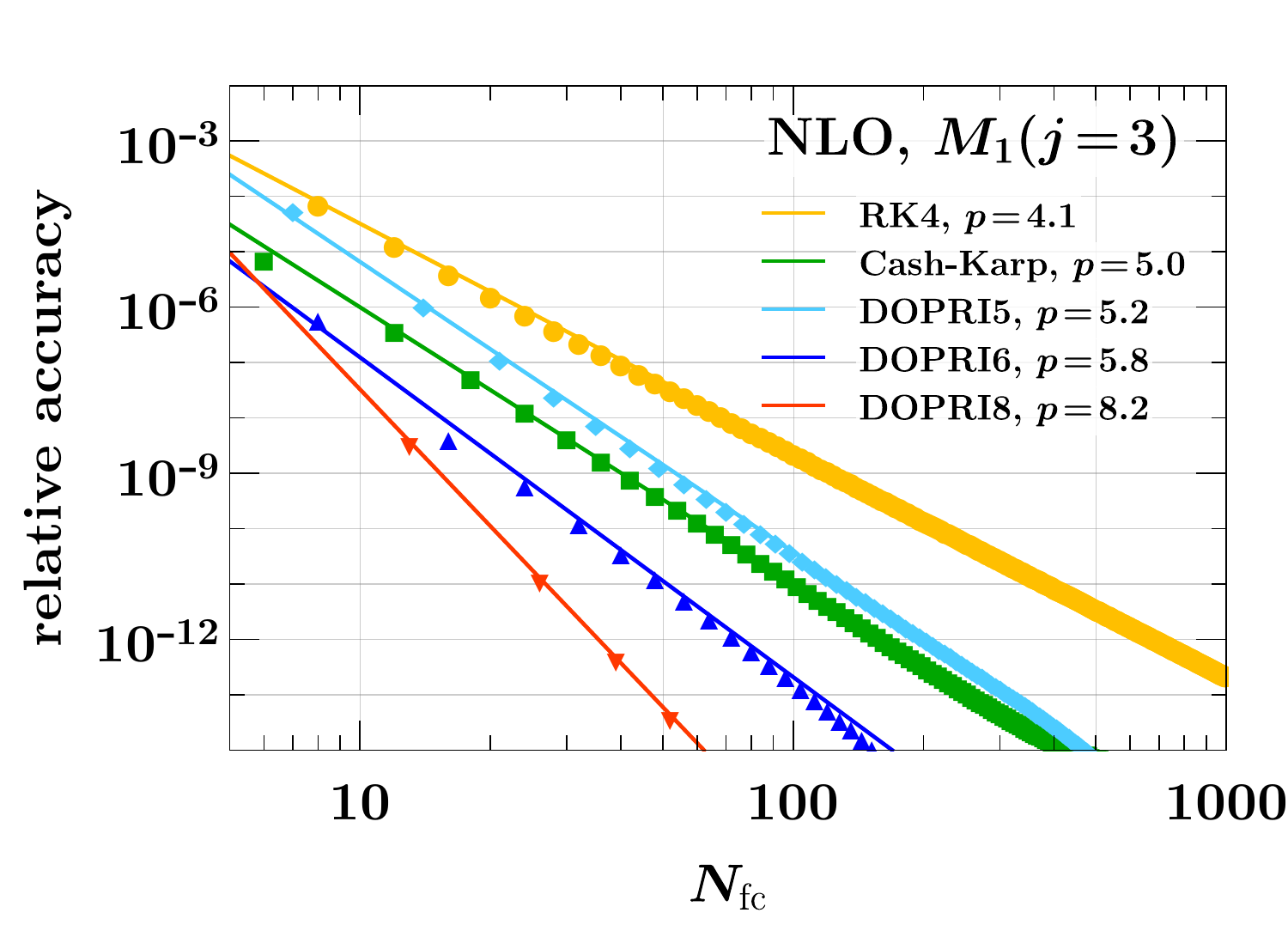}%
\caption{%
Relative accuracy of different Runge-Kutta methods as a function of the total
number $N_{\text{fc}} = s \ms N_{\text{steps}}$ of calls to the function
$F(t,y)$ in \eq{initial_value_problem}.  The accuracy is evaluated for the
truncated Mellin moment $M(j)$ of the sample PDF $f_1(x)$, evolved from $t_0 =
0.7$ to $t_f = 1.7$.  The lines connecting the points correspond to a fitted
power law $(N_{\text{fc}})^{-p}$ with $p$ reported in the legend.
}
\label{fig:runge_kutta_scaling}
\end{figure}

To monitor the accuracy of evolution, we consider either the evolved PDF at a
given $x$ or the truncated Mellin moment \eqref{eq:mellin-moment} with a given
moment index $j$.  The results are similar in both cases.
In \fig{runge_kutta_scaling}, we show the accuracy for the moment with $j=3$ as
a function of $N_{\text{fc}} = s \ms N_{\text{steps}}$, where $s$ is the number
of stages of the RK method and $N_{\text{steps}}$ the number of RK steps between
$t_0$ to $t_f$.  Hence, $N_{\text{fc}}$ is the total number of calls to the
function $F(t,y)$ during evolution.\footnote{Note that $F(t,y)$ is vector valued
in the discretized DGLAP equation \protect\eqref{eq:dglap_discretized}.  Its
evaluation accounts for the bulk of the computational cost of evolution.}
The accuracy is determined from the difference between the solution at the
selected $N_{\text{fc}}$ and the solution at a value of $N_{\text{fc}}$ in the
region where the result does not significantly change any more.

For each RK method, we can identify a region in which the accuracy follows a
power law $(N_{\text{fc}})^{-p}$.  We determine the corresponding value $p$ by a
fit and find it in reasonable agreement with the order of the method.  This
indicates that in the corresponding region of $N_{\text{fc}}$ the error estimate
in \eq{rk_error} is indeed applicable.

Our investigation shows the significant advantage of RK methods with high order
$p$ for solving the DGLAP equations.  For the example shown in the figure, the
DOPRI8 method yields a relative accuracy below $10^{-8}$ with a single step
(corresponding to 13 function calls), and the limits of machine precision are
reached with 50 to 100 function calls.

A similar conclusion can be drawn for the computation of $\alpha_s(\mu)$ from
\eq{running_rge}.  To illustrate this, we evolve $\alpha_s(M_Z) = 0.118$ down to
$\mu_0 = 2.25 \GeV$ with $n_F = 5$.  We use the DOPRI8 method and impose a
maximum step size $h_{\text{max}} = 0.2$.  The evolved value $\alpha_s(\mu_0)$
is then inserted into the exact analytic expression of $\mu(\alpha_s)$ at the
appropriate perturbative order.  Both at NLO and at NNLO, we find that the
resulting value of $\mu$ agrees with $\mu_0$ at the level of $2 \times
10^{-15}$.

\addcontentsline{toc}{section}{References}
\bibliographystyle{jhep}
\bibliography{chilipdf}

\providecommand{\href}[2]{#2}\begingroup\raggedright\begin{thebibliography}{10}

\bibitem{Buckley:2014ana}
A.~Buckley, J.~Ferrando, S.~Lloyd, K.~Nordstr{\"o}m, B.~Page, M.~R{\"u}fenacht
  et~al., \emph{{LHAPDF6: parton density access in the LHC precision era}},
  \href{https://doi.org/10.1140/epjc/s10052-015-3318-8}{\emph{Eur. Phys. J.}
  {\bfseries C75} (2015) 132}
  [\href{https://arxiv.org/abs/1412.7420}{{\ttfamily 1412.7420}}].

\bibitem{Yndurain:1977wz}
F.~J. Yndurain, \emph{{Reconstruction of the Deep Inelastic Structure Functions
  from their Moments}},
  \href{https://doi.org/10.1016/0370-2693(78)90062-X}{\emph{Phys. Lett. B}
  {\bfseries 74} (1978) 68}.

\bibitem{Parisi:1978jv}
G.~Parisi and N.~Sourlas, \emph{{A Simple Parametrization of the $Q^2$
  Dependence of the Quark Distributions in {QCD}}},
  \href{https://doi.org/10.1016/0550-3213(79)90448-6}{\emph{Nucl. Phys. B}
  {\bfseries 151} (1979) 421}.

\bibitem{Furmanski:1981ja}
W.~Furmanski and R.~Petronzio, \emph{{A Method of Analyzing the Scaling
  Violation of Inclusive Spectra in Hard Processes}},
  \href{https://doi.org/10.1016/0550-3213(82)90398-4}{\emph{Nucl. Phys. B}
  {\bfseries 195} (1982) 237}.

\bibitem{Kobayashi:1994hy}
R.~Kobayashi, M.~Konuma and S.~Kumano, \emph{{FORTRAN program for a numerical
  solution of the nonsinglet Altarelli-Parisi equation}},
  \href{https://doi.org/10.1016/0010-4655(94)00159-Y}{\emph{Comput. Phys.
  Commun.} {\bfseries 86} (1995) 264}
  [\href{https://arxiv.org/abs/hep-ph/9409289}{{\ttfamily hep-ph/9409289}}].

\bibitem{Chyla:1986eb}
J.~Chyla and J.~Rames, \emph{{On Methods of Analyzing Scaling Violation in Deep
  Inelastic Scattering}}, \href{https://doi.org/10.1007/BF01559606}{\emph{Z.
  Phys. C} {\bfseries 31} (1986) 151}.

\bibitem{Blumlein:1989pd}
J.~Bl{\"u}mlein, M.~Klein, G.~Ingelman and R.~R{\"u}ckl, \emph{{Testing {QCD}
  Scaling Violations in the {HERA} Energy Range}},
  \href{https://doi.org/10.1007/BF01549682}{\emph{Z. Phys. C} {\bfseries 45}
  (1990) 501}.

\bibitem{Krivokhizhin:1990ct}
V.~G. Krivokhizhin, S.~P. Kurlovich, R.~Lednicky, S.~Nemecek, V.~V. Sanadze,
  I.~A. Savin et~al., \emph{{Next-to-leading order QCD analysis of structure
  functions with the help of Jacobi polynomials}},
  \href{https://doi.org/10.1007/BF01554485}{\emph{Z. Phys. C} {\bfseries 48}
  (1990) 347}.

\bibitem{Bonvini:2010tp}
M.~Bonvini, S.~Forte and G.~Ridolfi, \emph{{Soft gluon resummation of Drell-Yan
  rapidity distributions: Theory and phenomenology}},
  \href{https://doi.org/10.1016/j.nuclphysb.2011.01.023}{\emph{Nucl. Phys. B}
  {\bfseries 847} (2011) 93} [\href{https://arxiv.org/abs/1009.5691}{{\ttfamily
  1009.5691}}].

\bibitem{Bonvini:2014joa}
M.~Bonvini and S.~Marzani, \emph{{Resummed Higgs cross section at N$^{3}$LL}},
  \href{https://doi.org/10.1007/JHEP09(2014)007}{\emph{JHEP} {\bfseries 09}
  (2014) 007} [\href{https://arxiv.org/abs/1405.3654}{{\ttfamily 1405.3654}}].

\bibitem{Pumplin:2009bb}
J.~Pumplin, \emph{{Parametrization dependence and $\Delta \chi^2$ in parton
  distribution fitting}},
  \href{https://doi.org/10.1103/PhysRevD.82.114020}{\emph{Phys. Rev. D}
  {\bfseries 82} (2010) 114020}
  [\href{https://arxiv.org/abs/0909.5176}{{\ttfamily 0909.5176}}].

\bibitem{Glazov:2010bw}
A.~Glazov, S.~Moch and V.~Radescu, \emph{{Parton Distribution Uncertainties
  using Smoothness Prior}},
  \href{https://doi.org/10.1016/j.physletb.2010.11.025}{\emph{Phys. Lett. B}
  {\bfseries 695} (2011) 238}
  [\href{https://arxiv.org/abs/1009.6170}{{\ttfamily 1009.6170}}].

\bibitem{Martin:2012da}
A.~D. Martin, A.~J. T.~M. Mathijssen, W.~J. Stirling, R.~S. Thorne, B.~J.~A.
  Watt and G.~Watt, \emph{{Extended Parameterisations for MSTW PDFs and their
  effect on Lepton Charge Asymmetry from W Decays}},
  \href{https://doi.org/10.1140/epjc/s10052-013-2318-9}{\emph{Eur. Phys. J. C}
  {\bfseries 73} (2013) 2318}
  [\href{https://arxiv.org/abs/1211.1215}{{\ttfamily 1211.1215}}].

\bibitem{Harland-Lang:2014zoa}
L.~A. Harland-Lang, A.~D. Martin, P.~Motylinski and R.~S. Thorne, \emph{{Parton
  distributions in the LHC era: MMHT 2014 PDFs}},
  \href{https://doi.org/10.1140/epjc/s10052-015-3397-6}{\emph{Eur. Phys. J.}
  {\bfseries C75} (2015) 204}
  [\href{https://arxiv.org/abs/1412.3989}{{\ttfamily 1412.3989}}].

\bibitem{Dulat:2017prg}
F.~Dulat, B.~Mistlberger and A.~Pelloni, \emph{{Differential Higgs production
  at N$^{3}$LO beyond threshold}},
  \href{https://doi.org/10.1007/JHEP01(2018)145}{\emph{JHEP} {\bfseries 01}
  (2018) 145} [\href{https://arxiv.org/abs/1710.03016}{{\ttfamily
  1710.03016}}].

\bibitem{Miyama:1995bd}
M.~Miyama and S.~Kumano, \emph{{Numerical solution of $Q^2$ evolution equations
  in a brute force method}},
  \href{https://doi.org/10.1016/0010-4655(96)00013-6}{\emph{Comput. Phys.
  Commun.} {\bfseries 94} (1996) 185}
  [\href{https://arxiv.org/abs/hep-ph/9508246}{{\ttfamily hep-ph/9508246}}].

\bibitem{Ratcliffe:2000kp}
P.~G. Ratcliffe, \emph{{A matrix approach to numerical solution of the DGLAP
  evolution equations}},
  \href{https://doi.org/10.1103/PhysRevD.63.116004}{\emph{Phys. Rev. D}
  {\bfseries 63} (2001) 116004}
  [\href{https://arxiv.org/abs/hep-ph/0012376}{{\ttfamily hep-ph/0012376}}].

\bibitem{Pascaud:2001bi}
C.~Pascaud and F.~Zomer, \emph{{A Fast and precise method to solve the
  Altarelli-Parisi equations in x space}},
  \href{https://arxiv.org/abs/hep-ph/0104013}{{\ttfamily hep-ph/0104013}}.

\bibitem{Dasgupta:2001eq}
M.~Dasgupta and G.~Salam, \emph{{Resummation of the jet broadening in DIS}},
  \href{https://doi.org/10.1007/s100520200915}{\emph{Eur. Phys. J. C}
  {\bfseries 24} (2002) 213}
  [\href{https://arxiv.org/abs/hep-ph/0110213}{{\ttfamily hep-ph/0110213}}].

\bibitem{DelDebbio:2007ee}
{\scshape NNPDF} collaboration, L.~Del~Debbio, S.~Forte, J.~I. Latorre,
  A.~Piccione and J.~Rojo, \emph{{Neural network determination of parton
  distributions: The Nonsinglet case}},
  \href{https://doi.org/10.1088/1126-6708/2007/03/039}{\emph{JHEP} {\bfseries
  03} (2007) 039} [\href{https://arxiv.org/abs/hep-ph/0701127}{{\ttfamily
  hep-ph/0701127}}].

\bibitem{Weinzierl:2002mv}
S.~Weinzierl, \emph{{Fast evolution of parton distributions}},
  \href{https://doi.org/10.1016/S0010-4655(02)00584-2}{\emph{Comput. Phys.
  Commun.} {\bfseries 148} (2002) 314}
  [\href{https://arxiv.org/abs/hep-ph/0203112}{{\ttfamily hep-ph/0203112}}].

\bibitem{Vogt:2004ns}
A.~Vogt, \emph{{Efficient evolution of unpolarized and polarized parton
  distributions with QCD-PEGASUS}},
  \href{https://doi.org/10.1016/j.cpc.2005.03.103}{\emph{Comput. Phys. Commun.}
  {\bfseries 170} (2005) 65}
  [\href{https://arxiv.org/abs/hep-ph/0408244}{{\ttfamily hep-ph/0408244}}].

\bibitem{Candido:2022tld}
A.~Candido, F.~Hekhorn and G.~Magni, \emph{{EKO: Evolution Kernel Operators}},
  \href{https://arxiv.org/abs/2202.02338}{{\ttfamily 2202.02338}}.

\bibitem{Cafarella:2008du}
A.~Cafarella, C.~Coriano and M.~Guzzi, \emph{{Precision Studies of the NNLO
  DGLAP Evolution at the LHC with CANDIA}},
  \href{https://doi.org/10.1016/j.cpc.2008.06.004}{\emph{Comput. Phys. Commun.}
  {\bfseries 179} (2008) 665}
  [\href{https://arxiv.org/abs/0803.0462}{{\ttfamily 0803.0462}}].

\bibitem{Salam:2008qg}
G.~P. Salam and J.~Rojo, \emph{{A Higher Order Perturbative Parton Evolution
  Toolkit (HOPPET)}},
  \href{https://doi.org/10.1016/j.cpc.2008.08.010}{\emph{Comput. Phys. Commun.}
  {\bfseries 180} (2009) 120}
  [\href{https://arxiv.org/abs/0804.3755}{{\ttfamily 0804.3755}}].

\bibitem{Botje:2010ay}
M.~Botje, \emph{{QCDNUM: Fast QCD Evolution and Convolution}},
  \href{https://doi.org/10.1016/j.cpc.2010.10.020}{\emph{Comput. Phys. Commun.}
  {\bfseries 182} (2011) 490}
  [\href{https://arxiv.org/abs/1005.1481}{{\ttfamily 1005.1481}}].

\bibitem{Bertone:2013vaa}
V.~Bertone, S.~Carrazza and J.~Rojo, \emph{{APFEL: A PDF Evolution Library with
  QED corrections}},
  \href{https://doi.org/10.1016/j.cpc.2014.03.007}{\emph{Comput. Phys. Commun.}
  {\bfseries 185} (2014) 1647}
  [\href{https://arxiv.org/abs/1310.1394}{{\ttfamily 1310.1394}}].

\bibitem{Bertone:2017gds}
V.~Bertone, \emph{{APFEL++: A new PDF evolution library in C++}},
  \href{https://doi.org/10.22323/1.297.0201}{\emph{PoS} {\bfseries DIS2017}
  (2018) 201} [\href{https://arxiv.org/abs/1708.00911}{{\ttfamily
  1708.00911}}].

\bibitem{Procura:2014cba}
M.~Procura, W.~J. Waalewijn and L.~Zeune, \emph{{Resummation of
  Double-Differential Cross Sections and Fully-Unintegrated Parton Distribution
  Functions}}, \href{https://doi.org/10.1007/JHEP02(2015)117}{\emph{JHEP}
  {\bfseries 02} (2015) 117} [\href{https://arxiv.org/abs/1410.6483}{{\ttfamily
  1410.6483}}].

\bibitem{Lustermans:2019plv}
G.~Lustermans, J.~K.~L. Michel, F.~J. Tackmann and W.~J. Waalewijn,
  \emph{{Joint two-dimensional resummation in $q_{T}$ and $0$-jettiness at
  NNLL}}, \href{https://doi.org/10.1007/JHEP03(2019)124}{\emph{JHEP} {\bfseries
  03} (2019) 124} [\href{https://arxiv.org/abs/1901.03331}{{\ttfamily
  1901.03331}}].

\bibitem{Lustermans:2019cau}
G.~Lustermans, J.~K.~L. Michel and F.~J. Tackmann, \emph{{Generalized Threshold
  Factorization with Full Collinear Dynamics}},
  \href{https://arxiv.org/abs/1908.00985}{{\ttfamily 1908.00985}}.

\bibitem{Gaunt:2014xxa}
J.~R. Gaunt and M.~Stahlhofen, \emph{{The Fully-Differential Quark Beam
  Function at NNLO}},
  \href{https://doi.org/10.1007/JHEP12(2014)146}{\emph{JHEP} {\bfseries 12}
  (2014) 146} [\href{https://arxiv.org/abs/1409.8281}{{\ttfamily 1409.8281}}].

\bibitem{Gaunt:2020xlc}
J.~R. Gaunt and M.~Stahlhofen, \emph{{The fully-differential gluon beam
  function at NNLO}},
  \href{https://doi.org/10.1007/JHEP07(2020)234}{\emph{JHEP} {\bfseries 07}
  (2020) 234} [\href{https://arxiv.org/abs/2004.11915}{{\ttfamily
  2004.11915}}].

\bibitem{Hornig:2017pud}
A.~Hornig, D.~Kang, Y.~Makris and T.~Mehen, \emph{{Transverse Vetoes with
  Rapidity Cutoff in SCET}},
  \href{https://doi.org/10.1007/JHEP12(2017)043}{\emph{JHEP} {\bfseries 12}
  (2017) 043} [\href{https://arxiv.org/abs/1708.08467}{{\ttfamily
  1708.08467}}].

\bibitem{Michel:2018hui}
J.~K.~L. Michel, P.~Pietrulewicz and F.~J. Tackmann, \emph{{Jet Veto
  Resummation with Jet Rapidity Cuts}},
  \href{https://doi.org/10.1007/JHEP04(2019)142}{\emph{JHEP} {\bfseries 04}
  (2019) 142} [\href{https://arxiv.org/abs/1810.12911}{{\ttfamily
  1810.12911}}].

\bibitem{Bonvini:2015pxa}
M.~Bonvini, A.~S. Papanastasiou and F.~J. Tackmann, \emph{{Resummation and
  matching of b-quark mass effects in $ b\overline{b}H $ production}},
  \href{https://doi.org/10.1007/JHEP11(2015)196}{\emph{JHEP} {\bfseries 11}
  (2015) 196} [\href{https://arxiv.org/abs/1508.03288}{{\ttfamily
  1508.03288}}].

\bibitem{Pietrulewicz:2017gxc}
P.~Pietrulewicz, D.~Samitz, A.~Spiering and F.~J. Tackmann,
  \emph{{Factorization and Resummation for Massive Quark Effects in Exclusive
  Drell-Yan}}, \href{https://doi.org/10.1007/JHEP08(2017)114}{\emph{JHEP}
  {\bfseries 08} (2017) 114}
  [\href{https://arxiv.org/abs/1703.09702}{{\ttfamily 1703.09702}}].

\bibitem{Moult:2016fqy}
I.~Moult, L.~Rothen, I.~W. Stewart, F.~J. Tackmann and H.~X. Zhu,
  \emph{{Subleading Power Corrections for N-Jettiness Subtractions}},
  \href{https://doi.org/10.1103/PhysRevD.95.074023}{\emph{Phys. Rev. D}
  {\bfseries 95} (2017) 074023}
  [\href{https://arxiv.org/abs/1612.00450}{{\ttfamily 1612.00450}}].

\bibitem{Moult:2017jsg}
I.~Moult, L.~Rothen, I.~W. Stewart, F.~J. Tackmann and H.~X. Zhu, \emph{{N
  -jettiness subtractions for $gg\to H$ at subleading power}},
  \href{https://doi.org/10.1103/PhysRevD.97.014013}{\emph{Phys. Rev. D}
  {\bfseries 97} (2018) 014013}
  [\href{https://arxiv.org/abs/1710.03227}{{\ttfamily 1710.03227}}].

\bibitem{Ebert:2018lzn}
M.~A. Ebert, I.~Moult, I.~W. Stewart, F.~J. Tackmann, G.~Vita and H.~X. Zhu,
  \emph{{Power Corrections for N-Jettiness Subtractions at ${\cal
  O}(\alpha_s)$}}, \href{https://doi.org/10.1007/JHEP12(2018)084}{\emph{JHEP}
  {\bfseries 12} (2018) 084}
  [\href{https://arxiv.org/abs/1807.10764}{{\ttfamily 1807.10764}}].

\bibitem{Boughezal:2016zws}
R.~Boughezal, X.~Liu and F.~Petriello, \emph{{Power Corrections in the
  N-jettiness Subtraction Scheme}},
  \href{https://doi.org/10.1007/JHEP03(2017)160}{\emph{JHEP} {\bfseries 03}
  (2017) 160} [\href{https://arxiv.org/abs/1612.02911}{{\ttfamily
  1612.02911}}].

\bibitem{Boughezal:2018mvf}
R.~Boughezal, A.~Isgr\`o and F.~Petriello, \emph{{Next-to-leading-logarithmic
  power corrections for $N$-jettiness subtraction in color-singlet
  production}}, \href{https://doi.org/10.1103/PhysRevD.97.076006}{\emph{Phys.
  Rev. D} {\bfseries 97} (2018) 076006}
  [\href{https://arxiv.org/abs/1802.00456}{{\ttfamily 1802.00456}}].

\bibitem{Bhattacharya:2018vph}
A.~Bhattacharya, I.~Moult, I.~W. Stewart and G.~Vita, \emph{{Helicity Methods
  for High Multiplicity Subleading Soft and Collinear Limits}},
  \href{https://doi.org/10.1007/JHEP05(2019)192}{\emph{JHEP} {\bfseries 05}
  (2019) 192} [\href{https://arxiv.org/abs/1812.06950}{{\ttfamily
  1812.06950}}].

\bibitem{Ebert:2018gsn}
M.~A. Ebert, I.~Moult, I.~W. Stewart, F.~J. Tackmann, G.~Vita and H.~X. Zhu,
  \emph{{Subleading power rapidity divergences and power corrections for
  q$_{T}$}}, \href{https://doi.org/10.1007/JHEP04(2019)123}{\emph{JHEP}
  {\bfseries 04} (2019) 123}
  [\href{https://arxiv.org/abs/1812.08189}{{\ttfamily 1812.08189}}].

\bibitem{Ball:2016spl}
R.~D. Ball, E.~R. Nocera and J.~Rojo, \emph{{The asymptotic behaviour of parton
  distributions at small and large $x$}},
  \href{https://doi.org/10.1140/epjc/s10052-016-4240-4}{\emph{Eur. Phys. J. C}
  {\bfseries 76} (2016) 383}
  [\href{https://arxiv.org/abs/1604.00024}{{\ttfamily 1604.00024}}].

\bibitem{Diehl:2017wew}
M.~Diehl and J.~R. Gaunt, \emph{{Double parton scattering theory overview}},
  \href{https://doi.org/10.1142/9789813227767_0002}{\emph{Adv. Ser. Direct.
  High Energy Phys.} {\bfseries 29} (2018) 7}
  [\href{https://arxiv.org/abs/1710.04408}{{\ttfamily 1710.04408}}].

\bibitem{Carrazza:2020qwu}
S.~Carrazza, J.~M. Cruz-Martinez and M.~Rossi, \emph{{PDFFlow: Parton
  distribution functions on GPU}},
  \href{https://doi.org/10.1016/j.cpc.2021.107995}{\emph{Comput. Phys. Commun.}
  {\bfseries 264} (2021) 107995}
  [\href{https://arxiv.org/abs/2009.06635}{{\ttfamily 2009.06635}}].

\bibitem{Runge:1901}
C.~Runge, \emph{{\"Uber empirische Funktionen und die Interpolation zwischen
  \"aquidistanten Ordinaten}}, {\emph{Z. Math. Phys.} {\bfseries 46} (1901)
  224}.

\bibitem{Trefethen11sixmyths}
L.~N. Trefethen, ``{Six Myths of Polynomial Interpolation and Quadrature}.''
  \url{https://people.maths.ox.ac.uk/trefethen/mythspaper.pdf}, 2011.

\bibitem{Trefethen}
L.~N. Trefethen, \emph{{Approximation Theory and Approximation Practice}}.
  Society for Industrial and Applied Mathematics, 2012.

\bibitem{Trefethen:2008sia}
L.~N. Trefethen, \emph{{Is Gauss Quadrature Better than Clenshaw–Curtis?}},
  \href{https://doi.org/10.1137/060659831}{\emph{SIAM Review} {\bfseries 50}
  (2008) 67}.

\bibitem{Waldvogel:2006bit}
J.~Waldvogel, \emph{{Fast Construction of the Fej{\'e}r and Clenshaw–Curtis
  Quadrature Rules}},
  \href{https://doi.org/10.1007/s10543-006-0045-4}{\emph{Bit Numer. Math.}
  {\bfseries 46} (2006) 195}.

\bibitem{Gauss-Kronrod}
\emph{{Gauss-Kronrod quadrature formula}},  in \emph{{Encyclopedia of
  Mathematics}},
  \href{https://www.encyclopediaofmath.org/index.php/Gauss-Kronrod\_quadrature\_formula}{https://www.encyclopediaofmath.org/index.php/Gauss-Kronrod\_quadrature\_formula}.

\bibitem{Patterson:1968mat}
T.~N.~L. Patterson, \emph{{The optimum addition of points to quadrature
  formulae}}, \href{https://doi.org/10.1090/S0025-5718-68-99866-9}{\emph{Math.
  Comp.} {\bfseries 22} (1968) 847}.

\bibitem{Alekhin:2017kpj}
S.~Alekhin, J.~Bl{\"u}mlein, S.~Moch and R.~Placakyte, \emph{{Parton
  distribution functions, $\alpha_s$, and heavy-quark masses for LHC Run II}},
  \href{https://doi.org/10.1103/PhysRevD.96.014011}{\emph{Phys. Rev.}
  {\bfseries D96} (2017) 014011}
  [\href{https://arxiv.org/abs/1701.05838}{{\ttfamily 1701.05838}}].

\bibitem{Abramowicz:2015mha}
{\scshape H1, ZEUS} collaboration, H.~Abramowicz et~al., \emph{{Combination of
  measurements of inclusive deep inelastic ${e^{\pm }p}$ scattering cross
  sections and QCD analysis of HERA data}},
  \href{https://doi.org/10.1140/epjc/s10052-015-3710-4}{\emph{Eur. Phys. J.}
  {\bfseries C75} (2015) 580}
  [\href{https://arxiv.org/abs/1506.06042}{{\ttfamily 1506.06042}}].

\bibitem{Jimenez-Delgado:2014twa}
P.~Jimenez-Delgado and E.~Reya, \emph{{Delineating parton distributions and the
  strong coupling}},
  \href{https://doi.org/10.1103/PhysRevD.89.074049}{\emph{Phys. Rev.}
  {\bfseries D89} (2014) 074049}
  [\href{https://arxiv.org/abs/1403.1852}{{\ttfamily 1403.1852}}].

\bibitem{Ball:2017nwa}
{\scshape NNPDF} collaboration, R.~D. Ball et~al., \emph{{Parton distributions
  from high-precision collider data}},
  \href{https://doi.org/10.1140/epjc/s10052-017-5199-5}{\emph{Eur. Phys. J. C}
  {\bfseries 77} (2017) 663}
  [\href{https://arxiv.org/abs/1706.00428}{{\ttfamily 1706.00428}}].

\bibitem{Hou:2019efy}
T.-J. Hou et~al., \emph{{New CTEQ global analysis of quantum chromodynamics
  with high-precision data from the LHC}},
  \href{https://doi.org/10.1103/PhysRevD.103.014013}{\emph{Phys. Rev. D}
  {\bfseries 103} (2021) 014013}
  [\href{https://arxiv.org/abs/1912.10053}{{\ttfamily 1912.10053}}].

\bibitem{Bailey:2020ooq}
S.~Bailey, T.~Cridge, L.~A. Harland-Lang, A.~D. Martin and R.~S. Thorne,
  \emph{{Parton distributions from LHC, HERA, Tevatron and fixed target data:
  MSHT20 PDFs}},
  \href{https://doi.org/10.1140/epjc/s10052-021-09057-0}{\emph{Eur. Phys. J. C}
  {\bfseries 81} (2021) 341}
  [\href{https://arxiv.org/abs/2012.04684}{{\ttfamily 2012.04684}}].

\bibitem{Ball:2021leu}
R.~D. Ball et~al., \emph{{The Path to Proton Structure at One-Percent
  Accuracy}},  \href{https://arxiv.org/abs/2109.02653}{{\ttfamily 2109.02653}}.

\bibitem{Dulat:2015mca}
S.~Dulat, T.-J. Hou, J.~Gao, M.~Guzzi, J.~Huston, P.~Nadolsky et~al.,
  \emph{{New parton distribution functions from a global analysis of quantum
  chromodynamics}},
  \href{https://doi.org/10.1103/PhysRevD.93.033006}{\emph{Phys. Rev. D}
  {\bfseries 93} (2016) 033006}
  [\href{https://arxiv.org/abs/1506.07443}{{\ttfamily 1506.07443}}].

\bibitem{Vogt:2004mw}
A.~Vogt, S.~Moch and J.~A.~M. Vermaseren, \emph{{The Three-loop splitting
  functions in QCD: The Singlet case}},
  \href{https://doi.org/10.1016/j.nuclphysb.2004.04.024}{\emph{Nucl. Phys.}
  {\bfseries B691} (2004) 129}
  [\href{https://arxiv.org/abs/hep-ph/0404111}{{\ttfamily hep-ph/0404111}}].

\bibitem{Ablinger:2017tan}
J.~Ablinger, A.~Behring, J.~Bl{\"u}mlein, A.~De~Freitas, A.~von Manteuffel and
  C.~Schneider, \emph{{The three-loop splitting functions $P_{qg}^{(2)}$ and
  $P_{gg}^{(2, N_F)}$}},
  \href{https://doi.org/10.1016/j.nuclphysb.2017.06.004}{\emph{Nucl. Phys. B}
  {\bfseries 922} (2017) 1} [\href{https://arxiv.org/abs/1705.01508}{{\ttfamily
  1705.01508}}].

\bibitem{Ball:2014uwa}
{\scshape NNPDF} collaboration, R.~D. Ball et~al., \emph{{Parton distributions
  for the LHC Run II}},
  \href{https://doi.org/10.1007/JHEP04(2015)040}{\emph{JHEP} {\bfseries 04}
  (2015) 040} [\href{https://arxiv.org/abs/1410.8849}{{\ttfamily 1410.8849}}].

\bibitem{Dokshitzer:1977sg}
Y.~L. Dokshitzer, \emph{{Calculation of the Structure Functions for Deep
  Inelastic Scattering and e+ e- Annihilation by Perturbation Theory in Quantum
  Chromodynamics.}}, {\emph{Sov. Phys. JETP} {\bfseries 46} (1977) 641}.

\bibitem{Gribov:1972ri}
V.~N. Gribov and L.~N. Lipatov, \emph{{Deep inelastic e p scattering in
  perturbation theory}}, {\emph{Sov. J. Nucl. Phys.} {\bfseries 15} (1972)
  438}.

\bibitem{Altarelli:1977zs}
G.~Altarelli and G.~Parisi, \emph{{Asymptotic Freedom in Parton Language}},
  \href{https://doi.org/10.1016/0550-3213(77)90384-4}{\emph{Nucl. Phys.}
  {\bfseries B126} (1977) 298}.

\bibitem{Moch:2004pa}
S.~Moch, J.~A.~M. Vermaseren and A.~Vogt, \emph{{The Three loop splitting
  functions in QCD: The Nonsinglet case}},
  \href{https://doi.org/10.1016/j.nuclphysb.2004.03.030}{\emph{Nucl. Phys.}
  {\bfseries B688} (2004) 101}
  [\href{https://arxiv.org/abs/hep-ph/0403192}{{\ttfamily hep-ph/0403192}}].

\bibitem{Chetyrkin:1997sg}
K.~Chetyrkin, B.~A. Kniehl and M.~Steinhauser, \emph{{Strong coupling constant
  with flavor thresholds at four loops in the MS scheme}},
  \href{https://doi.org/10.1103/PhysRevLett.79.2184}{\emph{Phys. Rev. Lett.}
  {\bfseries 79} (1997) 2184}
  [\href{https://arxiv.org/abs/hep-ph/9706430}{{\ttfamily hep-ph/9706430}}].

\bibitem{Buza:1996wv}
M.~Buza, Y.~Matiounine, J.~Smith and W.~van Neerven, \emph{{Charm
  electroproduction viewed in the variable flavor number scheme versus fixed
  order perturbation theory}},
  \href{https://doi.org/10.1007/BF01245820}{\emph{Eur. Phys. J. C} {\bfseries
  1} (1998) 301} [\href{https://arxiv.org/abs/hep-ph/9612398}{{\ttfamily
  hep-ph/9612398}}].

\bibitem{Behring:2014eya}
A.~Behring, I.~Bierenbaum, J.~Blümlein, A.~De~Freitas, S.~Klein and
  F.~Wißbrock, \emph{{The logarithmic contributions to the $O(\alpha^3_s)$
  asymptotic massive Wilson coefficients and operator matrix elements in deeply
  inelastic scattering}},
  \href{https://doi.org/10.1140/epjc/s10052-014-3033-x}{\emph{Eur. Phys. J. C}
  {\bfseries 74} (2014) 3033}
  [\href{https://arxiv.org/abs/1403.6356}{{\ttfamily 1403.6356}}].

\bibitem{Giele:2002hx}
W.~Giele et~al., \emph{{The QCD / SM working group: Summary report}},  in
  \emph{{Physics at TeV colliders. Proceedings, Euro Summer School, Les
  Houches, France, May 21-June 1, 2001}}, pp.~275--426, 2002,
  \href{https://arxiv.org/abs/hep-ph/0204316}{{\ttfamily hep-ph/0204316}}.

\bibitem{Dittmar:2005ed}
M.~Dittmar et~al., \emph{{Working Group I: Parton distributions: Summary report
  for the HERA LHC Workshop Proceedings}},
  \href{https://arxiv.org/abs/hep-ph/0511119}{{\ttfamily hep-ph/0511119}}.

\bibitem{Salam:2019pri}
G.~P. Salam and A.~Vogt, \emph{private communication}.

\bibitem{RuizArriola:1998er}
E.~Ruiz~Arriola, \emph{{NLO evolution for large scale distances, positivity
  constraints and the low-energy model of the nucleon}},
  \href{https://doi.org/10.1016/S0375-9474(98)00489-8}{\emph{Nucl. Phys. A}
  {\bfseries 641} (1998) 461}.

\bibitem{Hairer:1993}
E.~Hairer, S.~P. N{\o}rsett and G.~Wanner, \emph{Solving Ordinary Differential
  Equations I: Nonstiff Problems}. Springer-Verlag, 1993.

\bibitem{Cash:1990}
J.~R. Cash and A.~H. Karp, \emph{A variable order runge-kutta method for
  initial value problems with rapidly varying right-hand sides},
  \href{https://doi.org/10.1145/79505.79507}{\emph{ACM Trans. Math. Softw.}
  {\bfseries 16} (1990) 201}.

\bibitem{DORMAND198019}
J.~Dormand and P.~Prince, \emph{A family of embedded runge-kutta formulae},
  \href{https://doi.org/https://doi.org/10.1016/0771-050X(80)90013-3}{\emph{Journal
  of Computational and Applied Mathematics} {\bfseries 6} (1980) 19 }.

\bibitem{PRINCE198167}
P.~Prince and J.~Dormand, \emph{High order embedded runge-kutta formulae},
  \href{https://doi.org/https://doi.org/10.1016/0771-050X(81)90010-3}{\emph{Journal
  of Computational and Applied Mathematics} {\bfseries 7} (1981) 67 }.

\end{thebibliography}\endgroup

\end{document}